\documentclass[12pt]{article}
\usepackage[english]{babel}
\usepackage[font=footnotesize]{caption}
\usepackage{graphicx}
\usepackage{epstopdf}
\usepackage[mathscr]{euscript}
\usepackage{enumerate}
\usepackage{dcolumn}
\usepackage{soul}
\usepackage{rotating}
\usepackage{dsfont}
\usepackage{tabularx}
\usepackage{lscape}
\usepackage{placeins}
\usepackage{longtable}
 \usepackage{setspace}
\usepackage{threeparttable}
\usepackage[centertags,sumlimits, intlimits, namelimits,]{amsmath}
\usepackage{amsfonts}
\usepackage[english]{babel}
\usepackage{amssymb}
\usepackage{appendix}
\usepackage{fancyhdr}
\usepackage{booktabs}
\usepackage{makecell}
\usepackage[normalem]{ulem}
\usepackage{natbib}
\usepackage[applemac]{inputenc}
\usepackage{url}
\usepackage{longtable}
\usepackage[small]{titlesec}
\usepackage{geometry}
\usepackage{eurosym}
\usepackage{comment}
\usepackage{adjustbox}
\usepackage{array}
\usepackage{booktabs}
\usepackage{multirow}
\newcolumntype{R}[2]{%
    >{\adjustbox{angle=#1,lap=\width-(#2)}\bgroup}%
    l%
    <{\egroup}%
}
\newcommand*\rot{\multicolumn{1}{R{90}{1em}}}% no optional argument here, please!

\usepackage{rotating}
\newcounter{daggerfootnote}
\newcommand*{\daggerfootnote}[1]{%
    \setcounter{daggerfootnote}{\value{footnote}}%
    \renewcommand*{\thefootnote}{\fnsymbol{footnote}}%
    \footnote[2]{#1}%
    \setcounter{footnote}{\value{daggerfootnote}}%
    \renewcommand*{\thefootnote}{\arabic{footnote}}%
    }

\usepackage{xcolor}
\definecolor{mycolor}{rgb}{0.55, 0.0, 0.0}
\usepackage[hidelinks]{hyperref}\hypersetup{colorlinks=true,linkcolor=mycolor,citecolor=mycolor,filecolor=magenta,urlcolor=mycolor,pdfpagemode = UseNone}

\def\sym#1{\ifmmode^{#1}\else\(^{#1}\)\fi}
\renewcommand\abstract{{\noindent\bfseries Abstract\\[1.5ex]}}
\def\keywordname{{\bfseries Keywords}}%
\def\keywords#1{\par\addvspace\medskipamount{\rightskip=0pt plus1cm
\def\and{\ifhmode\unskip\nobreak\fi\ $\cdot$
}\noindent\keywordname\enspace\ignorespaces#1\par}}
\def\JELname{{\bfseries JEL Classification}\enspace}
\def\JEL#1{\par\addvspace\medskipamount{\rightskip=0pt plus1cm
\def\and{\ifhmode\unskip\nobreak\fi\ $\cdot$
}\noindent\JELname\ignorespaces#1\par}}
\def\ackname{Acknowledgments}%
\def\acknowledgement{\par\addvspace{17pt}\small\rmfamily
\trivlist\if!\ackname!\item[]\else
\item[\hskip\labelsep
{\bfseries\ackname}]\fi}
\def\endacknowledgement{\endtrivlist\addvspace{6pt}}

\renewcommand{\tablename}{TABLE}
\renewcommand\thetable{\arabic{table}}
\def\fnum@table{\tablename\nobreakspace\thetable}
\renewcommand{\sectionmark}[1]{\markboth{#1}{}}
\newcommand\blfootnote[1]{%
  \begingroup
  \renewcommand\thefootnote{}\footnote{#1}%
  \addtocounter{footnote}{-1}%
  \endgroup
}

\begin{document}

%-------------------------------------------
\title{\vspace{-0mm} Individual characteristics associated with risk and time preferences: A multi country representative survey% 
%--------Anonymize--------------------------
\footnote{%Comments are solicited.
Corresponding author: Thomas Meissner (\href{mailto: meissnet@gmail.com}{meissnet@gmail.com}). This research was funded by the European Union's Horizon 2020 Framework Program under the project BRISKEE -- Behavioral Response to Investment Risks in Energy Efficiency (project number 649875).}
%--------Anonymize--------------------------
}
\author{
\normalsize\begin{tabular}{c}
%--------Anonymize--------------------------
Thomas Meissner\footnote{Maastricht University, School of Business and Economics, Tongersestraat 53, 6211 LM Maastricht, The Netherlands}\\
Xavier Gassmann\footnote{CEREN, EA 7477, Burgundy School of Business - Universite Bourgogne Franche-Comte, 29 rue Sambin, Dijon 21000, France}\\
Corinne Faure\footnote{Grenoble Ecole de Management, 38000 Grenoble, France}\\
Joachim Schleich\footnote{Grenoble Ecole de Management, 38000 Grenoble, France \& Fraunhofer Institute for Systems and Innovation Research, 76139 Karlsruhe, Germany}\\
[2ex]
%--------Anonymize--------------------------
%{\date{\begin{small}\today\end{small}\\[6ex]PRELIMINARY VERSION\\[2ex]Please do not cite or circulate.\\Comments are solicited.\vspace*{5mm}
%}
%}
\end{tabular}
}
%-------------------------------------------
\maketitle
%-------------------------------------------
\bigskip
\bigskip

\renewcommand{\baselinestretch}{1} \normalsize
\begin{abstract}
\noindent 
This paper empirically analyzes how individual characteristics are associated with risk aversion, loss aversion, time discounting, and present bias. To this end, we conduct a large-scale demographically representative survey across eight European countries. We elicit preferences using incentivized multiple price lists and jointly estimate preference parameters to account for their structural dependencies. Our findings suggest that preferences are linked to a variety of individual characteristics such as age, gender, and income as well as some personal values. We also report evidence on the relationship between cognitive ability and preferences. Incentivization, stake size, and the order of presentation of binary choices matter, underlining the importance of controlling for these factors when eliciting economic preferences.
\JEL{D01 \and D03 \and C91}
\end{abstract}
%-------------------------------------------
%\renewcommand{\baselinestretch}{1.45} \normalsize
%\setlength{\footnotesep}{0.75\baselineskip}
%-------------------------------------------
\newpage
\onehalfspacing

\section{Introduction}

Economic preferences, such as risk aversion and time preferences, have been found to predict a wide range of individual decisions, such as savings \citep[e.g.][]{Bradford2017}, environmental choices \citep[e.g.][]{Bartczak2015}, and investments in health \citep[e.g.][]{Galizzi2018} or in retirement funds \citep[e.g.][]{Goda2019}. For policymakers, it is particularly important to identify individual characteristics associated with such preferences, so that policies can be designed for the appropriate target groups (for instance, offering upfront subsidies to socio-demographic groups known to discount the future highly or warranties to socio-demographic groups known to be particularly risk averse).

Relationships between individual characteristics and economic preferences have been studied extensively in empirical research. The results reported in these studies are often inconsistent, however, making it difficult to derive clear insights and policies. The inconsistencies may stem from a variety of factors. First, studies make use of vastly different methods to elicit and estimate preferences; some studies use incentivized experimental methods \citep[e.g.][]{Boschini2019, Haridon2019}, while others rely on self-reported measures \citep[e.g.][]{Falk2018, Gorlitz2020}. Second, studies make use of different samples: while some studies utilize large scale, demographically representative samples, many rely on small samples consisting predominantly of students.\footnote{Recent exceptions
include \cite{Falk2018} and \cite{Bouchouicha2019}, who use multi-country representative samples, \citet{Rieger2015} and \citet{Wang2016}, who use large student samples, or \citet{Rosch2017}, who employs a large sample of farmers.} 
Finally, studies differ in how they account for structural dependencies between different domains of preferences. \citet{Andersen2008}, for instance, argue that the curvature of utility should be taken into account when estimating discount rates. \citet{Abdellaoui2007} show that failing to account for loss aversion can introduce bias in the estimated parameter of risk aversion. %{\tm Mention that in this paper we test a bunch of different structural assumptions} %Similarly, not accounting for probability weighting may affect estimates of risk preferences \citep{Bouchouicha2019, Abdellaoui2019}.\footnote{We discuss the implications for our estimation in Section~\ref{sec:robustness_utility}.}

In this paper, we start with a broad review of the empirical literature on the relationships between the most studied individual characteristics with risk aversion, loss aversion, time discounting and present bias.\footnote{To simplify the exposition, in the remainder of this paper we will categorize risk aversion and loss aversion as risk preferences, and time discounting and present bias as time preferences.} This literature review enables us to identify relationships for which patterns of findings are clear, ambiguous, inconsistent, or missing.  As our main contribution, we then present results from a large-scale multi-country study (with over 12,000 respondents) covering demographically representative samples in eight European countries, eliciting risk aversion, loss aversion, time discounting, and present bias, using state-of-the-art methods for elicitation and estimation as well as a wide range of robustness checks. Preferences are elicited using Multiple Price List (MPL) designs, as introduced by \citet{Holt2002} for risk preferences, and by \citet{Coller1999} for time preferences. Multiple price lists are incentive-compatible and easy to explain and understand; they also make it possible to elicit risk aversion, loss aversion, time discounting, and present bias using the same design. We use real monetary incentives and account for stake and order effects.\footnote{Specifically, we use a between-subjects random incentivized system (BRIS). See Section \ref{sec:incentives} for more details.} In our preferred specification, we jointly estimate preference parameters to account for their structural dependencies; we also conduct a variety of alternative estimations to examine the robustness of the findings to different specifications, including different ways of modeling the structural dependencies between risk and time preferences. This study employs a rich set of individual characteristics with a wide range of socio-demographic characteristics as well as psychological characteristics such as cognitive reflection \citep{Frederick2005} and cultural values \citep{Schwartz2012}, allowing us to analyze how these characteristics are related to risk and time preferences.

Overall, this paper aims at obtaining a better understanding of relationships between individual characteristics and risk and time preferences. To our knowledge, this is the first effort to elicit risk aversion, loss aversion, time discounting, and present bias jointly using multi-country representative samples. The findings contribute to the lively discussion on how risk and time preferences are associated with individual characteristics. Our literature review contributes to the literature through the identification of relationships for which more knowledge is needed (either because previous results are ambiguous or inconsistent or because these relationships have rarely or never been studied previously). The findings from our large-sample multi-country survey then provide valuable orientation for these relationships.
Table \ref{tab:sumfind} summarizes our main findings. To highlight some of the results, we find that age and gender correlate with all considered preferences. Income appears to be negatively correlated with risk and loss aversion. We also find robust negative correlations between cognitive ability and risk aversion and time discounting. Interestingly, loss aversion appears to be positively correlated with cognitive ability. 

Additional findings suggest that design features appear to have significant effects on the estimated parameters: incentivized respondents appear to be less risk averse, present biased, and loss averse, but they tend to discount the future to a greater extent. Stake size as well as order of presentation are also significantly related to elicited preference parameters. In line with mounting evidence on this topic, we find that controlling for decision noise is important:\footnote{See e.g. \cite{Andersson2016, Gillen2019, Andersson2020}.} Failing to do so may lead to spurious correlations between preference parameters and individual characteristics that are correlated with decision noise.

\begin{table}
\centering
\scriptsize

     \caption{Correlations between individual characteristics and preferences \label{tab:sumfind}}
  \begin{tabular*}{\textwidth}{l @{\extracolsep{\fill}} ccccccccc}
        \toprule
Preferences	& 	Age&	Male &	Income&	Education&	Children &	Couple &	Cognitive  \\
	&  &		 &	&	&	 &	 &	ability \\
\midrule
Risk aversion		&	+	& -		&-		&	n.s.		&	n.s.		&	n.s.		&-		\\
Loss aversion		&	-		&	-		&	-		&	n.s.		&	n.s.		&	n.s.		&   +		\\
Time discounting		&	-		&	+	&	n.s.		&	-		&	+		&	n.s.		&	-		\\
Present	bias	&	-		&	+	&	n.s.		&	n.s.		&	n.s.		&	n.s.		&	n.s.		\\

  \bottomrule
    \end{tabular*}
    
\begin{tablenotes}
\item ``+'' and ``-'' refer to significantly positive and negative associations (at the 5\% level). ``n.s.'' refers to non-significant associations.
\end{tablenotes}

    \end{table}

In the following, in Section~\ref{sec:lit} we first systematically review the empirical literature on the relation between individual characteristics and risk and time preferences. We present the theoretical framework used in Section~\ref{sec:theory} and the survey design in Section~\ref{sec:design}. In Section~\ref{sec:results} we report the findings from the joint estimation of the preference parameters as well as results obtained from a series of robustness checks. We discuss the implications of our findings in Section~\ref{sec:conclusion}.

\section{Individual characteristics and preferences: a literature overview}
\label{sec:lit}
In this section, we first present a systematic overview of empirical studies focusing on the relation between risk and time preferences and the following individual characteristics: age, gender, income (or wealth), education, having children (or household size), living as a couple (or being married), and cognitive ability. 

Given the vast number of empirical studies reporting the correlation of individual characteristics with risk and time preferences, a systematic search was necessary. We conducted this search in two separate steps.\footnote{More details on the search criteria can be found in Appendix~\ref{sec:search}.} In a first step, to ensure that the studies that are most comparable to ours were included, we conducted a focused search for studies using demographically representative samples and relying on experimental methods to elicit preferences.\footnote{To elicit economic preferences, the main experimental designs are the Multiple Price List, Random Lottery Pair, Ordered Lottery Selection, Becker Degroot Marschak, and Trade-off designs (\citet{Harrison2008}). We also included the Convex Time Budget of \citet{Andreoni2012}.} This initial focused search yielded 437 results on the Google Scholar library database.
In a second step, we more broadly searched the literature for studies on risk and time preferences without restrictions on the elicitation methods used (therefore also including stated preferences) nor on the sampling (therefore also including non-demographically representative samples). This search yielded 16,800 results on the Google Scholar library database. For feasibility reasons, we decided to retain the first 1,000 results only (ordered by relevance).\footnote{According to the Google Scholar website: "Google Scholar aims to rank documents the way researchers do, weighing the full text of each document, where it was published, who it was written by, as well as how often and how recently it has been cited in other scholarly literature." We acknowledge that this classification may vary over time and may not reflect the relevance of a particular study in a field.}

After removing 283 duplicates, this two-step procedure left us with 1,154 studies to be manually screened. At that stage, we eliminated 1,018 %note double check the number at the end
studies that did not present empirical results (for instance review papers), and studies that did not include associations of risk and time preferences with individual characteristics. We also added eight relevant studies that had not been found through the search processes. Ultimately, we retained 144 %note double check the number at the end
studies in this review (listed in Appendix Table \ref{tab:overlit}).

For each study, we specify the sample size, whether the sample was demographically representative, and whether preference elicitation was incentivized. Finally, we indicate which preferences were considered in the study. For practical reasons we used p-values (here with a 10\% threshold) to assess the findings in the literature on the correlation between preferences and covariates. While incentives have been an important concern in the literature over the past two decades \citep[e.g.][]{Camerer1999, BranasGarza2020}, we observe that large-sample surveys typically do not use incentivization for preference elicitation; we also observe that large-scale studies such as \citet{Dohmen2010} tend to rely on scales rather than experimental elicitation methods. Among the studies retained, about 40\% rely on representative samples, the remaining studies often use student samples or special population samples (e.g. homeowners, farmers).  The majority of studies focus on risk aversion and to a lesser extent on time discounting; only a few studies have considered present bias and loss aversion. Further, most studies consider one preference domain only. Finally, we note that none of the studies considering several preference domains estimate risk and time preferences parameters jointly.\footnote{Studies that elicit risk aversion and loss aversion often estimate these two preferences jointly.} We summarize the findings of this literature review in Table~\ref{tab:overlit} without any judgment. For example, we do not distinguish between findings according to the methods used to elicit and estimate economic preferences.\footnote{This choice allowed us to present a comprehensive overview of the existing literature on preferences and their covariates. This would not have been possible had we excluded or focused on specific elicitation methods. Including different elicitation methods implies however that studies retained in the overview may not be directly comparable.} Similarly, even though omitted variable bias may explain differences across studies because some of the selected individual characteristics may be correlated but not included as covariates in some of the studies, we do not account for these differences and report the results as they appear in the published articles.

Overall, Table~\ref{tab:overlit} makes clear that there are no previous demographically representative studies incorporating risk aversion, loss aversion, time discounting and present bias.\footnote{ \citet{Tanaka2010} incorporates these four individual preferences on a sample of farmers in Vietnam and is therefore not demographically representative. \cite{Breuer2020} also elicit these four preferences on a student sample.} Our study fills this gap. Specifically, because we use state-of-the-art preference elicitation methods with demographically representative samples of the population in multiple countries, and because we include (and jointly estimate) time and risk preferences, loss aversion, and present bias, and include a wide range of individual characteristics, our study should yield valuable results regarding the individual characteristics associated with these preferences.

In Table~\ref{tab:litterature} we summarize the correlations found between individual characteristics and the preference parameters in the 144 papers retained in this overview. In particular, we summarize correlations between risk aversion, loss aversion, time discounting, and the main individual characteristics included in our study. We report in this table the number of studies that investigated this relationship and document a statistically significant (positive or negative) correlation at the 10\% level or a non-significant correlation. We separately identify the number of studies for each category that used a demographically representative sample, because most non-representative studies have been conducted with student samples which typically exhibit little variation for many socio-demographic variables. In addition, to ensure that small or under-powered studies are not driving the conclusions, Appendix Table~\ref{tab:litteraturesample} reports the number of observations finding specific relations. Together, these tables can be used to identify, for each relationship between individual characteristics and preferences, the extent to which the literature has found consistent, conflicting, or ambiguous (i.e. non-significant) results. 

\begin{table}
\centering
\scriptsize
\begin{threeparttable}[ht]
     \caption{Correlations between individual characteristics and preferences reported in the literature: number of studies}
  \begin{tabular*}{\textwidth}{c @{\extracolsep{\fill}} rccccccccc}
        \toprule
Preferences	& Relationship &	Age&	Male &	Income&	Education&	Children &	Couple &	Cognitive  \\
	&  &	&	 &	&	&	 &	 &	ability \\
\midrule
Risk	&	Positive	&	63	/	28	*	&	3	/	0		&	9	/	7		&	11	/	6		&	10	/	6		&	11	/	7		&	0	/	0		\\
aversion	&	n.s.	&	55	/	18		&	49	/	23		&	46	/	14		&	42	/	18	*	&	19	/	9	*	&	25	/	15	*	&	9	/	5		\\
	&	Negative	&	22	/	11		&	86	/	37	*	&	41	/	21	*	&	34	/	18		&	3	/	2		&	8	/	3		&	12	/	4	*	\\ \\
Loss	&	Positive	&	5	/	3		&	1	/	1		&	2	/	1		&	4	/	2		&	0	/	0		&	1	/	1		&	1	/	1	*	\\
aversion	&	n.s.	&	8	/	2		&	10	/	1		&	2	/	1		&	6	/	4	*	&	2	/	1		&	2	/	1	*	&	1	/	0		\\
	&	Negative	&	5	/	3	*	&	8	/	6	*	&	7	/	4	*	&	3	/	2		&	1	/	1	*	&	1	/	1		&	2	/	1		\\ \\
Time	&	Positive	&	5	/	2		&	3	/	1	*	&	0	/	0		&	0	/	0		&	3	/	1	*	&	1	/	1		&	0	/	0		\\
discounting	&	n.s.	&	13	/	8		&	29	/	11		&	12	/	6	*	&	8	/	5		&	7	/	5		&	9	/	7	*	&	2	/	1		\\
	&	Negative	&	11	/	3	*	&	6	/	4		&	12	/	5		&	12	/	7	*	&	1	/	0		&	3	/	2		&	11	/	4	*	\\ \\
Present	&	Positive	&	1	/	0		&	0	/	0	*	&	0	/	0		&	1	/	0		&	0	/	0		&	1	/	0		&	0	/	0		\\
bias	&	n.s.	&	4	/	0		&	7	/	2		&	4	/	0	*	&	3	/	1	*	&	0	/	0	*	&	1	/	1	*	&	2	/	1	*	\\
	&	Negative	&	4	/	0	*	&	0	/	0		&	3	/	0		&	0	/	0		&	1	/	0		&	1	/	0		&	2	/	0		\\

  \bottomrule
    \end{tabular*}
    
\begin{tablenotes}
\item In each cell, we first report the number of studies finding a specific result (at the 10\% level for positively or negatively significant correlations) followed by the number of representative studies amongst these studies. For studies that reported multiple specifications, we included the results from the most sophisticated specification; further, to mitigate omitted variable bias, we chose the specification that used the largest set of the covariates of interest. Papers that reported results for different elicitation tasks or/and different sub-samples were treated as separate studies. We used wealth in the few cases where income was not reported, number of persons in household when having children was not reported. ``Couple'' summarizes people who live as a couple or are married. For cognitive ability, we considered studies reporting the following measures: Cognitive Reflection Test, Raven's Matrix, Intelligence Quotient or scores to specifically test mathematics ability. 
\item  Results from the present study are not included in the numbers reported but are indicated by *.
\end{tablenotes}
    \label{tab:litterature}
    \end{threeparttable}
    \end{table}

For age, statistically significant findings for number of studies and number of observations mostly suggest a positive correlation with risk aversion and a negative correlation with time discounting and present bias; still, a substantial share of studies find opposite or ambiguous results, making the association between age and economic preferences somewhat contested. Previous literature provides inconclusive evidence for the relation between age and loss aversion. \footnote{Note that many studies have considered non-linear relationships for age; however, this has not lead to consistent findings.} 

For gender, a majority of studies find men to be less risk averse than women, but a substantial number of studies report a non-significant correlation. In comparison, most demographically representative studies find men to be less loss averse than women, while analyses relying on non-representative samples typically fail to find a statistically significant relation between gender and loss aversion.\footnote{Note that \citet{bouchouicha2019gender} suggest that the relationships between gender and loss aversion might depend on the specification used for loss aversion.} The patterns of results for the relationship between gender and time preferences appear ambiguous, with the majority of studies reporting non-significant results. However, men are clearly found to be more patient than women when the number of observations is considered instead of the number of studies. Notably, no study so far finds a significant correlation between present bias and gender. 

For income, statistically significant results in the literature typically suggest a negative relation with risk aversion, but the majority of findings are ambiguous, especially when accounting for sample size. In comparison, the majority of the few studies on loss aversion find richer households to be less loss averse. For time discounting and present bias, about equal shares of studies (and observations) find either a negative relation with income or a no result; for both time preferences, none of the studies included in our review finds a positive correlation with income. 

Next, a majority of studies find ambiguous results for the relation between education and risk aversion, loss aversion and present bias. Focusing on the number of studies, most statistically significant results suggest a negative relation between education and risk aversion; there are numerous studies, however, finding this relation to be positive. When taking into account the number of observations, the relationship between education and risk aversion appears to be ambiguous. For time discounting, most studies suggest a positive association with education; the share of non-significant results is also quite high. 

In the majority of studies, having children or living as a couple/being married are not correlated with risk aversion, loss aversion, and time discounting. Statistically significant results mostly suggest a positive relation of risk aversion and having children or being married, especially when taking into account the number of observations and not only the number of studies. 

Finally, most previous studies (also when accounting for sample sizes) find participants with higher cognitive ability to be less risk averse and more patient, whereas the associations of cognitive ability with loss aversion and present bias have not yet received much attention and results are mixed. We also note that a high share of studies find an ambiguous association between cognitive ability and risk aversion. Relatedly, \citet{Andersson2016, Andersson2020} suggest that the negative correlation between cognitive ability and risk aversion found in many studies may be spurious; we focus on this issue in Section~\ref{sec:RA}. 
In general, the pattern of results in Table~\ref{tab:litterature} appears to be similar for studies that employ demographically representative samples and studies that do not employ such samples. It is worth noting that some relations have been little studied (especially relations with loss aversion and present bias) and that some conclusions are based on less than ten studies and less than 10,000 observations in total. 

Overall, this literature review highlights the need for a study that includes the four preference domains and a broad set of individual characteristics. We find that many of the results are either conflicting or ambiguous (non-significant). Further, some relationships have rarely been studied so that empirical evidence on these relationships is scarce or missing. We believe that using a joint estimation of four preferences parameters on large, incentivized, and representative multi-country samples provides valuable insights on the relations between individual characteristics and risk and time preferences.

\section{Theoretical framework: preferences over risk and time}
\label{sec:theory}

In this section we present the theoretical framework used to model and estimate preferences over risk and time. The framework is based on the commonly used Prospect Theory framework developed by \citet{Kahneman1979}.\footnote{Note that for practical reasons, we decided to not elicit probability weighting. All lottery choices were 50/50 gambles, conveyed in everyday language as coin flips. See Section~\ref{sec:robustness_utility} for a discussion.} Utility over gains and losses of wealth $x$, relative to the reference point $(x=0)$, is modelled as:

\begin{equation}
\label{eq:utility}
u(x)=
\begin{cases}
\frac{x^{1-\alpha}}{1-\alpha} & \mbox{ if } x\geq 0\\
\frac{-\lambda(-x)^{1-\alpha}}{1-\alpha} & \mbox{ if } x < 0\\
\end{cases}
\end{equation}

where $\alpha$ is the parameter indicating relative risk aversion and $\lambda$ is the parameter determining loss aversion. Levels of $\alpha>0$ imply risk aversion, $\alpha=0$ implies risk neutrality, and $\alpha<0$ implies risk-seeking behavior. Levels of $\lambda>1$ imply loss aversion, $\lambda=1$ implies loss neutrality, and $\lambda<1$ implies loss-seeking behavior.

Preferences over choices that involve separate points in time were modelled using quasi-hyperbolic discounting, as proposed by \citet{Laibson1997}:

\begin{equation}
\label{eq:timepreferences}
U_t(x_t,...,x_T)=E_t\left[\frac{1}{(1+\delta)^t} u(x_t)+\frac{1}{(1+\gamma)}\sum^{T-t}_{k=1}\frac{1}{(1+\delta)^{t+k}} u(x_{t+k})\right],
\end{equation}

where $U_t(x_t,...,x_T)$ is the expected utility of a stream of wealth gains or losses $x_0,...,x_T$, $u(x_t )$ in gain/loss utility at time t as described in equation (1), $\delta$ is the discount rate and $\gamma$ is the present bias parameter. Here $\delta>0$ implies impatience, $\delta=0$ implies time neutrality, and $\delta<0$ implies patience, while $\gamma>0$ implies present bias, $\gamma=0$ implies no bias, and $\gamma<0$ implies future bias.\footnote{Our notation differs somewhat from the usual one, where the present bias parameter is typically written as $\beta=1/(1+\gamma)$. We have chosen this notation to achieve a uniform interpretation of coefficient estimates across preferences in Table~\ref{tab:main_full}: larger coefficients can always be interpreted to indicate \emph{increased} risk aversion, loss aversion, time discounting and present bias.} In our model, $t$ is expressed in years. Therefore, $\delta$ is the yearly discount rate.

\section{Survey design}
\label{sec:design}

Data was collected in July and August 2016 through computer-assisted web interviews (CAWI) conducted by IPSOS GmbH in eight European countries (France, Germany, Italy, Poland, Romania, Spain, Sweden, and the United Kingdom). Respondents were members of IPSOS household panels; quota sampling was used to ensure that for each country, the sample was representative of the population for gender, age (between 18 and 65 years), and region of origin. The sample consisted of 15,055 respondents with the following distribution: 2,000 respondents each from France, Italy, and the United Kingdom; 2,002 from Germany; 2,008 from Poland; 1,529 from Romania; 2,001 from Spain; and 1,515 from Sweden. We dropped 2,330 (113) observations for respondents who answered ``do not know'' to the income (education level) question. All following analyses are reported based on the remaining 12,612 respondents.
The surveys were carefully translated from the original English version into each of the target languages; back translation was used to check the quality of each translation; after discussion with the translators, discrepancies were adjusted for to ensure consistency in understanding across countries. Further, because not all countries in the survey use the same currency, the monetary amounts displayed to respondents throughout the survey were adjusted to keep similar purchasing power values across countries while also showing rounded numbers to ensure equivalent computing difficulty. In all Eurozone countries, the monetary amounts shown to respondents were identical; for Poland, Romania, Sweden and the UK, monetary amounts were multiplied by the following factors: Poland: 1\EUR{} = 3 PLN; Romania: 1\EUR{} = 3 RON; Sweden: 1\EUR{} = 10 SEK; UK: 1\EUR{} = 1\pounds.\footnote{The data used in this paper was collected as part of the Horizon 2020 BRISKEE project (\url{https://www.briskee-cheetah.eu}). The data is publicly available and several other publications have made use of it, mostly to investigate household adoption of energy efficient technologies \citep[see e.g.][]{Schleich2019}.}

\subsection{Preference elicitation}

Tables~\ref{tab:MPL1.1} -- \ref{tab:MPL3} contain the Multiple Price Lists (MPLs) used to elicit preferences. MPL1.1 and MPL1.2 involve secure gains at varying time frames: one at an earlier date and one at a later date. MPL2 contains 14 choices between two lotteries (one less risky lottery and one riskier lottery, whose expected value increases from upper to lower rows). Finally, MPL3 involves seven choices between two lotteries, including gains as well as losses as outcomes. Together, these price lists jointly identify risk aversion, loss aversion, time discounting, and present bias.

It should be noted that the earliest date at which respondents could receive money was one week into the future. This could be a concern, as one week could be too far in the future to be considered ``present'' for eliciting present bias. As explained in greater detail in the next subsection, this was because processing and shipping of payments took one week's time. In Section~\ref{sec:MLE} we discuss the implications of this procedure for our parameter estimates.

No unique switch point was enforced in our multiple price lists; i.e. respondents were free to switch back and forth between Option A and B. In our sample, 16.48\% of respondents have multiple switch points; this share is similar to previous studies \citep[see e.g.][]{ Harrison2005}.

\begin{table}[p]
\centering
\footnotesize
\caption{Multiple price list for eliciting time preferences (MPL1.1 \& MPL1.2) \label{tab:MPL1.1}}
\begin{tabular*}{\hsize}{@{\hskip\tabcolsep\extracolsep\fill}l*{4}{c}}
\toprule
        Line & Option A & Option B  & Implied $\delta^*$ \\
        \midrule
1 & Receive 98\euro{} in 6 months and one week  & Receive 100\euro{} in 12 months & 0.041\\
2 & Receive 94\euro{} in 6 months and one week  & Receive 100\euro{} in 12 months & 0.132\\
3 & Receive 90\euro{} in 6 months and one week  & Receive 100\euro{} in 12 months & 0.235\\
4 & Receive 86\euro{} in 6 months and one week  & Receive 100\euro{} in 12 months & 0.352\\
5 & Receive 80\euro{} in 6 months and one week  & Receive 100\euro{} in 12 months & 0.563\\
6 & Receive 70\euro{} in 6 months and one week  & Receive 100\euro{} in 12 months & 1.041\\
7 & Receive 55\euro{} in 6 months and one week  & Receive 100\euro{} in 12 months & 2.306\\
        \midrule
1 & Receive 98\euro{} in one week  & Receive 100\euro{} in 6 months &  0.041\\
2 & Receive 94\euro{} in one week  & Receive 100\euro{} in 6 months &  0.132\\
3 & Receive 90\euro{} in one week  & Receive 100\euro{} in 6 months &  0.235\\
4 & Receive 86\euro{} in one week  & Receive 100\euro{} in 6 months & 0.352\\
5 & Receive 80\euro{} in one week  & Receive 100\euro{} in 6 months & 0.563\\
6 & Receive 70\euro{} in one week  & Receive 100\euro{} in 6 months &  1.041\\
7 & Receive 55\euro{} in one week  & Receive 100\euro{} in 6 months &  2.306\\

\bottomrule
\multicolumn{4}{l}{\scriptsize *Yearly discount rate implied by indifference between Option A and B, assuming linear utility ($\alpha = 0$).}
\end{tabular*}
\end{table}

\begin{table}[p]
\footnotesize
\centering
\caption{Multiple price list for eliciting risk aversion (MPL2) \label{tab:MPL2}}
\begin{tabular*}{\hsize}{@{\hskip\tabcolsep\extracolsep\fill}l*{6}{c}}
\toprule
         & \multicolumn{2}{c}{Option A} & \multicolumn{2}{c}{Option B} & \\ \cmidrule(lr){2-3} \cmidrule(lr){4-5}

        Line & Coin shows Heads & Coin shows Tails & Coin shows Heads & Coin shows Tails & Implied $\alpha^*$\\
            \midrule
1 & 50\euro{} & 40\euro{} & 54\euro{} & 10\euro{} & -3.247\\
2 & 50\euro{} & 40\euro{} & 58\euro{} & 10\euro{} & -1.828\\
3 & 50\euro{} & 40\euro{} & 62\euro{} & 10\euro{} & -1.149\\
4 & 50\euro{} & 40\euro{} & 66\euro{} & 10\euro{} & -0.733\\
5 & 50\euro{} & 40\euro{} & 70\euro{} & 10\euro{} & -0.446\\
6 & 50\euro{} & 40\euro{} & 74\euro{} & 10\euro{} & -0.24\\
7 & 50\euro{} & 40\euro{} & 78\euro{} & 10\euro{} & -0.069\\
8 & 50\euro{} & 40\euro{} & 82\euro{} & 10\euro{} & 0.062\\
9 & 50\euro{} & 40\euro{} & 87\euro{} & 10\euro{} & 0.195\\
10 & 50\euro{} & 40\euro{} & 97\euro{} & 10\euro{} & 0.390\\
11 & 50\euro{} & 40\euro{} & 112\euro{} & 10\euro{} & 0.582\\
12 & 50\euro{} & 40\euro{} & 132\euro{} & 10\euro{} & 0.744\\
13 & 50\euro{} & 40\euro{} & 167\euro{} & 10\euro{} & 0.908\\
14 & 50\euro{} & 40\euro{} & 222\euro{} & 10\euro{} & 1.044\\

\bottomrule
\multicolumn{6}{l}{\scriptsize *Relative risk aversion parameter implied by indifference between Option A and B}
\end{tabular*}
\end{table}

\begin{table}[p]
\footnotesize
\centering
\caption{Multiple price list for eliciting loss aversion (MPL3) \label{tab:MPL3}}
\begin{tabular*}{\hsize}{@{\hskip\tabcolsep\extracolsep\fill}l*{6}{c}}
\toprule
         & \multicolumn{2}{c}{Option A} & \multicolumn{2}{c}{Option B} &  \\
        \cmidrule(lr){2-3} \cmidrule(lr){4-5}
    
        Line & Coin shows Heads & Coin shows Tails & Coin shows Heads & Coin shows Tails & Implied $\lambda^*$ \\
            \midrule
1 & +100\euro{} & -20\euro{} & +150\euro{} & -100\euro{} & 0.625\\
2 & +55\euro{} & -20\euro{} & +150\euro{} & -100\euro{}& 1.188\\
3 & +15\euro{} & -20\euro{} & +150\euro{} & -100\euro{} & 1.686\\
4 & +5\euro{} & -20\euro{} & +150\euro{} & -90\euro{} & 2.071\\
5 & +5\euro{} & -30\euro{} & +150\euro{} & -90\euro{} & 2.417\\
6 & +5\euro{} & -40\euro{} & +150\euro{} & -90\euro{}& 2.900\\
7 & +5\euro{} & -40\euro{} & +150\euro{} & -70\euro{} & 4.833\\

\bottomrule
\multicolumn{6}{l}{\scriptsize *Loss aversion parameter implied by indifference between Option A and B, assuming linear utility ($\alpha = 0$).}
\end{tabular*}
\end{table}

\subsection{Stakes, incentives, and order effects}

\label{sec:incentives}

Most respondents faced the MPLs with the monetary amounts shown in the previous subsection. In the remainder of this paper, we refer to these amounts as the baseline monetary amounts. In addition, we varied stake size: in the high-stake scenario (ca. 10\% of the total sample), the amounts shown were multiplied by 10 compared with amounts in the baseline treatment; in the low-stake scenario (ca. 7\% of the total sample), the amounts shown were divided by 10 compared with amounts in the baseline treatment.

In addition to stake levels, we also manipulated incentivization. The majority of the respondents (55.5\%) were incentivized, to mitigate and test for hypothetical bias, as highlighted by \citet{Charness2016}. Incentivized respondents were informed that they would have a 1\% chance of being picked and paid based on one of their actual choices in the MPLs. This type of incentive scheme is increasingly utilized in experimental economics and is also known as Between-subjects Random Incentivized System (BRIS).\footnote{See e.g. \cite{abdellaoui2008tractable} for an early example of BRIS incentivization. \cite{Clot2018} compare the impact of different incentive schemes on behavior in dictator games, and find no difference between BRIS and full incentivization, whereas no incentivization leads to less selfish play. Relevant for our study, \cite{BranasGarza2020} find that using BRIS leads to decreased levels of risk aversion compared to hypothetical or full incentives. So while an incentivization of the full sample is likely superior, we would argue that for large scale studies in which a full incentivization is infeasible (such as ours), BRIS appears to be better than no incentives at all.} Because of budget constraints, incentivization was implemented only for the baseline and low-stake scenarios. For each selected respondent, one choice was randomly designated as the pay-out-relevant choice. Respondents were informed that if a question from Table~\ref{tab:MPL3} (including monetary losses) was chosen as the pay-out-relevant choice, they would receive an additional 100 Euros (or an equivalent sum in Polish, Romanian, or Swedish currencies), regardless of the choice; losses would then be subtracted from these 100 Euros, and gains would be added.\footnote{ This procedure differs somewhat from the main approaches in the literature, where often either no incentives are used, or losses are subtracted from a show-up fee that every participant receives. Due to budgetary constraints, offering every participant 100 Euros as a show-up fee was infeasible. Our approach retains incentive compatibility and is viable for large-scale surveys.}

Overall, all respondents selected as winners were immediately informed and paid with a prepaid credit card (a MasterCard) sent to them by postal mail. Because it took one week to process and ship these cards, the earliest payment date was one week after survey completion.\footnote{Perceived payment reliability is an issue that may confound the elicitation of preferences, especially when an earlier payment may be deemed more reliable or may involve lower transaction costs. In our survey, the payment procedure was kept constant across all time horizons. Additionally, respondents were informed that the market research company would guarantee all payments as specified, and were provided with an email address that they could contact in case they had questions regarding the payment procedure. The survey drew from existing panels, consisting mostly of respondents who had previous experience with the market research company and their payment procedure, which should further alleviate issues of perceived payment reliability.} This was reflected in the time frame used in the time preference MPLs. To ensure comparability between the incentivized and non-incentivized conditions, non-incentivized respondents were also shown MPLs with an ``in one week'' formulation.

In total, 75 respondents among the roughly 7,500 incentivized respondents were randomly selected to be paid, with an average pay-off of 54.34 Euros; in total, about 4,000 Euros were paid out in incentives to respondents in addition to the regular study participation fee.

Finally, we randomized the order of presentation: about half of the respondents saw the MPLs as presented above; for the other half, Options A and B were reversed, such that Option B was presented in the left column of the table with Option A in the right column.

Overall, respondents were randomly assigned to one of the experimental conditions (unique combinations of stake level, incentivization or not, and AB or BA presentation order); to avoid confusion or experimenter demand effects, each respondent was assigned to the same combination across all four MPLs.

\subsection{Individual characteristics}

Our set of covariates included socio-demographic variables employed in previous studies, that is, age, gender, education level, income, and whether respondents had children, lived as a couple; in addition, we also asked whether participants lived in an urban area. To capture cultural differences, we included a ten-item subscale of the personal value questionnaire \citep{Knoppen2009} measuring the following individual values: self-direction, stimulation, hedonism, achievement, power, security, conformity, tradition, benevolence, and universalism \citep{Schwartz2012}. Finally, the score obtained on the standard cognitive reflection test \citep[CRT,][]{Frederick2005} was included to reflect cognitive ability. Table~\ref{tab:variables} provides a description of the variables used in our estimations.

The multivariate estimations also included country-specific dummies, with each country dummy equal to 1 if the respondent was from the specified country and 0 otherwise.
Appendix Table~\ref{tab:summarystats} reports the summary statistics for each variable in the sample across the eight countries.

\begin{table}[htbp]
\centering

\caption{Description of variables \label{tab:variables}}
\begin{tabularx}{\textwidth}{cX}
\toprule
Variable label & Variable description \\\midrule
Age & Respondents' age in years\\
Male & Code used: =1 if male; 0 if female\\
Education & Code used: =1 if higher education degree; 0 otherwise\\
Income & Respondents' income in 1,000 Euros per year\\
Children & Code used: =1 if the respondent has children; 0 otherwise\\
Couple & Code used: =1 if the respondent is in a couple; 0 otherwise\\
Urban & Code used: =1 if the respondent lives in the center or the suburban area of a major city; 0 otherwise\\
CRT & Number of correct answers to the three questions of the Cognitive Reflection Test \citep{Frederick2005}\\
Achievement & Code used: = 1 if the respondents answer 4-6 on a scale: 1-``Not like me at all'' to 6-``Very much like me''; 0 otherwise: \emph{``Being very successful is important to him/her; to have people recognize his/her achievements.''} \\ 
Benevolence & Code used: = 1 if the respondents answer 4-6 on a scale: 1-``Not like me at all'' to 6-``Very much like me''; 0 otherwise: \emph{``It is important for him/her to help other people nearby; to care for their well-being.''} On a scale: 1-``Not like me at all'' to 6-``Very much like me''.\\
Conformity & Code used: = 1 if the respondents answer 4-6 on a scale: 1-``Not like me at all'' to 6-``Very much like me''; 0 otherwise: \emph{``It is important to him/her to always behave properly; to avoid doing anything people would say is wrong.''}\\
Hedonism &Code used: = 1 if the respondents answer 4-6 on a scale: 1-``Not like me at all'' to 6-``Very much like me''; 0 otherwise: \emph{``It is important to him/her to have a good time; to spoil himself/herself.''} \\
Power & Code used: = 1 if the respondents answer 4-6 on a scale: 1-``Not like me at all'' to 6-``Very much like me''; 0 otherwise: \emph{``It is important to him/her to be rich; to have a lot of money and expensive things.''}\\
Security &Code used: = 1 if the respondents answer 4-6 on a scale: 1-``Not like me at all'' to 6-``Very much like me''; 0 otherwise: \emph{``Living in secure surroundings is important to him/her; to avoid anything that might be dangerous.''} \\
Stimulation & Code used: = 1 if the respondents answer 4-6 on a scale: 1-``Not like me at all'' to 6-``Very much like me''; 0 otherwise: \emph{``Adventure and taking risks are important to him/her; to have an exciting life.''} \\
Tradition & Code used: = 1 if the respondents answer 4-6 on a scale: 1-``Not like me at all'' to 6-``Very much like me''; 0 otherwise: \emph{``Tradition is important to him/her; to follow the customs handed down by his/her religion or family.''} \\
Universalism & Code used: = 1 if the respondents answer 4-6 on a scale: 1-``Not like me at all'' to 6-``Very much like me''; 0 otherwise: \emph{``Looking after the environment is important to him/her; to care for nature and save resources''}\\
\bottomrule

\end{tabularx}
\end{table}

\section{Results}
\label{sec:results}

\subsection{Aggregate maximum-likelihood estimation \label{sec:MLE}}

We estimate all preference parameters jointly, broadly following the maximum likelihood specification in \citet{Andersen2008} and \citet{Harrison2008}. Respondents repeatedly choose between two options, A and B. We denote expected utility as specified in Equation (2) as $U^A$ and $U^B$ for Options A and B, respectively.\footnote{For notational convenience, we drop the indices for time.}

We consider a random utility model in which individuals may make two types of errors in the decision-making process. First, they may make tremble errors, i.e. with some probability $\kappa$ they may randomize between both options. Second, they may make errors in evaluating the expected utility of lotteries.\footnote{For a discussion of this error structure see also \citet{Apesteguia2018} or \citet{Andersson2020}.} In particular, Options A and B are evaluated by their expected utility plus a stochastic component $\epsilon$, such that an individual (provided they did not randomize as a result of the tremble error), chooses Option B if $U^B+\epsilon^B \geq U^A+\epsilon^A$. Overall, the probability of choosing Option B therefore writes as follows:

\begin{equation}
\mbox{Prob}(B)=(1-\kappa)F\left(\frac{U^B-U^A}{\mu}\right)+\frac{\kappa}{2}=(1-\kappa)F(\Delta U)+\frac{\kappa}{2} ,
\end{equation}

where $F$ is the cumulative distribution function of $(\epsilon^A-\epsilon^B)$. In our main specification, we assume that $F$ follows a standard logistic distribution function with ${F(\xi)=(1-e^{-\xi})^{-1}}$. This specification is also commonly referred to as the Luce model \citep{Luce1960, Holt2002} or a Fechner error with logit link \citep{Drichoutis2014}.\footnote{Equation (3) is algebraically identical to $\mbox{Prob}(B)=(1-\kappa)\frac{\exp(U^B/\mu)}{\exp(U^A/\mu)+exp(U^B/\mu)}+\frac{\kappa}{2}$.}

We estimate six parameters: risk aversion $\alpha$, loss aversion $\lambda$, time discounting $\delta$, present bias $\gamma$, the tremble error term $\kappa$, and the Fechner error term $\mu$. The latter two parameters can be interpreted as follows: for $\kappa \rightarrow 0$, the tremble error has no effect on choice and for $\kappa \rightarrow 1$, choice approaches uniform randomization. Analogously, for $\mu\rightarrow 0$, the decision becomes deterministic (conditional on not choosing at random owing to the tremble error), and for $\mu \rightarrow \infty$, choice approaches uniform randomization.

The log-likelihood function aggregates over all choices and all respondents and writes as follows: 

\begin{equation}
\begin{aligned}
\ln\left(L(\alpha,\delta,\gamma,\lambda,\kappa,\mu)\right)= \\
\sum_j\sum_i \left[ \ln\left((1-\kappa)F(\Delta U)+\frac{\kappa}{2} \right)\times d_{ji}+\ln\left(1-(1-\kappa)F(\Delta U)-\frac{\kappa}{2} \right)\times (1-d_{ji}) \right],
\end{aligned}
\end{equation}

where $d_{ji}=1$ when Option B is chosen by individual $j$ in lottery choice $i$ and $d_{ji}=0$ when Option A is chosen. The log-likelihood function is maximized numerically using Stata's modified Newton-Raphson (NR) algorithm.\footnote{The results are robust to the utilization of different algorithms, such as Davidon-Fletcher-Powell (DFP) or Broyden-Fletcher-Goldfarb-Shanno (BFGS).} Standard errors are clustered at the individual level.

Table~\ref{tab:main} summarizes the resulting preference parameters estimated for the full sample. On average, respondents in our sample are risk averse, with a parameter of relative risk aversion $\alpha = 0.46$. To put this into perspective, the risk premium associated with a benchmark lottery that pays 0 or 100 Euro with equal probability $L=(0,0.5;100,0.5)$ would be 22.16 Euro. Respondents are on average loss averse, with parameter $\lambda = 1.93$, implying that losses are weighted roughly twice as much as equal-sized gains. They are impatient, with a yearly discount factor of $\delta = 0.28$, and slightly present-biased, with $\gamma = 0.01$. Both structural error parameters are significantly positive, indicating that choices are influenced by both Fechner and tremble errors.\footnote{Note that our estimate of the tremble error is relatively large. \cite{Andersson2020} find tremble errors as high as 0.304, though the magnitude does seem to depend on the specific MPL design. For our estimation, allowing the Fechner error to vary between MPLs appears to reduce the tremble error somewhat. As discussed in Section~\ref{sec:robustness_error}, this only has a negligible effect on the remaining estimates, so we decided to keep the simpler model as our main specification.}

\begin{table}
\caption{Preference parameter estimates\label{tab:main}}

\begin{tabular*}{\hsize}{@{\hskip\tabcolsep\extracolsep\fill}l*{1}{ccc}}
\toprule
                    &Point estimate&Standard Error&95\% Conf. Interval\\
\midrule
Risk aversion: $\alpha$&       0.460&       0.007& 0.447, 0.473\\
Loss aversion: $\lambda$&       1.934&       0.006& 1.922, 1.947\\
Discounting: $\delta$&       0.280&       0.005& 0.271, 0.289\\
Present bias: $\gamma$&       0.010&       0.001& 0.009, 0.012\\
Tremble error: $\kappa$&       0.448&       0.003& 0.442, 0.454\\
Fechner error: $\mu$&       0.682&       0.017& 0.650, 0.715\\
\midrule
N                   &      526925&            &            \\
Log. Likelihood     &     -308865&            &            \\
BIC                 &      617809&            &            \\
\bottomrule
\multicolumn{4}{l}{\footnotesize Standard errors (clustered at the subject level) in parentheses}\\
\end{tabular*}

\end{table}

Note that the point estimate of $\gamma$ should be interpreted with caution. In essence, $\gamma$ measures the difference in the discount rates elicited in MPL1.1 and MPL1.2. Because the temporal distance and the magnitude of the payments in both lists are kept constant, a positive $\gamma$ would indicate that respondents discount the future more in MPL1.2, where choices are closer to the present. This is how present bias is typically defined. However, as the earliest possible payment was one week into the future, our estimate of $\gamma$ is likely biased downwards compared to studies that have a smaller front-end delay.\footnote{See also \cite{Meier2015}, who faced a similar issue. They sent out checks as incentives, and acknowledge that the front-end delay may impact their estimates of present bias.}

In the following, we allow preference and error parameters to vary with individual characteristics as we test how preference parameters are related to these characteristics. To this end, we employ logistic models, where the six preference and error parameters are estimated jointly as linear functions of the individual characteristics outlined in section 3.3. The estimation combines data from all countries (totaling 441,420 choices made by 12,612 respondents). We include country dummies to capture country-specific effects, using the largest country, Germany, as the baseline. The results are presented in Table~\ref{tab:main_full}. In the following subsections, we present the findings about the covariates associated with each preference parameter.\footnote{While interesting in its own right, a discussion of how individual characteristics are related to decision-making errors is beyond the scope of this paper.} Note that when comparing our findings to the literature, we will not relate our findings to all studies reviewed in Section~\ref{sec:lit}; instead we will focus on a few exemplary studies, mainly those relying on demographically representative samples.

\begin{table}[htbp]
\caption{Main specification \label{tab:main_full}}
\scriptsize
\renewcommand{\footnotesize}{\scriptsize}
{
\def\sym#1{\ifmmode^{#1}\else\(^{#1}\)\fi}
\begin{tabular*}{\hsize}{@{\hskip\tabcolsep\extracolsep\fill}l*{6}{c}}
\toprule
                &\makecell{Risk aversion \\ $\alpha$}         &\makecell{Loss aversion \\ $\lambda$}         &\makecell{Discounting \\ $\delta$}         &\makecell{Present bias \\ $\gamma$}         &\makecell{Tremble error \\ $\kappa$}         &\makecell{Fechner error \\ $\mu$}         \\
\midrule
Age             &    0.002\sym{***}&   -0.006\sym{***}&   -0.001\sym{***}&   -0.000\sym{*}  &    0.001\sym{*}  &   -0.004\sym{***}\\
                &  (0.001)         &  (0.001)         &  (0.000)         &  (0.000)         &  (0.000)         &  (0.001)         \\
Male          &   -0.035\sym{**} &   -0.143\sym{***}&    0.018\sym{*}  &    0.003\sym{*}  &   -0.006         &    0.077\sym{**} \\
                &  (0.012)         &  (0.030)         &  (0.008)         &  (0.002)         &  (0.007)         &  (0.028)         \\
Education       &   -0.018         &    0.001         &   -0.032\sym{***}&    0.001         &   -0.020\sym{**} &    0.017         \\
                &  (0.013)         &  (0.031)         &  (0.009)         &  (0.002)         &  (0.007)         &  (0.032)         \\
Income          &   -0.001\sym{**} &   -0.002\sym{**} &   -0.000         &   -0.000         &   -0.001\sym{***}&    0.002\sym{*}  \\
                &  (0.000)         &  (0.001)         &  (0.000)         &  (0.000)         &  (0.000)         &  (0.001)         \\
Children        &   -0.013         &   -0.066         &    0.030\sym{**} &   -0.002         &    0.024\sym{**} &    0.028         \\
                &  (0.014)         &  (0.035)         &  (0.009)         &  (0.002)         &  (0.008)         &  (0.031)         \\
Couple          &   -0.013         &    0.016         &   -0.000         &    0.001         &    0.001         &    0.022         \\
                &  (0.014)         &  (0.035)         &  (0.009)         &  (0.002)         &  (0.008)         &  (0.031)         \\
Urban           &   -0.021         &   -0.009         &    0.008         &    0.003\sym{*}  &    0.020\sym{**} &    0.034         \\
                &  (0.012)         &  (0.029)         &  (0.008)         &  (0.002)         &  (0.007)         &  (0.028)         \\
CRT             &   -0.022\sym{***}&    0.028\sym{*}  &   -0.027\sym{***}&   -0.001         &   -0.063\sym{***}&    0.004         \\
                &  (0.006)         &  (0.014)         &  (0.004)         &  (0.001)         &  (0.003)         &  (0.013)         \\
Achievement     &   -0.012         &    0.005         &    0.008         &   -0.000         &    0.010         &    0.010         \\
                &  (0.016)         &  (0.034)         &  (0.010)         &  (0.002)         &  (0.008)         &  (0.040)         \\
Benevolence     &    0.000         &    0.035         &   -0.001         &    0.002         &   -0.006         &   -0.014         \\
                &  (0.015)         &  (0.039)         &  (0.010)         &  (0.002)         &  (0.008)         &  (0.033)         \\
Conformity      &    0.002         &    0.019         &   -0.004         &   -0.004\sym{*}  &    0.013         &   -0.010         \\
                &  (0.014)         &  (0.032)         &  (0.009)         &  (0.002)         &  (0.007)         &  (0.032)         \\
Hedonism        &   -0.022         &   -0.034         &    0.032\sym{***}&    0.002         &   -0.021\sym{**} &    0.102\sym{**} \\
                &  (0.015)         &  (0.032)         &  (0.009)         &  (0.002)         &  (0.007)         &  (0.036)         \\
Power           &    0.031         &    0.042         &   -0.005         &    0.000         &    0.053\sym{***}&   -0.073         \\
                &  (0.017)         &  (0.037)         &  (0.011)         &  (0.002)         &  (0.009)         &  (0.042)         \\
Security        &   -0.003         &    0.022         &   -0.005         &   -0.000         &   -0.014         &   -0.007         \\
                &  (0.015)         &  (0.035)         &  (0.010)         &  (0.002)         &  (0.008)         &  (0.036)         \\
Self direction  &   -0.001         &   -0.032         &    0.005         &   -0.002         &   -0.031\sym{***}&   -0.011         \\
                &  (0.014)         &  (0.034)         &  (0.009)         &  (0.002)         &  (0.008)         &  (0.033)         \\
Stimulation     &   -0.055\sym{***}&   -0.126\sym{***}&    0.023\sym{*}  &   -0.001         &    0.039\sym{***}&    0.125\sym{**} \\
                &  (0.016)         &  (0.030)         &  (0.011)         &  (0.002)         &  (0.008)         &  (0.044)         \\
Tradition       &    0.022         &   -0.034         &    0.002         &    0.000         &    0.017\sym{*}  &   -0.047         \\
                &  (0.013)         &  (0.031)         &  (0.009)         &  (0.002)         &  (0.007)         &  (0.032)         \\
Universalism    &   -0.012         &   -0.039         &   -0.011         &   -0.004         &   -0.013         &    0.007         \\
                &  (0.014)         &  (0.038)         &  (0.010)         &  (0.002)         &  (0.008)         &  (0.033)         \\
FR              &   -0.047         &    0.251\sym{***}&    0.018         &    0.006\sym{*}  &   -0.072\sym{***}&    0.093         \\
                &  (0.027)         &  (0.061)         &  (0.014)         &  (0.003)         &  (0.013)         &  (0.056)         \\
IT              &   -0.077\sym{**} &    0.194\sym{***}&    0.083\sym{***}&    0.016\sym{***}&   -0.063\sym{***}&    0.224\sym{***}\\
                &  (0.028)         &  (0.054)         &  (0.016)         &  (0.003)         &  (0.013)         &  (0.067)         \\
PL              &    0.199\sym{***}&    0.045         &    0.068\sym{**} &    0.004         &   -0.135\sym{***}&   -0.038         \\
                &  (0.029)         &  (0.067)         &  (0.022)         &  (0.004)         &  (0.016)         &  (0.052)         \\
RO              &    0.103\sym{*}  &   -0.092         &    0.245\sym{***}&   -0.018\sym{***}&   -0.033         &    0.187         \\
                &  (0.041)         &  (0.065)         &  (0.043)         &  (0.005)         &  (0.018)         &  (0.104)         \\
ES              &   -0.034         &   -0.001         &    0.075\sym{***}&    0.007         &   -0.018         &    0.204\sym{**} \\
                &  (0.028)         &  (0.053)         &  (0.018)         &  (0.004)         &  (0.013)         &  (0.069)         \\
SE              &   -0.110\sym{***}&    0.262\sym{***}&    0.019         &    0.003         &   -0.033\sym{**} &    0.243\sym{***}\\
                &  (0.027)         &  (0.055)         &  (0.014)         &  (0.003)         &  (0.013)         &  (0.073)         \\
UK              &    0.043         &    0.253\sym{***}&    0.031\sym{*}  &    0.003         &   -0.047\sym{***}&   -0.035         \\
                &  (0.027)         &  (0.062)         &  (0.015)         &  (0.003)         &  (0.013)         &  (0.051)         \\
ABreversed      &    0.039\sym{**} &    0.027         &   -0.019\sym{*}  &   -0.007\sym{***}&    0.023\sym{***}&   -0.100\sym{***}\\
                &  (0.012)         &  (0.028)         &  (0.008)         &  (0.002)         &  (0.006)         &  (0.028)         \\
Incentivized    &   -0.044\sym{**} &   -0.189\sym{***}&    0.064\sym{***}&   -0.022\sym{***}&   -0.014\sym{*}  &    0.084\sym{*}  \\
                &  (0.015)         &  (0.036)         &  (0.009)         &  (0.002)         &  (0.007)         &  (0.035)         \\
LowStakes       &   -0.048\sym{*}  &   -0.147\sym{**} &   -0.010         &    0.028\sym{***}&    0.015         &    0.102         \\
                &  (0.023)         &  (0.050)         &  (0.017)         &  (0.004)         &  (0.013)         &  (0.069)         \\
HighStakes      &    0.023         &    0.158\sym{*}  &   -0.035\sym{**} &   -0.007\sym{**} &    0.032\sym{**} &   -0.011         \\
                &  (0.024)         &  (0.073)         &  (0.013)         &  (0.003)         &  (0.012)         &  (0.048)         \\
Constant        &    0.525\sym{***}&    2.270\sym{***}&    0.256\sym{***}&    0.030\sym{***}&    0.490\sym{***}&    0.619\sym{***}\\
                &  (0.041)         &  (0.098)         &  (0.023)         &  (0.005)         &  (0.021)         &  (0.082)         \\
\midrule
N               &   443415         &                  &                  &                  &                  &                  \\
Log. Likelihood &  -253086         &                  &                  &                  &                  &                  \\
BIC             &   508513         &                  &                  &                  &                  &                  \\
\bottomrule
\multicolumn{7}{l}{\footnotesize Standard errors (clustered at the subject level) in parentheses}\\
\multicolumn{7}{l}{\footnotesize \sym{*} \(p<0.05\), \sym{**} \(p<0.01\), \sym{***} \(p<0.001\)}\\
\end{tabular*}
}

\renewcommand{\footnotesize}{\footnotesize}
\end{table}

\subsection{Individual characteristics associated with risk aversion}
\label{sec:RA}

Our results suggest that risk aversion is statistically significantly related with the socio-demographic variables \emph{Age}, \emph{Male}, and \emph{Income}.  More specifically, we find a positive correlation between \textit{Age} and risk aversion. Thus, in our sample, older respondents are, on average, more risk averse than younger respondents, ceteris paribus. While this finding is in line with the majority of previous studies \citep[e.g.][]{Schurer2015, Mata2016, Bouchouicha2019}, a substantial number of studies find a negative relation of age and risk aversion \citep[e.g.][]{Cardenas2013, Noussair2014, Andersson2016}. Consistent with the thrust of the literature  \citep[e.g.][]{Hansen2016, Lee2016, Falk2018}, men appear to be less risk averse than women. 

In line with the relative majority of the studies surveyed in our literature review (see Table~\ref{tab:litterature}) we fail to find a statistically significant relation between \emph{Education} and risk aversion. Yet, previous studies also report negative associations \citep[e.g.][]{Mata2016, Bouchouicha2019, Browne2020} or positive associations  \citep[e.g.][]{ Lee2016, Sepahvand2017, Chapman2018} between education and risk aversion.  
Similar to \cite{ Castillo2018}, \cite{Blake2019} and \cite{ Bouchouicha2019} for example, we find that respondents with higher \textit{Income} are less risk averse. The relative majority of previous studies failed to find a statistically significant relation between income and risk aversion. Whether participants have children, are married, or live in an urban environment, does not appear to be associated with risk aversion. These findings are in line with the thrust of the literature.

For cognitive ability, we find that respondents with higher CRT scores are less risk averse. There is some debate in the literature regarding the effects of cognitive ability on risk aversion. In line with our results, \citet{Chapman2018a}, \citet{Dohmen2010}, and \citet{Goda2019} find cognitive ability to be negatively correlated with risk aversion.\footnote{See also  \citet{Lilleholt2019} for a meta-study that also finds small but significantly negative effects.} \citet{Andersson2016} argue however that such results may be spurious, an artifact caused by the method used to elicit risk aversion. \citet{Dohmen2010} use an MPL design in which the risk-neutral switch point is located in the upper half of the MPL. \citet{Andersson2016} show that this may lead to a spurious (negative) correlation between cognitive ability and risk aversion.\footnote{Respondents with lower cognitive ability may make more mistakes, and since the risk-neutral switch point falls in the upper half of the MPL, there are more opportunities to err towards the safe than to the risky option. A risk-neutral respondent who chooses completely at random will thus appear risk averse.} \citet{Andersson2020} further argue that, after carefully controlling for decision errors in estimating preference parameters, the effects of cognitive ability on risk aversion disappear. To mitigate the effect of risk-neutral switch-point location in the MPL on the relationship between CRT and risk preferences, we designed our MPL in a symmetric way: the risk-neutral switch point falls in the middle of the MPL.\footnote{A risk-neutral respondent who chooses completely at random will thus appear risk neutral on average. Note that this also implies that respondents who are risk averse should now switch from the safe to the risky option somewhere in the lower half of our MPL. This implies that they have more room to err towards the risky option: a risk-averse respondent who chooses randomly will appear risk neutral.} Thus, given that respondents are on average risk averse, our design is biased in the opposite direction of our finding, such that increased randomness (correlated with cognitive ability) should lead to lower observed risk aversion.\footnote{This intuition is confirmed in Appendix~\ref{sec:simulation}, where we simulate choice data and test this explicitly. The findings of these simulations underline the robustness of our finding that risk aversion is negatively associated with cognitive ability and suggest that the effect is likely even larger than documented in Table~\ref{tab:main_full}.} In contrast to \citet{Andersson2016, Andersson2020}, in our study the effects of cognitive ability on risk aversion remain significant, even after controlling for the effects of cognitive ability on the decision error.\footnote{\cite{Andersson2020} use a random parameter model based on \cite{Apesteguia2018}. To our knowledge, this kind of random parameter model has not yet been extended to multi-parameter utility models. In their online Appendix, however, they use a random utility model with tremble error and (logit-link) Fechner error, and show that this is sufficient to control for spurious effects. Our maximum likelihood estimation is based on that model.} This may be because our analysis benefits from a substantially larger sample ($N=12,615$ vs. $N= 1,396$ in \citet{Andersson2020} and our study, respectively), which could lead to an increased chance to detect small effects. It should be noted, however, that our study is not directly comparable with \citet{Andersson2020}, owing to differences in the MPL design and measurement of cognitive ability. Also, while we believe that the relation between cognitive ability and risk aversion is robust (see also the simulation exercise in Appendix~\ref{sec:simulation}), we can, of course, not exclude the existence of other spurious effects.

Regarding the cultural values, we find that \emph{Stimulation} is negatively related to risk aversion which implies that individuals who like adventures are less risk averse.

Concerning country effects, risk aversion appears to be higher in Poland and Romania and lower in Italy and Sweden relative to the baseline country, Germany.

We also find order effects: respondents facing the riskier outcomes on the left side of the screen exhibit greater risk aversion. To our knowledge, these effects are rarely considered in the literature \citep{Andersen2006} and may suggest that randomization of order is important when eliciting preferences using multiple price lists.

Our results further suggest that incentivized participants are less risk averse than non-incentivized participants. While there is no consensus in the literature regarding the effects of incentivization on risk behavior in experiments, most existing studies have found that incentivization has either no effects \citep{Camerer1999, Rieger2015} or that it increases risk aversion \citep{Holt2002}. Recent evidence, however, suggests that the effect of incentivization on risk behavior may depend on whether all or only some (randomly selected) participants are paid. In line with our finding, \cite{BranasGarza2020} observe that a between-subject random incentive scheme (BRIS) leads to less risk aversion, compared to hypothetical incentives. Finally, in line with \citet{Holt2002} and \citet{Harrison2005} we find that when stakes are low (compared to the baseline) risk aversion decreases. %\footnote{Note that comparing the estimated coefficients in terms of economic significance is difficult, because the covariates are measured in different units. In Appendix Table~\ref{tab:l1t_fullz}, we present the findings of our main specification using z-scores for all covariates. This allows to interpret the coefficients as the change in the dependent variable in response to an increase in the covariate by one standard deviation.}

\subsection{Individual characteristics associated with loss aversion}

 Overall, we find that age, being male and income are negatively related with loss aversion. While for these characteristics, the scant literature on loss aversion mostly found a non-significant relation, our findings are consistent with \cite{Cardenas2013}, \cite{Chapman2018a} and \cite{Blake2019} for age, with \cite{Booij2009} and \cite{Cardenas2013} for gender, and with \cite{VonGaudecker2011}, \cite{Stephens2012} and \cite{ Blake2019} for income. For \textit{CRT}, we find a positive relation with loss aversion, which is consistent with \citet{Chapman2018a}. For \textit{Education}, \textit{Children}, \textit{Couple}, and \textit{Urban}, we find no statistically significant relation with loss aversion. Previous literature has rarely considered these characteristics, with the exception of education, for which the majority of studies also find a null result.

Loss aversion is also related with \emph{Stimulation}: respondents who like adventure and taking risks are less loss averse.

Relating to country effects, we find higher loss aversion in the samples from France, Italy, Sweden and the UK compared to the sample from Germany.

Unlike for the other preference parameters, we find no order effects for loss aversion. Finally, we find that loss aversion is lower for incentivized respondents, and that loss aversion increases with stake size. The latter result is consistent with \citet{Vieider2012}.

\subsection{Individual characteristics associated with time discounting}
The demographic variables \textit{Age}, \textit{Male}, \textit{Education}, and \textit{Children} are statistically significantly related to time discounting. We find older respondents to be more patient. The relative majority of studies in our literature review (see Table~\ref{tab:litterature}) did not find a statistically significant relation between \textit{Age} and time discounting. Our result is, however, consistent with numerous studies including \cite{Andersen2018}, \cite{Chapman2018a} and \cite{Falk2018}. While most studies fail to find gender to be statistically significantly related with time discounting, for our sample, male respondents are more impatient. Further, more highly educated respondents appear to be more patient, which is in line with the thrust of findings reported in the literature \citep[e.g.][]{BrudererEnzler2014, Chapman2018a, Goda2019}. Next, respondents with children are more impatient, consistent with what \citet{BrudererEnzler2014} document. The coefficients associated with \textit{Couple} and \textit{Urban} are not statistically significant.

In line with the literature \citep[e.g.][]{Chapman2018a, Falk2018, Goda2019}, respondents from our sample with high CRT scores are more patient.

Among the cultural values, only \emph{Hedonism} and \emph{Stimulation} are significantly correlated with time discounting. Respondents who like to have a good time, to spoil themselves, and those who like adventures and taking risks appear to be more impatient. These results make intuitive sense, because these two cultural values strongly emphasize immediate rewards, which is consistent with impatience.

Pertaining to country effects, discounting appears to be higher in Italy, Poland, Romania, Spain and the UK compared to Germany.

Further, we find significant relations between time discounting and the order in which the lotteries were shown, as well as with incentivization and stake levels. Specifically, respondents who faced lotteries' options in reverse order appear more patient than those who faced the lotteries as displayed in Tables \ref{tab:MPL1.1} through \ref{tab:MPL3}. This result highlights the importance of controlling for order effects when eliciting preferences using multiple price lists. Our findings further imply that incentivized respondents discount the future more than non-incentivized respondents. Finally, respondents who faced high stakes discount the future less than respondents who faced the amounts in the baseline treatment.

\subsection{Individual characteristics associated with present bias}

Only few representative studies have investigated the relationship between individual characteristics and present bias \citep[e.g.][]{ Pinger2017, Goda2019}. Our results suggest that older respondents are less present biased, which is consistent with what \citet{Wang2016}, \citet{Hunter2018} and \citet{Breuer2020} report. In contrast to the scant existing literature, we find a statistically significant relation between gender and present bias: men are more present biased than women. Further, respondents living in an urban environment appear to be more present biased. We find no evidence for present bias to be related to education, income, having children, living as a couple, cognitive ability, and most cultural values. Yet the finding for \emph{Conformity} suggests that respondents stating that it is important to them to always behave properly and to avoid doing anything people would say is wrong are less likely to be present biased. 

Respondents in France and Italy are more present biased and respondents in Romania are less present biased than those in the baseline country (Germany).

Respondents who faced binary choices in reverse order (more delayed outcomes on the left side of the screen) appear to be less present biased. Finally, respondents who were incentivized or faced higher stakes are less present biased whereas those facing lower stakes (compared with the moderate stakes baseline) are more present biased. These results are rather intuitive as they suggest that higher or more perceptible payments (for delayed outcomes) may contribute to reducing respondents' bias toward present payments.\footnote{Note that the effects of individual characteristics on present bias presented here are rather small. This may (at least in part) be explained by the fact that our estimate of present bias is likely biased downward, as discussed in Section~\ref{sec:MLE}.}

\subsection{Country-level estimates \label{sec:country}}

\begin{table}[t]
\footnotesize
\caption{Parameter estimates by country \label{tab:l1t_8countries}}
\begin{tabular*}{\hsize}{@{\hskip\tabcolsep\extracolsep\fill}l*{9}{c}}
\toprule
                    &\multicolumn{1}{c}{All countries}&\multicolumn{1}{c}{France}&\multicolumn{1}{c}{Germany}&\multicolumn{1}{c}{Italy}&\multicolumn{1}{c}{Poland}&\multicolumn{1}{c}{Romania}&\multicolumn{1}{c}{Spain}&\multicolumn{1}{c}{Sweden}&\multicolumn{1}{c}{UK}\\
\midrule
$\alpha$            &       0.460&       0.420&       0.433&       0.379&       0.683&       0.551&       0.425&       0.341&       0.498\\
                    &     (0.007)&     (0.016)&     (0.018)&     (0.017)&     (0.019)&     (0.032)&     (0.018)&     (0.017)&     (0.017)\\
$\lambda$           &       1.934&       1.976&       1.925&       1.952&       1.812&       1.598&       1.599&       1.971&       1.984\\
                    &     (0.006)&     (0.008)&     (0.014)&     (0.013)&     (0.049)&     (0.039)&     (0.016)&     (0.011)&     (0.011)\\
$\delta$            &       0.280&       0.221&       0.231&       0.325&       0.329&       0.514&       0.294&       0.238&       0.225\\
                    &     (0.005)&     (0.008)&     (0.010)&     (0.013)&     (0.022)&     (0.041)&     (0.013)&     (0.010)&     (0.010)\\
$\gamma$            &       0.010&       0.012&       0.008&       0.025&       0.009&      -0.016&       0.014&       0.006&       0.011\\
                    &     (0.001)&     (0.002)&     (0.002)&     (0.003)&     (0.003)&     (0.005)&     (0.003)&     (0.002)&     (0.002)\\
$\kappa$            &       0.448&       0.402&       0.474&       0.428&       0.343&       0.457&       0.462&       0.421&       0.415\\
                    &     (0.003)&     (0.007)&     (0.008)&     (0.008)&     (0.010)&     (0.014)&     (0.008)&     (0.009)&     (0.007)\\
$\mu$               &       0.682&       0.619&       0.623&       0.893&       0.593&       0.851&       0.787&       0.877&       0.530\\
                    &     (0.017)&     (0.037)&     (0.041)&     (0.059)&     (0.037)&     (0.089)&     (0.051)&     (0.061)&     (0.031)\\
\midrule
N                   &      526925&       70000&       70070&       70000&       70280&       53515&       70035&       53025&       70000\\
\bottomrule
\multicolumn{10}{l}{\footnotesize Standard errors (clustered at the subject level) in parentheses}\\
\end{tabular*}

\end{table}

Our data allows us to compare preferences and their relations with individual characteristics between countries. To this end, we first estimate preference parameters separately for each country. Table~\ref{tab:l1t_8countries} illustrates the resulting point estimates and standard errors of the six preference and error parameters.

Most variance across countries can be found in time discounting $\delta$ and risk aversion $\alpha$. Estimates for the parameter of relative risk aversion vary between $\alpha = 0.34$ (Sweden) and $\alpha=0.68$ (Poland). Using our benchmark lottery $L=(0,0.5;100,0.5)$, these parameters would generate risk premiums of 13.02 and 36.08 Euro in Sweden and Poland respectively. This implies that differences in risk-taking behavior could be quite large. Romanians have by far the highest discount rate $\delta=0.51$, and the French appear to discount the future least $\delta=0.22$. Loss aversion and present bias exhibit lower overall normalized variances between countries, although respondents from Spain and Romania appear to be markedly less loss averse. Romania is the only country where respondents appear to be future biased on average.\footnote{Note that some of the results in this section appear to be at odds with the estimated coefficients of the country dummies in Section~\ref{sec:MLE}. This is to be expected, as the analysis in this section (unlike the analysis in Section~\ref{sec:MLE}) does not control for observed heterogeneity.}

To investigate the associations between individual characteristics and preference parameters across countries, we estimate country-level models with a reduced set of covariates. Ideally we would use the same set of covariates as in our main specification; however with the drastically reduced sample sizes, the models fail to converge reliably in the different countries when using the full set of covariates.  
Estimation results can be found in Appendix Tables~\ref{tab:l1t_reduced}--\ref{tab:l1t_UK}. Although a country-by-country comparison of the relations between all individual characteristics and the four preference parameters is beyond the scope of this paper, we observe that due to decreased power, many correlations between preferences and individual characteristics become insignificant at the country level. Nevertheless, the  estimates appear consistent across countries: none of the individual country models exhibits a significant effect that is opposite in sign to a significant effect in the model that includes all countries.

\subsection{Robustness Checks}

Our findings on preferences and their correlation with covariates may be sensitive to the specification of intertemporal utility, the selected error process, and the measurement of preferences. Yet, there appears to be no consensus in the literature regarding these assumptions. 
We therefore analyze the robustness of our results pertaining to i) exclusion of multiple switchers, ii) alternative utility specifications, iii) alternative error-process specifications, and iv) alternative measures of preferences, such as the number of times a specific option was chosen, or switch points. For brevity, we relegate details to Appendix \ref{sec:robustness}. Summing up, our main findings are largely robust. Among other findings, we document that controlling for noise in the decision making process is important. Simulation exercises confirm that using measures of preference and estimation methods that fail to account for decision noise may lead to spurious correlations between preferences and individual characteristics that are correlated with decision noise. This result is in line with recent findings \citep{Andersson2016,Andersson2020}.

\section{Discussion and conclusion}
\label{sec:conclusion}

In this paper, we empirically analyze the relationships between individual characteristics and risk aversion, loss aversion, time discounting, and present bias in a large-scale demographically representative sample in eight EU countries. %We jointly estimate preferences to avoid bias from neglecting structural relations between these preferences, such as the well-documented curvature bias when eliciting time preferences \citet{Andersen2008}.
We cautiously control for decision errors in estimating preference parameters to avoid spurious results in cases where individual characteristics may be correlated with decision errors. Our findings provide in particular guidance on relationships that previous literature has never or only rarely analysed such as the relation between individual characteristics and loss aversion or present bias. Regarding present bias and loss aversion, this study uses the largest sample to date to analyze relations between individual characteristics and these preferences. Likewise, most likely because of the large sample size, our study finds statistically significant associations for several relations where a majority or relative majority of the studies included in our systematic literature review find a no result such as the relation of income with risk aversion, of gender with loss aversion, of age and having children with time discounting, and of age and gender with present bias. More specifically, our findings suggest that risk aversion is negatively correlated with income. We also find evidence suggesting the existence of a negative relationship between risk aversion and cognitive ability (as measured by CRT), a relationship that has been subject to debate in the literature. We further find that time discounting is negatively associated with age. In addition, our results suggest that men discount the future to a greater extent and are also more present biased than women. The latter finding is not yet established in the literature. Finally, we find that older respondents and males are less loss averse.

Few studies have tested whether cultural values are related to economic preferences. In our study, \emph{Stimulation} appears to consistently correlate with risk and time preferences. Respondents who identify to a greater extent with the statement ``Adventure and taking risks are important to him/her; to have an exciting life'' are less risk averse and loss averse, and also discount the future to a greater extent. These links have remained largely unstudied in the literature, and may be a fruitful area for future research.

Our study also makes methodological contributions that may prove useful for researchers interested in using MPLs to elicit economic preferences: we find that the order of presentation of the two options as well as stake size have statistically significant effects on elicited preference parameters. While it may be difficult to control for stake size in some settings, order of presentation can easily be randomized. Incentivization appears to be important, as it is significantly associated with all elicited preference parameters. Finally, our study highlights the importance of controlling for the effects of observable heterogeneity on decision noise. 

\begin{comment}
\section*{Statements and Declarations}

The authors declare that they have no conflict of interest.

\end{comment}
\newpage

\begin{singlespace}
{\footnotesize
\bibliographystyle{ecta}
\bibliography{bib}

\begin{thebibliography}{176}
\newcommand{\enquote}[1]{``#1''}
\expandafter\ifx\csname natexlab\endcsname\relax\def\natexlab#1{#1}\fi

\bibitem[\protect\citeauthoryear{Abdellaoui, Bleichrodt, and
  l'Haridon}{Abdellaoui et~al.}{2008}]{abdellaoui2008tractable}
\textsc{Abdellaoui, M., H.~Bleichrodt, and O.~l'Haridon} (2008): \enquote{A
  tractable method to measure utility and loss aversion under prospect theory,}
  \emph{Journal of Risk and uncertainty}, 36, 245--266.

\bibitem[\protect\citeauthoryear{Abdellaoui, Bleichrodt, l'Haridon, and
  Paraschiv}{Abdellaoui et~al.}{2013}]{Abdellaoui2013}
\textsc{Abdellaoui, M., H.~Bleichrodt, O.~l'Haridon, and C.~Paraschiv} (2013):
  \enquote{Is there one unifying concept of utility? An experimental comparison
  of utility under risk and utility over time,} \emph{Management Science}, 59,
  2153--2169.

\bibitem[\protect\citeauthoryear{Abdellaoui, Bleichrodt, and
  Paraschiv}{Abdellaoui et~al.}{2007}]{Abdellaoui2007}
\textsc{Abdellaoui, M., H.~Bleichrodt, and C.~Paraschiv} (2007): \enquote{Loss
  aversion under prospect theory: A parameter-free measurement,}
  \emph{Management Science}, 53, 1659--1674.

\bibitem[\protect\citeauthoryear{Abdellaoui, Kemel, Panin, and
  Vieider}{Abdellaoui et~al.}{2019}]{Abdellaoui2019}
\textsc{Abdellaoui, M., E.~Kemel, A.~Panin, and F.~M. Vieider} (2019):
  \enquote{Measuring time and risk preferences in an integrated framework,}
  \emph{Games and Economic Behavior}, 115, 459--469.

\bibitem[\protect\citeauthoryear{Albert and Duffy}{Albert and
  Duffy}{2012}]{Albert2012}
\textsc{Albert, S.~M. and J.~Duffy} (2012): \enquote{Differences in risk
  aversion between young and older adults,} \emph{Neuroscience and
  Neuroeconomics}, 1, 3--9.

\bibitem[\protect\citeauthoryear{Almenberg and Dreber}{Almenberg and
  Dreber}{2015}]{Almenberg2015}
\textsc{Almenberg, J. and A.~Dreber} (2015): \enquote{Gender, stock market
  participation and financial literacy,} \emph{Economics Letters}, 137,
  140--142.

\bibitem[\protect\citeauthoryear{Andersen, Harrison, Lau, and
  Rutstr{\"o}m}{Andersen et~al.}{2006}]{Andersen2006}
\textsc{Andersen, S., G.~W. Harrison, M.~I. Lau, and E.~E. Rutstr{\"o}m}
  (2006): \enquote{Elicitation using multiple price list formats,}
  \emph{Experimental Economics}, 9, 383--405.

\bibitem[\protect\citeauthoryear{Andersen, Harrison, Lau, and
  Rutstr{\"o}m}{Andersen et~al.}{2008}]{Andersen2008}
---\hspace{-.1pt}---\hspace{-.1pt}--- (2008): \enquote{Eliciting risk and time
  preferences,} \emph{Econometrica}, 76, 583--618.

\bibitem[\protect\citeauthoryear{Andersen, Harrison, Lau, and
  Rutstr{\"o}m}{Andersen et~al.}{2010}]{Andersen2010}
---\hspace{-.1pt}---\hspace{-.1pt}--- (2010): \enquote{Preference heterogeneity
  in experiments: Comparing the field and laboratory,} \emph{Journal of
  Economic Behavior \& Organization}, 73, 209--224.

\bibitem[\protect\citeauthoryear{Andersen, Harrison, Lau, and
  Rutstr{\"o}m}{Andersen et~al.}{2018}]{Andersen2018}
---\hspace{-.1pt}---\hspace{-.1pt}--- (2018): \enquote{Multiattribute utility
  theory, intertemporal utility, and correlation aversion,} \emph{International
  Economic Review}, 59, 537--555.

\bibitem[\protect\citeauthoryear{Anderson, Gibson, Luchtenberg, and
  Seiler}{Anderson et~al.}{2021}]{Anderson2021}
\textsc{Anderson, J.~T., S.~Gibson, K.~F. Luchtenberg, and M.~J. Seiler}
  (2021): \enquote{How {Much} {Are} {Borrowers} {Willing} to {Pay} to {Remove}
  {Uncertainty} {Surrounding} {Mortgage} {Defaults}?} \emph{The Journal of Real
  Estate Finance and Economics}, 1--23.

\bibitem[\protect\citeauthoryear{Andersson, Holm, Tyran, and
  Wengstr{\"o}m}{Andersson et~al.}{2016}]{Andersson2016}
\textsc{Andersson, O., H.~J. Holm, J.-R. Tyran, and E.~Wengstr{\"o}m} (2016):
  \enquote{Risk aversion relates to cognitive ability: Preferences or noise?}
  \emph{Journal of the European Economic Association}, 14, 1129--1154.

\bibitem[\protect\citeauthoryear{Andersson, Holm, Tyran, and
  Wengstr{\"o}m}{Andersson et~al.}{2020}]{Andersson2020}
---\hspace{-.1pt}---\hspace{-.1pt}--- (2020): \enquote{Robust inference in risk
  elicitation tasks,} \emph{Journal of Risk and Uncertainty}, 1--15.

\bibitem[\protect\citeauthoryear{Andreoni and Sprenger}{Andreoni and
  Sprenger}{2012}]{Andreoni2012}
\textsc{Andreoni, J. and C.~Sprenger} (2012): \enquote{Estimating time
  preferences from convex budgets,} \emph{American Economic Review}, 102,
  3333--56.

\bibitem[\protect\citeauthoryear{Apesteguia and Ballester}{Apesteguia and
  Ballester}{2018}]{Apesteguia2018}
\textsc{Apesteguia, J. and M.~A. Ballester} (2018): \enquote{Monotone
  stochastic choice models: The case of risk and time preferences,}
  \emph{Journal of Political Economy}, 126, 74--106.

\bibitem[\protect\citeauthoryear{Aycinena, Baltaduonis, and
  Rentschler}{Aycinena et~al.}{2014}]{Aycinena2014}
\textsc{Aycinena, D., R.~Baltaduonis, and L.~Rentschler} (2014): \enquote{Risk
  Preferences and Prenatal Exposure to Sex Hormones for Ladinos,} \emph{PloS
  one}, 9, 1--10.

\bibitem[\protect\citeauthoryear{Bacon, Conte, and Moffatt}{Bacon
  et~al.}{2014}]{Bacon2014}
\textsc{Bacon, P.~M., A.~Conte, and P.~G. Moffatt} (2014): \enquote{Assortative
  mating on risk attitude,} \emph{Theory and Decision}, 77, 389--401.

\bibitem[\protect\citeauthoryear{Bajtelsmit and Bernasek}{Bajtelsmit and
  Bernasek}{2001}]{Bajtelsmit2001}
\textsc{Bajtelsmit, V.~L. and A.~Bernasek} (2001): \emph{Risk preferences and
  the investment decisions of older Americans}, AARP, Public Policy Institute.

\bibitem[\protect\citeauthoryear{Bansback, Harrison, Sadatsfavi, Stiggelbout,
  and Whitehurst}{Bansback et~al.}{2016}]{Bansback2016}
\textsc{Bansback, N., M.~Harrison, M.~Sadatsfavi, A.~Stiggelbout, and D.~G.~T.
  Whitehurst} (2016): \enquote{Attitude to health risk in the Canadian
  population: a cross-sectional survey,} \emph{{CMAJ} Open}, 4, E284--E291.

\bibitem[\protect\citeauthoryear{Bartczak, Chilton, and Meyerhoff}{Bartczak
  et~al.}{2015}]{Bartczak2015}
\textsc{Bartczak, A., S.~Chilton, and J.~Meyerhoff} (2015): \enquote{Wildfires
  in Poland: The impact of risk preferences and loss aversion on environmental
  choices,} \emph{Ecological Economics}, 116, 300--309.

\bibitem[\protect\citeauthoryear{Bartke and Schwarze}{Bartke and
  Schwarze}{2008}]{Bartke2008}
\textsc{Bartke, S. and R.~Schwarze} (2008): \enquote{Risk-{Averse} by {Nation}
  or by {Religion}? {Some} {Insights} on the {Determinants} of {Individual}
  {Risk} {Attitudes},} \emph{Available at SSRN 1285520}.

\bibitem[\protect\citeauthoryear{Bateman, Stevens, and Lai}{Bateman
  et~al.}{2015}]{Bateman2015}
\textsc{Bateman, H., R.~Stevens, and A.~Lai} (2015): \enquote{Risk Information
  and Retirement Investment Choice Mistakes Under Prospect Theory,}
  \emph{Journal of Behavioral Finance}, 16, 279--296.

\bibitem[\protect\citeauthoryear{Beauchaine, Ben-David, and Sela}{Beauchaine
  et~al.}{2017}]{Beauchaine2017}
\textsc{Beauchaine, T.~P., I.~Ben-David, and A.~Sela} (2017):
  \enquote{Attention-deficit/hyperactivity disorder, delay discounting, and
  risky financial behaviors: A preliminary analysis of self-report data,}
  \emph{PloS one}, 12, e0176933.

\bibitem[\protect\citeauthoryear{Benjamin, Brown, and Shapiro}{Benjamin
  et~al.}{2013}]{Benjamin2013}
\textsc{Benjamin, D.~J., S.~A. Brown, and J.~M. Shapiro} (2013): \enquote{Who
  is "behavioral"? Cognitive ability and anomalous preferences,} \emph{Journal
  of the European Economic Association}, 11, 1231--1255.

\bibitem[\protect\citeauthoryear{Blake, Cannon, and Wright}{Blake
  et~al.}{2019}]{Blake2019}
\textsc{Blake, D.~P., E.~Cannon, and D.~Wright} (2019): \enquote{Quantifying
  Loss Aversion: Evidence from a UK Population Survey,} \emph{Working Paper}.

\bibitem[\protect\citeauthoryear{Bonsang and Dohmen}{Bonsang and
  Dohmen}{2015}]{Bonsang2015}
\textsc{Bonsang, E. and T.~Dohmen} (2015): \enquote{Risk attitude and cognitive
  aging,} \emph{Journal of Economic Behavior {\&} Organization}, 112, 112--126.

\bibitem[\protect\citeauthoryear{Booij and van~de Kuilen}{Booij and van~de
  Kuilen}{2009}]{Booij2009a}
\textsc{Booij, A.~S. and G.~van~de Kuilen} (2009): \enquote{A parameter-free
  analysis of the utility of money for the general population under prospect
  theory,} \emph{Journal of Economic Psychology}, 30, 651--666.

\bibitem[\protect\citeauthoryear{Booij and van Praag}{Booij and van
  Praag}{2009}]{Booij2009}
\textsc{Booij, A.~S. and B.~M. van Praag} (2009): \enquote{A simultaneous
  approach to the estimation of risk aversion and the subjective time discount
  rate,} \emph{Journal of Economic Behavior \& Organization}, 70, 374--388.

\bibitem[\protect\citeauthoryear{Booij, van Praag, and van~de Kuilen}{Booij
  et~al.}{2010}]{Booij2010}
\textsc{Booij, A.~S., B.~M.~S. van Praag, and G.~van~de Kuilen} (2010):
  \enquote{A parametric analysis of prospect theory's functionals for the
  general population,} \emph{Theory and Decision}, 68, 115--148.

\bibitem[\protect\citeauthoryear{Boschini, Dreber, von Essen, Muren, and
  Ranehill}{Boschini et~al.}{2019}]{Boschini2019}
\textsc{Boschini, A., A.~Dreber, E.~von Essen, A.~Muren, and E.~Ranehill}
  (2019): \enquote{Gender, risk preferences and willingness to compete in a
  random sample of the Swedish population,} \emph{Journal of Behavioral and
  Experimental Economics}, 83, 101467.

\bibitem[\protect\citeauthoryear{Bouchouicha, Deer, Eid, McGee, Schoch, Stojic,
  Ygosse-Battisti, and Vieider}{Bouchouicha
  et~al.}{2019}]{bouchouicha2019gender}
\textsc{Bouchouicha, R., L.~Deer, A.~G. Eid, P.~McGee, D.~Schoch, H.~Stojic,
  J.~Ygosse-Battisti, and F.~M. Vieider} (2019): \enquote{Gender effects for
  loss aversion: Yes, no, maybe?} \emph{Journal of Risk and Uncertainty}, 59,
  171--184.

\bibitem[\protect\citeauthoryear{Bouchouicha and Vieider}{Bouchouicha and
  Vieider}{2019}]{Bouchouicha2019}
\textsc{Bouchouicha, R. and F.~M. Vieider} (2019): \enquote{Growth,
  entrepreneurship, and risk-tolerance: a risk-income paradox,} \emph{Journal
  of Economic Growth}, 24, 257--282.

\bibitem[\protect\citeauthoryear{Boyle, Yu, Buchman, Laibson, and
  Bennett}{Boyle et~al.}{2011}]{Boyle2011}
\textsc{Boyle, P.~A., L.~Yu, A.~S. Buchman, D.~I. Laibson, and D.~A. Bennett}
  (2011): \enquote{Cognitive function is associated with risk aversion in
  community-based older persons,} \emph{BMC geriatrics}, 11, 1--8.

\bibitem[\protect\citeauthoryear{Bradford, Courtemanche, Heutel, McAlvanah, and
  Ruhm}{Bradford et~al.}{2017}]{Bradford2017}
\textsc{Bradford, D., C.~Courtemanche, G.~Heutel, P.~McAlvanah, and C.~Ruhm}
  (2017): \enquote{Time preferences and consumer behavior,} \emph{Journal of
  Risk and Uncertainty}, 55, 119--145.

\bibitem[\protect\citeauthoryear{Bra{\~n}as-Garza, Estepa~Mohedano, Jorrat,
  Orozco, and Rascon-Ramirez}{Bra{\~n}as-Garza et~al.}{2020}]{BranasGarza2020}
\textsc{Bra{\~n}as-Garza, P., L.~Estepa~Mohedano, D.~Jorrat, V.~Orozco, and
  E.~Rascon-Ramirez} (2020): \enquote{To pay or not to pay: Measuring risk
  preferences in lab and field,} \emph{Available at SSRN 3687825}.

\bibitem[\protect\citeauthoryear{Bra{\~n}as-Garza, Galizzi, and
  Nieboer}{Bra{\~n}as-Garza et~al.}{2018}]{BranasGarza2018}
\textsc{Bra{\~n}as-Garza, P., M.~M. Galizzi, and J.~Nieboer} (2018):
  \enquote{Experimental and self-reported measures of risk taking and digit
  ratio (2d: 4d): evidence from a large, systematic study,} \emph{International
  Economic Review}, 59, 1131--1157.

\bibitem[\protect\citeauthoryear{Breuer, Renerken, and Salzmann}{Breuer
  et~al.}{2020}]{Breuer2020}
\textsc{Breuer, W., T.~Renerken, and A.~J. Salzmann} (2020): \enquote{On the
  measurement of risk-taking and patience in financial decision-making,}
  \emph{Available at SSRN 2538482}.

\bibitem[\protect\citeauthoryear{Brick, Visser, and Burns}{Brick
  et~al.}{2012}]{Brick2012}
\textsc{Brick, K., M.~Visser, and J.~Burns} (2012): \enquote{Risk Aversion:
  Experimental Evidence from South African Fishing Communities,} \emph{American
  Journal of Agricultural Economics}, 94, 133--152.

\bibitem[\protect\citeauthoryear{Browne, Hofmann, Richter, Roth, and
  Steinorth}{Browne et~al.}{2020}]{Browne2020}
\textsc{Browne, M.~J., A.~Hofmann, A.~Richter, S.-M. Roth, and P.~Steinorth}
  (2020): \enquote{Peer effects in risk preferences: Evidence from Germany,}
  \emph{Annals of Operations Research}.

\bibitem[\protect\citeauthoryear{Browne, Jaeger, Richter, and Steinorth}{Browne
  et~al.}{2016}]{Browne2016}
\textsc{Browne, M.~J., V.~Jaeger, A.~Richter, and P.~Steinorth} (2016):
  \enquote{Family {Transitions} and {Risk} {Attitude},} \emph{Available at SSRN
  2710688}.

\bibitem[\protect\citeauthoryear{{Bruderer Enzler}, Diekmann, and
  Meyer}{{Bruderer Enzler} et~al.}{2014}]{BrudererEnzler2014}
\textsc{{Bruderer Enzler}, H., A.~Diekmann, and R.~Meyer} (2014):
  \enquote{Subjective discount rates in the general population and their
  predictive power for energy saving behavior,} \emph{Energy Policy}, 65,
  524--540.

\bibitem[\protect\citeauthoryear{Bucciol and Miniaci}{Bucciol and
  Miniaci}{2011}]{Bucciol2011}
\textsc{Bucciol, A. and R.~Miniaci} (2011): \enquote{Household portfolios and
  implicit risk preference,} \emph{Review of Economics and Statistics}, 93,
  1235--1250.

\bibitem[\protect\citeauthoryear{Burks, Carpenter, G{\"o}tte, and
  Rustichini}{Burks et~al.}{2009}]{Burks2009}
\textsc{Burks, S.~V., J.~P. Carpenter, L.~G{\"o}tte, and A.~Rustichini} (2009):
  \enquote{Cognitive skills affect economic preferences, strategic behavior,
  and job attachment,} \emph{Proceedings of the National Academy of Sciences},
  106, 7745--7750.

\bibitem[\protect\citeauthoryear{Burks, Carpenter, G{\"o}tte, and
  Rustichini}{Burks et~al.}{2012}]{Burks2012}
---\hspace{-.1pt}---\hspace{-.1pt}--- (2012): \enquote{Which measures of time
  preference best predict outcomes: Evidence from a large-scale field
  experiment,} \emph{Journal of Economic Behavior \& Organization}, 84,
  308--320.

\bibitem[\protect\citeauthoryear{Burks, Lewis, Kivi, Wiener, Anderson,
  G{\"o}tte, DeYoung, and Rustichini}{Burks et~al.}{2015}]{Burks2015}
\textsc{Burks, S.~V., C.~Lewis, P.~A. Kivi, A.~Wiener, J.~E. Anderson,
  L.~G{\"o}tte, C.~G. DeYoung, and A.~Rustichini} (2015): \enquote{Cognitive
  skills, personality, and economic preferences in collegiate success,}
  \emph{Journal of Economic Behavior \& Organization}, 115, 30--44.

\bibitem[\protect\citeauthoryear{Buser, Grimalda, Putterman, and van~der
  Weele}{Buser et~al.}{2020}]{Buser2020}
\textsc{Buser, T., G.~Grimalda, L.~Putterman, and J.~van~der Weele} (2020):
  \enquote{Overconfidence and gender gaps in redistributive preferences:
  Cross-Country experimental evidence,} \emph{Journal of Economic Behavior {\&}
  Organization}, 178, 267--286.

\bibitem[\protect\citeauthoryear{Busic-Sontic and Brick}{Busic-Sontic and
  Brick}{2018}]{BusicSontic2018}
\textsc{Busic-Sontic, A. and C.~Brick} (2018): \enquote{Personality Trait
  Effects on Green Household Installations,} \emph{Collabra: Psychology}, 4.

\bibitem[\protect\citeauthoryear{Butler, Guiso, and Jappelli}{Butler
  et~al.}{2013}]{Butler2013}
\textsc{Butler, J.~V., L.~Guiso, and T.~Jappelli} (2013): \enquote{The role of
  intuition and reasoning in driving aversion to risk and ambiguity,}
  \emph{Theory and Decision}, 77, 455--484.

\bibitem[\protect\citeauthoryear{Camerer and Hogarth}{Camerer and
  Hogarth}{1999}]{Camerer1999}
\textsc{Camerer, C.~F. and R.~M. Hogarth} (1999): \enquote{The effects of
  financial incentives in experiments: A review and capital-labor-production
  framework,} \emph{Journal of risk and uncertainty}, 19, 7--42.

\bibitem[\protect\citeauthoryear{Cardenas and Carpenter}{Cardenas and
  Carpenter}{2013}]{Cardenas2013}
\textsc{Cardenas, J.~C. and J.~Carpenter} (2013): \enquote{Risk attitudes and
  economic well-being in Latin America,} \emph{Journal of Development
  Economics}, 103, 52--61.

\bibitem[\protect\citeauthoryear{Castillo, Petrie, Cotla, and Torero}{Castillo
  et~al.}{2018}]{Castillo2018}
\textsc{Castillo, M., R.~Petrie, C.~R. Cotla, and M.~Torero} (2018):
  \enquote{Risk preferences and decision quality of the poor,} \emph{Working
  Paper}.

\bibitem[\protect\citeauthoryear{Chapman, Dean, Ortoleva, Snowberg, and
  Camerer}{Chapman et~al.}{2018{\natexlab{a}}}]{Chapman2018}
\textsc{Chapman, J., M.~Dean, P.~Ortoleva, E.~Snowberg, and C.~Camerer}
  (2018{\natexlab{a}}): \enquote{Econographics,} \emph{Working Paper}.

\bibitem[\protect\citeauthoryear{Chapman, Snowberg, Wang, and Camerer}{Chapman
  et~al.}{2018{\natexlab{b}}}]{Chapman2018a}
\textsc{Chapman, J., E.~Snowberg, S.~Wang, and C.~Camerer}
  (2018{\natexlab{b}}): \enquote{Loss attitudes in the US population: Evidence
  from dynamically optimized sequential experimentation (DOSE),} \emph{Working
  Paper}.

\bibitem[\protect\citeauthoryear{Charness, Gneezy, and Halladay}{Charness
  et~al.}{2016}]{Charness2016}
\textsc{Charness, G., U.~Gneezy, and B.~Halladay} (2016): \enquote{Experimental
  methods: Pay one or pay all,} \emph{Journal of Economic Behavior \&
  Organization}, 131, 141--150.

\bibitem[\protect\citeauthoryear{Clot, Grolleau, and Ibanez}{Clot
  et~al.}{2018}]{Clot2018}
\textsc{Clot, S., G.~Grolleau, and L.~Ibanez} (2018): \enquote{Shall we pay
  all? An experimental test of Random Incentivized Systems,} \emph{Journal of
  Behavioral and Experimental Economics}, 73, 93--98.

\bibitem[\protect\citeauthoryear{Clot, Stanton, and Willinger}{Clot
  et~al.}{2017}]{Clot2017}
\textsc{Clot, S., C.~Y. Stanton, and M.~Willinger} (2017): \enquote{Are
  impatient farmers more risk-averse? Evidence from a lab-in-the-field
  experiment in rural Uganda,} \emph{Applied Economics}, 49, 156--169.

\bibitem[\protect\citeauthoryear{Coller and Williams}{Coller and
  Williams}{1999}]{Coller1999}
\textsc{Coller, M. and M.~B. Williams} (1999): \enquote{Eliciting individual
  discount rates,} \emph{Experimental Economics}, 2, 107--127.

\bibitem[\protect\citeauthoryear{Dasgupta, Mani, Sharma, and Singhal}{Dasgupta
  et~al.}{2016}]{Dasgupta2016}
\textsc{Dasgupta, U., S.~Mani, S.~Sharma, and S.~Singhal} (2016):
  \enquote{Caste differences in behaviour and personality: Evidence from
  India,} \emph{WIDER Working Paper}.

\bibitem[\protect\citeauthoryear{Dave, Eckel, Johnson, and Rojas}{Dave
  et~al.}{2010}]{Dave2010}
\textsc{Dave, C., C.~C. Eckel, C.~A. Johnson, and C.~Rojas} (2010):
  \enquote{Eliciting risk preferences: When is simple better?} \emph{Journal of
  Risk and Uncertainty}, 41, 219--243.

\bibitem[\protect\citeauthoryear{Delavande, Ganguli, and Mengel}{Delavande
  et~al.}{2018}]{Delavande2018}
\textsc{Delavande, A., J.~Ganguli, and F.~Mengel} (2018): \enquote{Attitudes to
  uncertainty and household decisions,} \emph{Working Paper}.

\bibitem[\protect\citeauthoryear{Di~Falco and Vieider}{Di~Falco and
  Vieider}{2018}]{DiFalco2018}
\textsc{Di~Falco, S. and F.~M. Vieider} (2018): \enquote{Assimilation in the
  risk preferences of spouses,} \emph{Economic Inquiry}, 56, 1809--1816.

\bibitem[\protect\citeauthoryear{Dittrich and Leipold}{Dittrich and
  Leipold}{2014}]{Dittrich2014}
\textsc{Dittrich, M. and K.~Leipold} (2014): \enquote{Gender differences in
  time preferences,} \emph{Economics Letters}, 122, 413--415.

\bibitem[\protect\citeauthoryear{Dohmen, Falk, Huffman, and Sunde}{Dohmen
  et~al.}{2010}]{Dohmen2010}
\textsc{Dohmen, T., A.~Falk, D.~Huffman, and U.~Sunde} (2010): \enquote{Are
  Risk Aversion and Impatience Related to Cognitive Ability?} \emph{American
  Economic Review}, 100, 1238--1260.

\bibitem[\protect\citeauthoryear{Dohmen, Falk, Huffman, Sunde, Schupp, and
  Wagner}{Dohmen et~al.}{2011}]{Dohmen2011}
\textsc{Dohmen, T., A.~Falk, D.~Huffman, U.~Sunde, J.~Schupp, and G.~G. Wagner}
  (2011): \enquote{Individual risk attitudes: Measurement, determinants, and
  behavioral consequences,} \emph{Journal of the European Economic
  Association}, 9, 522--550.

\bibitem[\protect\citeauthoryear{Donkers, Melenberg, and Van~Soest}{Donkers
  et~al.}{2001}]{Donkers2001}
\textsc{Donkers, B., B.~Melenberg, and A.~Van~Soest} (2001):
  \enquote{Estimating risk attitudes using lotteries: A large sample approach,}
  \emph{Journal of Risk and Uncertainty}, 22, 165--195.

\bibitem[\protect\citeauthoryear{Drichoutis and Lusk}{Drichoutis and
  Lusk}{2014}]{Drichoutis2014}
\textsc{Drichoutis, A.~C. and J.~L. Lusk} (2014): \enquote{Judging statistical
  models of individual decision making under risk using in-and out-of-sample
  criteria,} \emph{PloS one}, 9, e102269.

\bibitem[\protect\citeauthoryear{Drichoutis and Nayga}{Drichoutis and
  Nayga}{2021}]{Drichoutis2021}
\textsc{Drichoutis, A.~C. and R.~M. Nayga} (2021): \enquote{On the stability of
  risk and time preferences amid the COVID-19 pandemic,} \emph{Experimental
  Economics}, 1--36.

\bibitem[\protect\citeauthoryear{Eckel, Johnson, and Montmarquette}{Eckel
  et~al.}{2005}]{Eckel2005}
\textsc{Eckel, C., C.~Johnson, and C.~Montmarquette} (2005): \enquote{Saving
  decisions of the working poor: Short-and long-term horizons,} \emph{Research
  in experimental economics}, 10, 219--260.

\bibitem[\protect\citeauthoryear{Faff, Hallahan, and McKenzie}{Faff
  et~al.}{2011}]{Faff2011}
\textsc{Faff, R., T.~Hallahan, and M.~McKenzie} (2011): \enquote{Women and risk
  tolerance in an aging world,} \emph{International Journal of Accounting \&
  Information Management}, 19, 100--117.

\bibitem[\protect\citeauthoryear{Falk, Becker, Dohmen, Enke, Huffman, and
  Sunde}{Falk et~al.}{2018}]{Falk2018}
\textsc{Falk, A., A.~Becker, T.~Dohmen, B.~Enke, D.~Huffman, and U.~Sunde}
  (2018): \enquote{Global evidence on economic preferences,} \emph{The
  Quarterly Journal of Economics}, 133, 1645--1692.

\bibitem[\protect\citeauthoryear{Fan, Orhun, and Turjeman}{Fan
  et~al.}{2020}]{Fan2020}
\textsc{Fan, Y., A.~Y. Orhun, and D.~Turjeman} (2020): \enquote{Heterogeneous
  Actions, Beliefs, Constraints and Risk Tolerance During the COVID-19
  Pandemic,} \emph{NBER Working Papers 27211}.

\bibitem[\protect\citeauthoryear{Fang, Hanna, and Chatterjee}{Fang
  et~al.}{2013}]{Fang2013}
\textsc{Fang, M.-C., S.~Hanna, and S.~Chatterjee} (2013): \enquote{The impact
  of immigrant status and racial/ethnic group on differences in responses to a
  risk aversion measure,} \emph{Journal of Financial Counseling and Planning},
  24, 76--89.

\bibitem[\protect\citeauthoryear{Fehr and Reichlin}{Fehr and
  Reichlin}{2021}]{Fehr2021}
\textsc{Fehr, D. and Y.~Reichlin} (2021): \enquote{Perceived Relative Wealth
  and Risk Taking,} \emph{Available at SSRN 3907737}.

\bibitem[\protect\citeauthoryear{Frederick}{Frederick}{2005}]{Frederick2005}
\textsc{Frederick, S.} (2005): \enquote{Cognitive reflection and decision
  making,} \emph{Journal of Economic perspectives}, 19, 25--42.

\bibitem[\protect\citeauthoryear{Fredslund, M{\o}rkbak, and
  Gyrd-Hansen}{Fredslund et~al.}{2018}]{Fredslund2018}
\textsc{Fredslund, E.~K., M.~R. M{\o}rkbak, and D.~Gyrd-Hansen} (2018):
  \enquote{Different domains {\textendash} Different time preferences?}
  \emph{Social Science {\&} Medicine}, 207, 97--105.

\bibitem[\protect\citeauthoryear{Freeney and O'Connell}{Freeney and
  O'Connell}{2010}]{Freeney2010}
\textsc{Freeney, Y. and M.~O'Connell} (2010): \enquote{Wait for it:
  Delay-discounting and academic performance among an Irish adolescent sample,}
  \emph{Learning and individual differences}, 20, 231--236.

\bibitem[\protect\citeauthoryear{Frey, Richter, Schupp, Hertwig, and Mata}{Frey
  et~al.}{2021}]{Frey2021}
\textsc{Frey, R., D.~Richter, J.~Schupp, R.~Hertwig, and R.~Mata} (2021):
  \enquote{Identifying robust correlates of risk preference: A systematic
  approach using specification curve analysis.} \emph{Journal of Personality
  and Social Psychology}, 120, 538.

\bibitem[\protect\citeauthoryear{G{\"a}chter, Johnson, and
  Herrmann}{G{\"a}chter et~al.}{2021}]{Gachter2021}
\textsc{G{\"a}chter, S., E.~J. Johnson, and A.~Herrmann} (2021):
  \enquote{Individual-level loss aversion in riskless and risky choices,}
  \emph{Theory and Decision}, 1--26.

\bibitem[\protect\citeauthoryear{Galizzi, Harrison, and Miraldo}{Galizzi
  et~al.}{2018}]{Galizzi2018}
\textsc{Galizzi, M.~M., G.~W. Harrison, and M.~Miraldo} (2018):
  \emph{Experimental methods and behavioral insights in health economics:
  Estimating risk and time preferences in health}, Emerald Publishing Limited.

\bibitem[\protect\citeauthoryear{Galizzi, Machado, and Miniaci}{Galizzi
  et~al.}{2016}]{Galizzi2016}
\textsc{Galizzi, M.~M., S.~R. Machado, and R.~Miniaci} (2016):
  \enquote{Temporal stability, cross-validity, and external validity of risk
  preferences measures: Experimental evidence from a UK representative sample,}
  \emph{Working Paper}.

\bibitem[\protect\citeauthoryear{G{\"a}rtner, Mollerstrom, and
  Seim}{G{\"a}rtner et~al.}{2017}]{Gartner2017}
\textsc{G{\"a}rtner, M., J.~Mollerstrom, and D.~Seim} (2017):
  \enquote{Individual risk preferences and the demand for redistribution,}
  \emph{Journal of Public Economics}, 153, 49--55.

\bibitem[\protect\citeauthoryear{Giampietri, Bugin, and Trestini}{Giampietri
  et~al.}{2021}]{Giampietri2021}
\textsc{Giampietri, E., G.~Bugin, and S.~Trestini} (2021): \enquote{On the
  association between risk attitude and fruit and vegetable consumption:
  insights from university students in Italy,} \emph{Agricultural and Food
  Economics}, 9, 1--16.

\bibitem[\protect\citeauthoryear{Gillen, Snowberg, and Yariv}{Gillen
  et~al.}{2019}]{Gillen2019}
\textsc{Gillen, B., E.~Snowberg, and L.~Yariv} (2019): \enquote{Experimenting
  with measurement error: Techniques with applications to the caltech cohort
  study,} \emph{Journal of Political Economy}, 127, 1826--1863.

\bibitem[\protect\citeauthoryear{Gloede, Menkhoff, and Waibel}{Gloede
  et~al.}{2015}]{Gloede2015}
\textsc{Gloede, O., L.~Menkhoff, and H.~Waibel} (2015): \enquote{Shocks,
  Individual Risk Attitude, and Vulnerability to Poverty among Rural Households
  in Thailand and Vietnam,} \emph{World Development}, 71, 54--78.

\bibitem[\protect\citeauthoryear{Goda, Levy, Manchester, Sojourner, and
  Tasoff}{Goda et~al.}{2019}]{Goda2019}
\textsc{Goda, G.~S., M.~Levy, C.~F. Manchester, A.~Sojourner, and J.~Tasoff}
  (2019): \enquote{Predicting Retirement Savings Using Survey Measures Of
  Exponential - Growth Bias And Present Bias,} \emph{Economic Inquiry}, 57,
  1636--1658.

\bibitem[\protect\citeauthoryear{Golsteyn, Gr{\"o}nqvist, and Lindahl}{Golsteyn
  et~al.}{2014}]{Golsteyn2014}
\textsc{Golsteyn, B.~H., H.~Gr{\"o}nqvist, and L.~Lindahl} (2014):
  \enquote{Adolescent time preferences predict lifetime outcomes,} \emph{The
  Economic Journal}, 124, F739--F761.

\bibitem[\protect\citeauthoryear{G{\"o}rlitz and Tamm}{G{\"o}rlitz and
  Tamm}{2020}]{Gorlitz2020}
\textsc{G{\"o}rlitz, K. and M.~Tamm} (2020): \enquote{Parenthood, risk
  attitudes and risky behavior,} \emph{Journal of Economic Psychology}, 79,
  102189.

\bibitem[\protect\citeauthoryear{Guenther, Galizzi, and Sanders}{Guenther
  et~al.}{2021}]{Guenther2021}
\textsc{Guenther, B., M.~M. Galizzi, and J.~G. Sanders} (2021):
  \enquote{Heterogeneity in Risk-Taking During the COVID-19 Pandemic: Evidence
  From the UK Lockdown,} \emph{Frontiers in Psychology}, 12, 852.

\bibitem[\protect\citeauthoryear{Halko, Kaustia, and Alanko}{Halko
  et~al.}{2012}]{Halko2012}
\textsc{Halko, M.-L., M.~Kaustia, and E.~Alanko} (2012): \enquote{The gender
  effect in risky asset holdings,} \emph{Journal of Economic Behavior {\&}
  Organization}, 83, 66--81.

\bibitem[\protect\citeauthoryear{Hansen, Jacobsen, and Lau}{Hansen
  et~al.}{2016}]{Hansen2016}
\textsc{Hansen, J.~V., R.~H. Jacobsen, and M.~I. Lau} (2016):
  \enquote{Willingness to pay for insurance in Denmark,} \emph{Journal of Risk
  and Insurance}, 83, 49--76.

\bibitem[\protect\citeauthoryear{Hardeweg, Menkhoff, and Waibel}{Hardeweg
  et~al.}{2013}]{Hardeweg2013}
\textsc{Hardeweg, B., L.~Menkhoff, and H.~Waibel} (2013):
  \enquote{Experimentally validated survey evidence on individual risk
  attitudes in rural Thailand,} \emph{Economic Development and Cultural
  Change}, 61, 859--888.

\bibitem[\protect\citeauthoryear{Harrati}{Harrati}{2014}]{Harrati2014}
\textsc{Harrati, A.} (2014): \enquote{Characterizing the genetic influences on
  risk aversion,} \emph{Biodemography and Social Biology}, 60, 185--198.

\bibitem[\protect\citeauthoryear{Harrison, Johnson, McInnes, and
  Rutstr{\"o}m}{Harrison et~al.}{2005}]{Harrison2005}
\textsc{Harrison, G.~W., E.~Johnson, M.~M. McInnes, and E.~E. Rutstr{\"o}m}
  (2005): \enquote{Risk aversion and incentive effects: Comment,}
  \emph{American Economic Review}, 95, 897--901.

\bibitem[\protect\citeauthoryear{Harrison, Lau, and Rutstr{\"o}m}{Harrison
  et~al.}{2007}]{Harrison2007}
\textsc{Harrison, G.~W., M.~I. Lau, and E.~E. Rutstr{\"o}m} (2007):
  \enquote{Estimating risk attitudes in Denmark: A field experiment,}
  \emph{Scandinavian Journal of Economics}, 109, 341--368.

\bibitem[\protect\citeauthoryear{Harrison, Lau, and Williams}{Harrison
  et~al.}{2002}]{Harrison2002}
\textsc{Harrison, G.~W., M.~I. Lau, and M.~B. Williams} (2002):
  \enquote{Estimating individual discount rates in Denmark: A field
  experiment,} \emph{American economic review}, 92, 1606--1617.

\bibitem[\protect\citeauthoryear{Harrison et~al.}{Harrison
  et~al.}{2008}]{Harrison2008}
\textsc{Harrison, G.~W. et~al.} (2008): \emph{Risk aversion in experiments},
  Emerald Group Publishing.

\bibitem[\protect\citeauthoryear{Hofmeyr, Monterosso, Dean, Morales, Bilder,
  Sabb, and London}{Hofmeyr et~al.}{2016}]{Hofmeyr2016}
\textsc{Hofmeyr, A., J.~Monterosso, A.~C. Dean, A.~M. Morales, R.~M. Bilder,
  F.~W. Sabb, and E.~D. London} (2016): \enquote{Mixture models of delay
  discounting and smoking behavior,} \emph{The American Journal of Drug and
  Alcohol Abuse}, 43, 271--280.

\bibitem[\protect\citeauthoryear{Holden and Tilahun}{Holden and
  Tilahun}{2019}]{Holden2019}
\textsc{Holden, S. and M.~Tilahun} (2019): \enquote{Gender Differences in Risk
  Tolerance, Trust and Trustworthiness: Are They Related?} \emph{CLTS Working
  Papers}.

\bibitem[\protect\citeauthoryear{Holm, Opper, and Nee}{Holm
  et~al.}{2013}]{Holm2013}
\textsc{Holm, H.~J., S.~Opper, and V.~Nee} (2013): \enquote{Entrepreneurs under
  uncertainty: An economic experiment in China,} \emph{Management Science}, 59,
  1671--1687.

\bibitem[\protect\citeauthoryear{Holt and Laury}{Holt and
  Laury}{2002}]{Holt2002}
\textsc{Holt, C.~A. and S.~K. Laury} (2002): \enquote{Risk aversion and
  incentive effects,} \emph{American economic review}, 92, 1644--1655.

\bibitem[\protect\citeauthoryear{Hopland, Matsen, and Str{\o}m}{Hopland
  et~al.}{2016}]{Hopland2016}
\textsc{Hopland, A.~O., E.~Matsen, and B.~Str{\o}m} (2016): \enquote{Income and
  choice under risk,} \emph{Journal of Behavioral and Experimental Finance},
  12, 55--64.

\bibitem[\protect\citeauthoryear{Horn and Kiss}{Horn and Kiss}{2019}]{Horn2019}
\textsc{Horn, D. and H.~J. Kiss} (2019): \enquote{Gender differences in risk
  aversion and patience: evidence from a representative survey,}
  \emph{Available at SSRN 3313749}.

\bibitem[\protect\citeauthoryear{Horn and Kiss}{Horn and
  Kiss}{2020{\natexlab{a}}}]{Horn2020}
---\hspace{-.1pt}---\hspace{-.1pt}--- (2020{\natexlab{a}}): \enquote{Do
  individuals with children value the future more?} \emph{Available at SSRN
  3540540}.

\bibitem[\protect\citeauthoryear{Horn and Kiss}{Horn and
  Kiss}{2020{\natexlab{b}}}]{Horn2020a}
---\hspace{-.1pt}---\hspace{-.1pt}--- (2020{\natexlab{b}}): \enquote{Time
  preferences and their life outcome correlates: Evidence from a representative
  survey,} \emph{PloS one}, 15, e0236486.

\bibitem[\protect\citeauthoryear{Howard and Roe}{Howard and
  Roe}{2011}]{Howard2011}
\textsc{Howard, G. and B.~Roe} (2011): \enquote{Comparing the Risk Attitudes of
  U.S. and German Farmers,} 2011 International Congress, August 30-September 2,
  2011, Zurich, Switzerland 114528, European Association of Agricultural
  Economists.

\bibitem[\protect\citeauthoryear{Huffman, Maurer, and Mitchell}{Huffman
  et~al.}{2019}]{Huffman2019}
\textsc{Huffman, D., R.~Maurer, and O.~S. Mitchell} (2019): \enquote{Time
  discounting and economic decision-making in the older population,} \emph{The
  Journal of the Economics of Ageing}, 14, 100121.

\bibitem[\protect\citeauthoryear{Hunter, Tang, Hutchinson, Chilton, Holmes, and
  Kee}{Hunter et~al.}{2018}]{Hunter2018}
\textsc{Hunter, R.~F., J.~Tang, G.~Hutchinson, S.~Chilton, D.~Holmes, and
  F.~Kee} (2018): \enquote{Association between time preference, present-bias
  and physical activity: implications for designing behavior change
  interventions,} \emph{{BMC} Public Health}, 18.

\bibitem[\protect\citeauthoryear{Hvide and Lee}{Hvide and
  Lee}{2015}]{Hvide2015}
\textsc{Hvide, H.~K. and J.~H. Lee} (2015): \enquote{Does source of income
  affect risk and intertemporal choices?} \emph{Working Paper}.

\bibitem[\protect\citeauthoryear{Hwang}{Hwang}{2017}]{Hwang2017}
\textsc{Hwang, I.~D.} (2017): \enquote{Behavioral aspects of household
  portfolio choice: Effects of loss aversion on life insurance uptake and
  savings,} \emph{Bank of Korea WP}, 8.

\bibitem[\protect\citeauthoryear{Ioannou and Sadeh}{Ioannou and
  Sadeh}{2016}]{Ioannou2016}
\textsc{Ioannou, C.~A. and J.~Sadeh} (2016): \enquote{Time preferences and risk
  aversion: Tests on domain differences,} \emph{Journal of Risk and
  Uncertainty}, 53, 29--54.

\bibitem[\protect\citeauthoryear{Ionescu and Turlea}{Ionescu and
  Turlea}{2011}]{Ionescu2011}
\textsc{Ionescu, I.~O. and E.~Turlea} (2011): \enquote{The Financial Auditor's
  Risk Behaviour-The Influence of Age on Risk Behaviour in A Financial Audit
  Context,} \emph{Accounting and Management Information Systems}, 10, 444.

\bibitem[\protect\citeauthoryear{Jarmolowicz, Bickel, Carter, Franck, and
  Mueller}{Jarmolowicz et~al.}{2012}]{Jarmolowicz2012}
\textsc{Jarmolowicz, D.~P., W.~K. Bickel, A.~E. Carter, C.~T. Franck, and E.~T.
  Mueller} (2012): \enquote{Using crowdsourcing to examine relations between
  delay and probability discounting,} \emph{Behavioural Processes}, 91,
  308--312.

\bibitem[\protect\citeauthoryear{Jin, He, Gong, Xu, and He}{Jin
  et~al.}{2017}]{Jin2017}
\textsc{Jin, J., R.~He, H.~Gong, X.~Xu, and C.~He} (2017): \enquote{Farmers'
  Risk Preferences in Rural China: Measurements and Determinants,}
  \emph{International Journal of Environmental Research and Public Health}, 14,
  713.

\bibitem[\protect\citeauthoryear{Johnson, G{\"a}chter, and Herrmann}{Johnson
  et~al.}{2006}]{Johnson2006}
\textsc{Johnson, E.~J., S.~G{\"a}chter, and A.~Herrmann} (2006):
  \enquote{{Exploring the Nature of Loss Aversion},} \emph{Available at SSRN
  892336}.

\bibitem[\protect\citeauthoryear{Jung}{Jung}{2017}]{Jung2017}
\textsc{Jung, S.} (2017): \enquote{The gender wage gap and sample selection via
  risk attitudes,} \emph{International Journal of Manpower}, 38, 318--335.

\bibitem[\protect\citeauthoryear{Jung and Treibich}{Jung and
  Treibich}{2015}]{Jung2015}
\textsc{Jung, S. and C.~Treibich} (2015): \enquote{Is Self-Reported Risk
  Aversion Time Variant?} \emph{Revue d'{\'e}conomie politique}, 125, 547--570.

\bibitem[\protect\citeauthoryear{Kahneman and Tversky}{Kahneman and
  Tversky}{1979}]{Kahneman1979}
\textsc{Kahneman, D. and A.~Tversky} (1979): \enquote{Prospect Theory: An
  Analysis of Decision under Risk,} \emph{Econometrica}, 47, 263--292.

\bibitem[\protect\citeauthoryear{Kam}{Kam}{2012}]{Kam2012}
\textsc{Kam, C.~D.} (2012): \enquote{Risk Attitudes and Political
  Participation,} \emph{American Journal of Political Science}, 56, 817--836.

\bibitem[\protect\citeauthoryear{Kettlewell}{Kettlewell}{2019}]{Kettlewell2019}
\textsc{Kettlewell, N.} (2019): \enquote{Risk preference dynamics around life
  events,} \emph{Journal of Economic Behavior {\&} Organization}, 162, 66--84.

\bibitem[\protect\citeauthoryear{Khachatryan, Dreber, von Essen, and
  Ranehill}{Khachatryan et~al.}{2015}]{Khachatryan2015}
\textsc{Khachatryan, K., A.~Dreber, E.~von Essen, and E.~Ranehill} (2015):
  \enquote{Gender and preferences at a young age: Evidence from Armenia,}
  \emph{Journal of Economic Behavior {\&} Organization}, 118, 318--332.

\bibitem[\protect\citeauthoryear{Kim and Lee}{Kim and Lee}{2012}]{Kim2012}
\textsc{Kim, Y.-I. and J.~Lee} (2012): \enquote{Estimating risk aversion using
  individual-level survey data,} \emph{The Korean Economic Review}, 28,
  221--239.

\bibitem[\protect\citeauthoryear{Knoppen and Saris}{Knoppen and
  Saris}{2009}]{Knoppen2009}
\textsc{Knoppen, D. and W.~Saris} (2009): \enquote{Do we have to combine values
  in the Schwartz' human values scale? A comment on the Davidov studies,} in
  \emph{Survey Research Methods}, vol.~3, 91--103.

\bibitem[\protect\citeauthoryear{Kristjanpoller and Olson}{Kristjanpoller and
  Olson}{2015}]{Kristjanpoller2015}
\textsc{Kristjanpoller, W.~D. and J.~E. Olson} (2015): \enquote{Choice of
  Retirement Funds in Chile: Are Chilean Women More Risk Averse than Men?}
  \emph{Sex Roles}, 72, 50--67.

\bibitem[\protect\citeauthoryear{Laban~Peryman}{Laban~Peryman}{2015}]{Peryman2015}
\textsc{Laban~Peryman, J.} (2015): \enquote{Culture and Risk-Sharing Networks,}
  \emph{Available at SSRN 2912006}.

\bibitem[\protect\citeauthoryear{Laibson}{Laibson}{1997}]{Laibson1997}
\textsc{Laibson, D.} (1997): \enquote{Golden eggs and hyperbolic discounting,}
  \emph{The Quarterly Journal of Economics}, 112, 443--478.

\bibitem[\protect\citeauthoryear{Lampi and Nordblom}{Lampi and
  Nordblom}{2013}]{Lampi2013}
\textsc{Lampi, E. and K.~Nordblom} (2013): \enquote{Risk-taking middle-borns: A
  study on birth-order and risk preferences,} \emph{Journal of Communications
  Research}, 5.

\bibitem[\protect\citeauthoryear{Le, Miller, Slutske, and Martin}{Le
  et~al.}{2010}]{Le2010}
\textsc{Le, A.~T., P.~W. Miller, W.~S. Slutske, and N.~G. Martin} (2010):
  \enquote{Are attitudes towards economic risk heritable? Analyses using the
  Australian twin study of gambling,} \emph{Twin Research and Human Genetics},
  13, 330--339.

\bibitem[\protect\citeauthoryear{Lee and Kang}{Lee and Kang}{2016}]{Lee2016}
\textsc{Lee, S.~H. and H.~G. Kang} (2016): \enquote{Integrated framework for
  the external cost assessment of nuclear power plant accident considering risk
  aversion: The Korean case,} \emph{Energy Policy}, 92, 111--123.

\bibitem[\protect\citeauthoryear{l'Haridon and Vieider}{l'Haridon and
  Vieider}{2019}]{Haridon2019}
\textsc{l'Haridon, O. and F.~M. Vieider} (2019): \enquote{All over the map: A
  worldwide comparison of risk preferences,} \emph{Quantitative Economics}, 10,
  185--215.

\bibitem[\protect\citeauthoryear{Lilleholt}{Lilleholt}{2019}]{Lilleholt2019}
\textsc{Lilleholt, L.} (2019): \enquote{Cognitive ability and risk aversion: A
  systematic review and meta analysis.} \emph{Judgment \& Decision Making}, 14.

\bibitem[\protect\citeauthoryear{Luce}{Luce}{1960}]{Luce1960}
\textsc{Luce, R.~D.} (1960): \enquote{Individual choice behavior, a theoretical
  analysis,} \emph{Bull. Amer. Math. Soc}, 66, 259--260.

\bibitem[\protect\citeauthoryear{Mart{\'{\i}}n-Fern{\'{a}}ndez,
  Medina-Palomino, Ariza-Cardiel, Polentinos-Castro, and
  Rutkowski}{Mart{\'{\i}}n-Fern{\'{a}}ndez et~al.}{2018}]{MartinFernandez2018}
\textsc{Mart{\'{\i}}n-Fern{\'{a}}ndez, J., H.~J. Medina-Palomino,
  G.~Ariza-Cardiel, E.~Polentinos-Castro, and A.~Rutkowski} (2018):
  \enquote{Health condition and risk attitude in the Dutch population: an
  exploratory approach,} \emph{Health, Risk {\&} Society}, 20, 126--146.

\bibitem[\protect\citeauthoryear{Mata, Josef, and Hertwig}{Mata
  et~al.}{2016}]{Mata2016}
\textsc{Mata, R., A.~K. Josef, and R.~Hertwig} (2016): \enquote{Propensity for
  risk taking across the life span and around the globe,} \emph{Psychological
  science}, 27, 231--243.

\bibitem[\protect\citeauthoryear{Meier and Sprenger}{Meier and
  Sprenger}{2015}]{Meier2015}
\textsc{Meier, S. and C.~D. Sprenger} (2015): \enquote{Temporal stability of
  time preferences,} \emph{Review of Economics and Statistics}, 97, 273--286.

\bibitem[\protect\citeauthoryear{Meissner and Pfeiffer}{Meissner and
  Pfeiffer}{2022}]{meissner2022measuring}
\textsc{Meissner, T. and P.~Pfeiffer} (2022): \enquote{Measuring preferences
  over the temporal resolution of consumption uncertainty,} \emph{Journal of
  Economic Theory}, 200, 105379.

\bibitem[\protect\citeauthoryear{Melesse and Cecchi}{Melesse and
  Cecchi}{2017}]{Melesse2017}
\textsc{Melesse, M.~B. and F.~Cecchi} (2017): \enquote{Does Market Experience
  Attenuate Risk Aversion? Evidence from Landed Farm Households in Ethiopia,}
  \emph{World Development}, 98, 447--466.

\bibitem[\protect\citeauthoryear{Menkhoff and Sakha}{Menkhoff and
  Sakha}{2017}]{Menkhoff2017}
\textsc{Menkhoff, L. and S.~Sakha} (2017): \enquote{Estimating risky behavior
  with multiple-item risk measures,} \emph{Journal of Economic Psychology}, 59,
  59--86.

\bibitem[\protect\citeauthoryear{Nebout, Cavillon, and Ventelou}{Nebout
  et~al.}{2018}]{Nebout2018}
\textsc{Nebout, A., M.~Cavillon, and B.~Ventelou} (2018): \enquote{Comparing
  {GPs}' risk attitudes for their own health and for their patients' : a
  troubling discrepancy?} \emph{{BMC} Health Services Research}, 18.

\bibitem[\protect\citeauthoryear{Newell and Siikam{\"a}ki}{Newell and
  Siikam{\"a}ki}{2015}]{Newell2015}
\textsc{Newell, R.~G. and J.~Siikam{\"a}ki} (2015): \enquote{Individual Time
  Preferences and Energy Efficiency,} \emph{American Economic Review}, 105,
  196--200.

\bibitem[\protect\citeauthoryear{Neyse, Johannesson, and Dreber}{Neyse
  et~al.}{2020}]{Neyse2020}
\textsc{Neyse, L., M.~Johannesson, and A.~Dreber} (2020): \enquote{2D: 4D does
  not predict economic preferences: Evidence from a large, representative
  sample,} \emph{Working Paper}.

\bibitem[\protect\citeauthoryear{Noussair, Trautmann, and Van~de
  Kuilen}{Noussair et~al.}{2014}]{Noussair2014}
\textsc{Noussair, C.~N., S.~T. Trautmann, and G.~Van~de Kuilen} (2014):
  \enquote{Higher order risk attitudes, demographics, and financial decisions,}
  \emph{Review of Economic Studies}, 81, 325--355.

\bibitem[\protect\citeauthoryear{Oechssler, Roider, and Schmitz}{Oechssler
  et~al.}{2009}]{Oechssler2009}
\textsc{Oechssler, J., A.~Roider, and P.~W. Schmitz} (2009): \enquote{Cognitive
  abilities and behavioral biases,} \emph{Journal of Economic Behavior \&
  Organization}, 72, 147--152.

\bibitem[\protect\citeauthoryear{Osama}{Osama}{2017}]{Osama2017}
\textsc{Osama, M.} (2017): \enquote{Individual Risk Attitudes with evidence
  from experimentally- validated Survey (The Egypt Labor Market Panel Survey
  (ELMPS)),} \emph{Journal of Economics and Finance}, 8, 32--37.

\bibitem[\protect\citeauthoryear{Parouty, Le, Krooshof, and Postma}{Parouty
  et~al.}{2014}]{Parouty2014}
\textsc{Parouty, M. B.~Y., H.~H. Le, D.~Krooshof, and M.~J. Postma} (2014):
  \enquote{Differential Time Preferences for Money and Quality of Life,}
  \emph{{PharmacoEconomics}}, 32, 411--419.

\bibitem[\protect\citeauthoryear{Pham, Liebenehm, and Waibel}{Pham
  et~al.}{2017}]{Pham2017}
\textsc{Pham, H.~D., S.~Liebenehm, and H.~Waibel} (2017):
  \enquote{Experimentally validated general risk attitude among different
  ethnic groups in Vietnam,} \emph{TVSEP Working Paper}.

\bibitem[\protect\citeauthoryear{Picazo-Tadeo and Wall}{Picazo-Tadeo and
  Wall}{2011}]{PicazoTadeo2011}
\textsc{Picazo-Tadeo, A.~J. and A.~Wall} (2011): \enquote{Production risk, risk
  aversion and the determination of risk attitudes among Spanish rice
  producers,} \emph{Agricultural Economics}, 42, 451--464.

\bibitem[\protect\citeauthoryear{Pinger et~al.}{Pinger
  et~al.}{2017}]{Pinger2017}
\textsc{Pinger, P.~R. et~al.} (2017): \enquote{Predicting experimental choice
  behavior and life outcomes from a survey measure of present bias,}
  \emph{Economics Bulletin}, 37, 2162--2172.

\bibitem[\protect\citeauthoryear{Pinjisakikool}{Pinjisakikool}{2017}]{Pinjisakikool2017}
\textsc{Pinjisakikool, T.} (2017): \enquote{The Influence of Personality Traits
  on Households' Financial Risk Tolerance and Financial Behaviour,}
  \emph{Journal of Interdisciplinary Economics}, 30, 32--54.

\bibitem[\protect\citeauthoryear{Qiu, Colson, and Grebitus}{Qiu
  et~al.}{2014}]{Qiu2014}
\textsc{Qiu, Y., G.~Colson, and C.~Grebitus} (2014): \enquote{Risk preferences
  and purchase of energy-efficient technologies in the residential sector,}
  \emph{Ecological Economics}, 107, 216--229.

\bibitem[\protect\citeauthoryear{Rieger, Wang, and Hens}{Rieger
  et~al.}{2015}]{Rieger2015}
\textsc{Rieger, M.~O., M.~Wang, and T.~Hens} (2015): \enquote{Risk Preferences
  Around the World,} \emph{Management Science}, 61, 637--648.

\bibitem[\protect\citeauthoryear{Rosch}{Rosch}{2017}]{Rosch2017}
\textsc{Rosch, S.} (2017): \enquote{Risk Attitudes of US Agricultural
  Producers,} 2017 Annual Meeting, July 30-August 1, Chicago, Illinois 258025,
  Agricultural and Applied Economics Association.

\bibitem[\protect\citeauthoryear{Rutledge, Smittenaar, Zeidman, Brown, Adams,
  Lindenberger, Dayan, and Dolan}{Rutledge et~al.}{2016}]{Rutledge2016}
\textsc{Rutledge, R.~B., P.~Smittenaar, P.~Zeidman, H.~R. Brown, R.~A. Adams,
  U.~Lindenberger, P.~Dayan, and R.~J. Dolan} (2016): \enquote{Risk Taking for
  Potential Reward Decreases across the Lifespan,} \emph{Current Biology}, 26,
  1634--1639.

\bibitem[\protect\citeauthoryear{Schleich, Gassmann, Meissner, and
  Faure}{Schleich et~al.}{2019}]{Schleich2019}
\textsc{Schleich, J., X.~Gassmann, T.~Meissner, and C.~Faure} (2019):
  \enquote{A large-scale test of the effects of time discounting, risk
  aversion, loss aversion, and present bias on household adoption of
  energy-efficient technologies,} \emph{Energy Economics}, 80, 377--393.

\bibitem[\protect\citeauthoryear{Schneider and Sutter}{Schneider and
  Sutter}{2020}]{Schneider2020}
\textsc{Schneider, S.~O. and M.~Sutter} (2020): \enquote{Higher Order Risk
  Preferences: New Experimental Measures, Determinants and Field Behavior,}
  \emph{MPI Collective Goods Discussion Paper}.

\bibitem[\protect\citeauthoryear{Schr{\"o}der and Freedman}{Schr{\"o}der and
  Freedman}{2020}]{Schroeder2020}
\textsc{Schr{\"o}der, D. and G.~G. Freedman} (2020): \enquote{Decision making
  under uncertainty: the relation between economic preferences and
  psychological personality traits,} \emph{Theory and Decision}, 1--23.

\bibitem[\protect\citeauthoryear{Schurer}{Schurer}{2015}]{Schurer2015}
\textsc{Schurer, S.} (2015): \enquote{Lifecycle patterns in the socioeconomic
  gradient of risk preferences,} \emph{Journal of Economic Behavior {\&}
  Organization}, 119, 482--495.

\bibitem[\protect\citeauthoryear{Schwartz}{Schwartz}{2012}]{Schwartz2012}
\textsc{Schwartz, S.~H.} (2012): \enquote{An overview of the Schwartz theory of
  basic values,} \emph{Online Readings in Psychology and Culture}, 2,
  2307--0919.

\bibitem[\protect\citeauthoryear{Sepahvand and Shahbazian}{Sepahvand and
  Shahbazian}{2017}]{Sepahvand2017}
\textsc{Sepahvand, M.~H. and R.~Shahbazian} (2017): \enquote{Individual's risk
  attitudes in sub-Saharan Africa: determinants and reliability of
  self-reported risk in Burkina Faso,} \emph{Working Paper}.

\bibitem[\protect\citeauthoryear{Stan{\v{e}}k and
  Kr{\v{c}}{\'{a}}l}{Stan{\v{e}}k and Kr{\v{c}}{\'{a}}l}{2018}]{Stanek2018}
\textsc{Stan{\v{e}}k, R. and O.~Kr{\v{c}}{\'{a}}l} (2018): \enquote{Time
  preferences, cognitive abilities and intrinsic motivation to exert effort,}
  \emph{Applied Economics Letters}, 26, 1033--1037.

\bibitem[\protect\citeauthoryear{Stanton, Mullette-Gillman, McLaurin, Kuhn,
  LaBar, Platt, and Huettel}{Stanton et~al.}{2011}]{Stanton2011}
\textsc{Stanton, S.~J., O.~A. Mullette-Gillman, R.~E. McLaurin, C.~M. Kuhn,
  K.~S. LaBar, M.~L. Platt, and S.~A. Huettel} (2011): \enquote{Low- and
  High-Testosterone Individuals Exhibit Decreased Aversion to Economic Risk,}
  \emph{Psychological Science}, 22, 447--453.

\bibitem[\protect\citeauthoryear{Stephens and Tyran}{Stephens and
  Tyran}{2012}]{Stephens2012}
\textsc{Stephens, T.~A. and J.-R. Tyran} (2012): \enquote{'At Least I Didn't
  Lose Money'-Nominal Loss Aversion Shapes Evaluations of Housing
  Transactions,} \emph{Univ. of Copenhagen Dept. of Economics Discussion
  Paper}.

\bibitem[\protect\citeauthoryear{Sutter, Kocher, Gl{\"a}tzle-R{\"u}tzler, and
  Trautmann}{Sutter et~al.}{2013}]{Sutter2013}
\textsc{Sutter, M., M.~G. Kocher, D.~Gl{\"a}tzle-R{\"u}tzler, and S.~T.
  Trautmann} (2013): \enquote{Impatience and uncertainty: Experimental
  decisions predict adolescents' field behavior,} \emph{American Economic
  Review}, 103, 510--31.

\bibitem[\protect\citeauthoryear{Tanaka, Camerer, and Nguyen}{Tanaka
  et~al.}{2010}]{Tanaka2010}
\textsc{Tanaka, T., C.~F. Camerer, and Q.~Nguyen} (2010): \enquote{Risk and
  Time Preferences: Linking Experimental and Household Survey Data from
  Vietnam,} \emph{American Economic Review}, 100, 557--571.

\bibitem[\protect\citeauthoryear{van~der Heijden, Klein, M{\"u}ller, and
  Potters}{van~der Heijden et~al.}{2011}]{VanderHeijden2011}
\textsc{van~der Heijden, E., T.~J. Klein, W.~M{\"u}ller, and J.~J.~J. Potters}
  (2011): \enquote{Nudges and impatience: Evidence from a large scale
  experiment,} \emph{Netspar Discussion Paper}.

\bibitem[\protect\citeauthoryear{Vendrik and Woltjer}{Vendrik and
  Woltjer}{2007}]{Vendrik2007}
\textsc{Vendrik, M.~C. and G.~B. Woltjer} (2007): \enquote{Happiness and loss
  aversion: Is utility concave or convex in relative income?} \emph{Journal of
  Public Economics}, 91, 1423--1448.

\bibitem[\protect\citeauthoryear{Vieider}{Vieider}{2012}]{Vieider2012}
\textsc{Vieider, F.~M.} (2012): \enquote{Moderate stake variations for risk and
  uncertainty, gains and losses: Methodological implications for comparative
  studies,} \emph{Economics Letters}, 117, 718--721.

\bibitem[\protect\citeauthoryear{Vieider, Beyene, Bluffstone, Dissanayake,
  Gebreegziabher, Martinsson, and Mekonnen}{Vieider et~al.}{2018}]{Vieider2018}
\textsc{Vieider, F.~M., A.~Beyene, R.~Bluffstone, S.~Dissanayake,
  Z.~Gebreegziabher, P.~Martinsson, and A.~Mekonnen} (2018): \enquote{Measuring
  risk preferences in rural Ethiopia,} \emph{Economic Development and Cultural
  Change}, 66, 417--446.

\bibitem[\protect\citeauthoryear{Vieider, Martinsson, Nam, and Truong}{Vieider
  et~al.}{2019}]{Vieider2019}
\textsc{Vieider, F.~M., P.~Martinsson, P.~K. Nam, and N.~Truong} (2019):
  \enquote{Risk preferences and development revisited,} \emph{Theory and
  Decision}, 86, 1--21.

\bibitem[\protect\citeauthoryear{Vieider, Villegas-Palacio, Martinsson, and
  Mej{\'\i}a}{Vieider et~al.}{2016}]{Vieider2016}
\textsc{Vieider, F.~M., C.~Villegas-Palacio, P.~Martinsson, and M.~Mej{\'\i}a}
  (2016): \enquote{Risk taking for oneself and others: A structural model
  approach,} \emph{Economic Inquiry}, 54, 879--894.

\bibitem[\protect\citeauthoryear{Von~Gaudecker, Van~Soest, and
  Wengstrom}{Von~Gaudecker et~al.}{2011}]{VonGaudecker2011}
\textsc{Von~Gaudecker, H.-M., A.~Van~Soest, and E.~Wengstrom} (2011):
  \enquote{Heterogeneity in risky choice behavior in a broad population,}
  \emph{American Economic Review}, 101, 664--94.

\bibitem[\protect\citeauthoryear{Wahl, Kirchler, and Walla}{Wahl
  et~al.}{2020}]{Wahl2020}
\textsc{Wahl, I., E.~Kirchler, and P.~Walla} (2020): \enquote{{RIsk}
  {SCreening} on the Financial Market ({RISC}-{FM}): A tool to assess
  investors' financial risk tolerance,} \emph{Cogent Psychology}, 7, 1714108.

\bibitem[\protect\citeauthoryear{Wakker}{Wakker}{2008}]{Wakker2008}
\textsc{Wakker, P.~P.} (2008): \enquote{Explaining the characteristics of the
  power (CRRA) utility family,} \emph{Health Economics}, 17, 1329--1344.

\bibitem[\protect\citeauthoryear{Wang, Rieger, and Hens}{Wang
  et~al.}{2016}]{Wang2016}
\textsc{Wang, M., M.~O. Rieger, and T.~Hens} (2016): \enquote{How time
  preferences differ: Evidence from 53 countries,} \emph{Journal of Economic
  Psychology}, 52, 115--135.

\bibitem[\protect\citeauthoryear{W{\"a}rneryd}{W{\"a}rneryd}{1996}]{Warneryd1996}
\textsc{W{\"a}rneryd, K.-E.} (1996): \enquote{Risk attitudes and risky
  behavior,} \emph{Journal of Economic Psychology}, 17, 749--770.

\bibitem[\protect\citeauthoryear{Yin}{Yin}{2020}]{Yin2020}
\textsc{Yin, Q.} (2020): \enquote{Financial Risk Tolerance Among Same-sex and
  Mixed-sex Couples,} \emph{Working Paper}.

\bibitem[\protect\citeauthoryear{Yoong, Hung, Barcellos, Carvalho, and
  Clift}{Yoong et~al.}{2019}]{Yoong2019}
\textsc{Yoong, J.~K., A.~A. Hung, S.~H. Barcellos, L.~Carvalho, and J.~Clift}
  (2019): \enquote{Disparities in Minority Retirement Savings Behavior: Survey
  and Experimental Evidence from A Nationally-Representative Sample of US
  Households,} \emph{Working Paper}.

\end{thebibliography}
}
\end{singlespace}

\newpage
\begin{appendices}
\renewcommand{\theequation}{\thesection.{}\arabic{equation}}
\renewcommand\thefigure{\thesection.\arabic{figure}}    
\renewcommand\thetable{\thesection.\arabic{table}}  
\setcounter{table}{0}

\section{Related Literature}

\subsection{Search criteria}
\label{sec:search}
In the first step of our literature search, we used the following criteria to search for representative studies that make use of the main established experimental elicitation methods \citep{Harrison2008}:\footnote{we excluded the Trade-Off design as the words were insufficiently discriminant, leading to too many unrelated results.}
\begin{itemize}
\item (`` representative sample'') AND (``risk preferences'' OR ``loss aversion'' OR ``time preferences'' OR ``present bias'') AND (``multiple price list'')
\item (`` representative sample'') AND (``risk preferences'' OR ``loss aversion'' OR ``time preferences'' OR ``present bias'') AND (``random lottery pair'')
\item (`` representative sample'') AND (``risk preferences'' OR ``loss aversion'' OR ``time preferences'' OR ``present bias'') AND (``ordered lottery selection'')
\item (`` representative sample'') AND (``risk preferences'' OR ``loss aversion'' OR ``time preferences'' OR ``present bias'') AND (``Becker Degroot Marschak'')
\item (`` representative sample'') AND (``risk preferences'' OR ``loss aversion'' OR ``time preferences'' OR ``present bias'') AND (``Convex Time Budget'')
\end{itemize}
\noindent
In the second step we broadened the search criteria considerably: \begin{itemize}
    \item (``large sample'' OR ``representative sample'') AND (``risk attitude'' OR ``risk preferences'' OR ``loss aversion'' OR ``time preferences'' OR ``present bias'')
\end{itemize}
Both steps were last updated on September $27^{th}$, 2021. 

\subsection{Overview of the literature}
\label{sec:overview}

\newpage
\begin{center}
\begin{scriptsize}
       \begin{longtable}[ht]{cccccccc}
  \caption{An overview of the literature \label{tab:overlit}}\\          
   \toprule
 Articles & Sample Size\footnotemark & \rot{Representative} & \rot{Incentives} & \rot{Risk Aversion} & \rot{Loss Aversion} & \rot{Time discounting} & \rot{Present bias} \\
 
  \midrule
  \endfirsthead
  \multicolumn{8}{l@{}}{continued \ldots}\\
     \toprule
 Articles & Sample Size\footnotemark[\value{footnote}] & \rot{Representative} & \rot{Incentives} & \rot{Risk Aversion} & \rot{Loss Aversion} & \rot{Discounting} & \rot{Present bias} \\
  \midrule
  \endhead
\bottomrule
\multicolumn{8}{r@{}}{continued \ldots}\\
\endfoot
\bottomrule
\endlastfoot
\citet{Albert2012}	&	60	&	 	&	$\surd$	&	$\surd$	&	 	&	$\surd$	&	 	\\
\citet{Almenberg2015} 	&	1132	&	$\surd$	&	 	&	$\surd$	&	 	&	 	&	 	\\
\citet{Andersen2010}	&	343	&	$\surd$	&	$\surd$	&	$\surd$	&	 	&	$\surd$	&	 	\\
\citet{Andersen2018}	&	413	&	$\surd$	&	$\surd$	&	$\surd$	&	 	&	$\surd$	&	 	\\
\citet{Anderson2021}	&	1817	&		&		&	$\surd$	&		&		&		\\
\citet{Andersson2016} 	&	895 \& 1386	&	$\surd$	&	$\surd$	&	$\surd$	&	 	&	 	&	 	\\
\citet{Andersson2020} 	&	1396	&	$\surd$	&	$\surd$	&	$\surd$	&	 	&	 	&	 	\\
\citet{Aycinena2014}	&	184	&	 	&	$\surd$	&	$\surd$	&	 	&	 	&	 	\\
\citet{Bacon2014}	&	7761$^\dagger$	&	$\surd$	&	 	&	$\surd$	&	 	&	 	&	 	\\
\citet{Bajtelsmit2001}	&	10412	&	 	&	 	&	$\surd$	&	 	&	 	&	 	\\
\citet{Bansback2016}	&	6780	&	$\surd$	&	 	&	$\surd$	&	 	&	 	&	 	\\
\citet{Bartke2008}	&	19211	&	$\surd$	&	 	&	$\surd$	&	 	&	 	&	 	\\
\citet{Bateman2015}	&	1199	&	$\surd$	&	 	&	$\surd$	&	$\surd$	&	 	&	 	\\
\citet{Beauchaine2017}	&	542	&	 	&	 	&	 	&	 	&	$\surd$	&	 	\\
\citet{Benjamin2013}	&	173	&	 	&	$\surd$	&	$\surd$	&	 	&	$\surd$	&	 	\\
\citet{Blake2019}	&	4016	&	$\surd$	&	 	&	$\surd$	&	$\surd$	&	 	&	 	\\
\citet{Bonsang2015}	&	12032	&	 	&	 	&	$\surd$	&	 	&	 	&	 	\\
\citet{Booij2010}	&	814 \& 438	&	$\surd$	&	 	&	$\surd$	&	 	&	 	&	 	\\
\citet{Booij2009}	&	1767	&	 	&	 	&	$\surd$	&	 	&	$\surd$	&	 	\\
\citet{Booij2009a}	&	1935	&	$\surd$	&	 	&	 	&	$\surd$	&	 	&	 	\\
\citet{Boschini2019}	&	926	&	$\surd$	&	$\surd$	&	$\surd$	&	 	&	 	&	 	\\
\citet{bouchouicha2019gender}	&	2939	&		&	$\surd$ 	&	$\surd$	&	$\surd$	&	 	\\
\citet{Bouchouicha2019}	&	96175	&	$\surd$	&	 	&	$\surd$	&	 	&	 	&	 	\\

\citet{Boyle2011}	&	369	&	 	&	 	&	$\surd$	&	 	&	 	&	 	\\
\citet{BranasGarza2018}	&	704	&	 	&	 $\surd$	&	$\surd$	&	 	&	 	&	 	\\
\citet{Breuer2020}	&	711 \& 327 \& 962	&	 	& $\surd$	 	&	$\surd$	&	$\surd$	&	$\surd$	&	$\surd$	\\
\citet{Brick2012}	&	2416	&	 	&	 $\surd$	&	$\surd$	&	 	&	 	&	 	\\
\citet{Browne2016}	&	7339	&	$\surd$	&	 	&	$\surd$	&	 	&	 	&	 	\\
\citet{Browne2020}	&	17980	&	$\surd$	&	 	&	 	&	 	&	 	&	 	\\
\citet{BrudererEnzler2014}	&	2892 \& 1407	&	$\surd$	&	 	&	 	&	 	&	$\surd$	&	 	\\
\citet{Bucciol2011}	&	4095	&	$\surd$	&	 	&	$\surd$	&	 	&	 	&	 	\\
\citet{Burks2009}	&	1012	&	 	&	$\surd$	&	$\surd$	&	 	&	$\surd$	&	$\surd$	\\
\citet{Burks2012}	&	958	&	 	&	$\surd$	&	 	&	 	&	$\surd$	&	$\surd$	\\
\citet{Burks2015}	&	100	&	 	&	$\surd$	&	$\surd$	&	 	&	$\surd$	&	$\surd$	\\
\citet{Buser2020}	&	2565	&	 	&	$\surd$	&	$\surd$	&	 	&	 	&	 	\\
\citet{BusicSontic2018}	&	3468	&	 	&	 	&	$\surd$	&	 	&	 	&	 	\\
\citet{Butler2013}	&	1686	&	 	&	$\surd$	&	$\surd$	&	 	&	 	&	 	\\
\citet{Cardenas2013}	&	3085	&	$\surd$	&	$\surd$	&	$\surd$	&	$\surd$	&	 	&	 	\\
\citet{Castillo2018}	&	9674	&	$\surd$	&	$\surd$	&	$\surd$	&	 	&	 	&	 	\\
\citet{Chapman2018}	&	1000	&	$\surd$	&	$\surd$	&	$\surd$	&	 	&	$\surd$	&	 	\\
\citet{Chapman2018a}	&	2000 \& 1740	&	$\surd$	&	$\surd$	&	$\surd$	&	$\surd$	&	$\surd$	&	 	\\
\citet{Clot2017}	&	186	&	 	&	$\surd$	&	$\surd$	&	 	&	 	&	 	\\
\citet{Dasgupta2016}	&	1872	&	 	&	$\surd$	&	$\surd$	&	 	&	 	&	 	\\
\citet{Dave2010}	&	881	&	$\surd$	&	$\surd$	&	$\surd$	&	 	&	 	&	 	\\
\citet{Delavande2018}	&	843 \& 780	&	 	&	$\surd$	&	$\surd$	&	$\surd$	&	 	&	 	\\
\citet{DiFalco2018}	&	657	&	 	&	$\surd$	&	$\surd$	&	 	&	 	&	 	\\
\citet{Dittrich2014}	&	1019	&	$\surd$	&	 	&	 	&	 	&	$\surd$	&	 	\\
\citet{Dohmen2010}	&	349 \& 384 \& 749	&	$\surd$	&	$\surd$	&	$\surd$	&	 	&	$\surd$	&	 	\\
\citet{Dohmen2011}	&	14773	& $\surd$	 	&	 	&	$\surd$	&	 	&	 	&	 	\\
\citet{Donkers2001}	&	2593	&	$\surd$	&	 	&	$\surd$	&	 	&	 	&	 	\\
\citet{Drichoutis2021} & 1008 &  & 	$\surd$ & 	$\surd$ & & 	$\surd$ & \\
\citet{Eckel2005}	&	256	&	 	&	$\surd$	&	$\surd$	&	 	&	$\surd$	&	 	\\
\citet{Faff2011}	&	15916	&	 	&	 	&	$\surd$	&	 	&	 	&	 	\\
\citet{Falk2018} 	&	77445	&	$\surd$	&	 	&	$\surd$	&	 	&	$\surd$	&	 	\\
\citet{Fan2020}	&	5500	&	$\surd$	&	 	&	$\surd$	&	 	&	 	&	 	\\
\citet{Fang2013}	&	2956	&	 	&	 	&	$\surd$	&	 	&	 	&	 	\\
\citet{Fehr2021} & 446 & 	$\surd$ & 	$\surd$ & 	$\surd$ & & & \\
\citet{Frederick2005}	&	3428	&	 	&	 	&	$\surd$	&	 	&	 	&	 	\\
\citet{Fredslund2018}	&	947	&	 	&	 	&	 	&	 	&	$\surd$	&	$\surd$	\\
\citet{Freeney2010}	&	1131	&	 	&	 	&	 	&	 	&	$\surd$	&	 	\\
\citet{Frey2021} & 916 & 	$\surd$ & 	$\surd$ & 	$\surd$ & & & \\
\citet{Gachter2021} & 326 &  & 	$\surd$ & & 	$\surd$ & & \\
\citet{Galizzi2016}	&	448 \& 280$^\dagger$		&	$\surd$	&	$\surd$	&	$\surd$	&	 	&	 	&	 	\\
\citet{Gartner2017}	&	1365	&	$\surd$	&	 	&	$\surd$	&	 	&	 	&	 	\\
\citet{Giampietri2021} & 284  & & & $\surd$ & & & \\
\citet{Gloede2015}	&	1994 \& 1843	&	 	&	 	&	$\surd$	&	 	&	 	&	 	\\
\citet{Goda2019}	&	2317	&	$\surd$	&	$\surd$ 	&	 	&	 	&	$\surd$	&	$\surd$	\\\citet{Golsteyn2014}	&	11907	&	 	&	 	&	 	&	 	&	$\surd$	&	 	\\
\citet{Gorlitz2020}	&	203687	&	 	&	 	&	$\surd$	&	 	&	 	&	 	\\
\citet{Guenther2021}	&	955 \& 299$^\dagger$		&	$\surd$	&	$\surd$	&	$\surd$	&		&		&		\\
\citet{Halko2012}	&	328	&	 	&	 	&	$\surd$	&	 	&	 	&	 	\\
\citet{Hansen2016}	&	253	&	$\surd$	&	$\surd$	&	$\surd$	&	 	&	$\surd$	&	 	\\
\citet{Hardeweg2013}	&	929	&	 	&	$\surd$	&	$\surd$	&	 	&	 	&	 	\\
\citet{Harrati2014}	&	10455	&$\surd$	 	&	 	&	$\surd$	&	 	&	 	&	 	\\
\citet{Harrison2002}	&	268	&	$\surd$	&	$\surd$	&	 	&	 	&	$\surd$	&	 	\\
\citet{Harrison2007}	&	253	&	$\surd$	&	$\surd$	&	$\surd$	&	 	&	 	&	 	\\
\citet{Hofmeyr2016}	&	1205	&	 	&	 $\surd$	&	 	&	 	&	$\surd$	&	$\surd$	\\
\citet{Holden2019}	&	1138	&	 	& $\surd$	 	&	$\surd$	&	 	&	 	&	 	\\
\citet{Holm2013}	&	362 \& 344	&	 	&	$\surd$	&	$\surd$	&	 	&	 	&	 	\\
\citet{Hopland2016}	&	203	&	 	&	$\surd$	&	$\surd$	&	 	&	 	&	 	\\
\citet{Horn2019}	&	932	&	$\surd$	&	 	&	$\surd$	&	 	&	 	&	 	\\
\citet{Horn2020}	&	149	&	 	&	 	&		&		&	$\surd$		&		\\
\citet{Horn2020a}	&	955	&	 	&	 	&	 	&	 	&	$\surd$	&	 	\\
\citet{Howard2011}	&	2416 \& 2327 \& 694	&	 	&	 	&	$\surd$	&	 	&	 	&	 	\\
\citet{Huffman2019}	&	591	&	 	&	 	&	 	&	 	&	$\surd$	&	 	\\
\citet{Hunter2018}	&	176	&	 	&	$\surd$	&	$\surd$	&	 	&	$\surd$	&	$\surd$	\\
\citet{Hvide2015}	&	78 \& 59	&	 	&	$\surd$	&	$\surd$	&	 	&	$\surd$	&	 	\\
\citet{Hwang2017}	&	1698	&	 	&	 	&	 	&	$\surd$	&	 	&	 	\\
\citet{Ioannou2016}	&	118	&	 	&	$\surd$	&	$\surd$	&	 	&	$\surd$	&	 	\\
\citet{Ionescu2011}	&	352	&	 	&	 	&	$\surd$	&	 	&	 	&	 	\\
\citet{Jarmolowicz2012}	&	904	&	 	&	 	&	 	&	 	&	$\surd$	&	 	\\
\citet{Jin2017}	&	200$^\dagger$		&	 	&	 	&	$\surd$	&	 	&	 	&	 	\\
\citet{Johnson2006}	&	360	&	 	&	$\surd$	&	 	&	$\surd$	&	 	&	 	\\
\citet{Jung2015}	&	3186	&	$\surd$	&	 	&	$\surd$	&	 	&	 	&	 	\\
\citet{Jung2017}	&	11453	&	$\surd$	&	 	&	$\surd$	&	 	&	 	&	 	\\
\citet{Kam2012}	&	2186 \& 1709	&	$\surd$	&	 	&	$\surd$	&	 	&	 	&	 	\\
\citet{Kettlewell2019}	&	4810	&	$\surd$	&	 	&	$\surd$	&	 	&	 	&	 	\\
\citet{Khachatryan2015}	&	762	&	 	&	 $\surd$	&	$\surd$	&	 	&	 	&	 	\\
\citet{Kim2012}	&	7553	&	$\surd$	&	 	&	$\surd$	&	 	&	 	&	 	\\
\citet{Kristjanpoller2015}	&	1788	&	 	&	 	&	$\surd$	&	 	&	 	&	 	\\
\citet{Peryman2015} 	&	552$^\dagger$		&	 	&	 	&	$\surd$	&	 	&	$\surd$	&	 	\\
\citet{Lampi2013}	&	2155	&	 	&	 	&	$\surd$	&	 	&	 	&	 	\\
\citet{Le2010}	&	4738	&	 	&	 	&	$\surd$	&	 	&	 	&	 	\\
\citet{Lee2016}	&	1086	&	 	&	 	&	$\surd$	&	 	&	 	&	 	\\
\citet{Haridon2019}	&	2939	&	 	&	$\surd$	&	$\surd$	&	$\surd$	&	 	&	 	\\
\citet{MartinFernandez2018}	&	1776 \& 1741 \& 1702$^\dagger$		&	$\surd$	&	 	&	$\surd$	&	 	&	 	&\\
\citet{Mata2016}	&	147118	&	$\surd$	& 	 	& $\surd$	 	&	 	&	 	&	 	\\
\citet{Meier2015}	&	1684	&	 	&	$\surd$	&	 	&	 	&	$\surd$	&	$\surd$	\\
\citet{Melesse2017}	&	532	&	 	&	$\surd$	&	$\surd$	&	 	&	 	&	 	\\
\citet{Menkhoff2017}	&	715$^\dagger$		&	 	&	$\surd$	&	$\surd$	&	 	&	 	&	 	\\
\citet{Nebout2018}	&	1519	&	 	&	 	&	$\surd$	&	 	&	 	&	 	\\
\citet{Newell2015}	&	1217	&	$\surd$	&	&	 	&	 	&	$\surd$	&	 	\\
\citet{Neyse2020}	&	3431	&	$\surd$	&	 	&	$\surd$	&	 	&	 	&	 	\\
\citet{Noussair2014}	&	1865	&	$\surd$	&	$\surd$	&	$\surd$	&	 	&	 	&	 	\\
\citet{Oechssler2009}	&	564	&	 	&	$\surd$	&	$\surd$	&	 	&	$\surd$	&	 	\\
\citet{Osama2017}	&	8411 \& 410	&	$\surd$	&	 	&	$\surd$	&	 	&	 	&	 	\\
\citet{Parouty2014}	&	847	&	$\surd$	&	 	&	 	&	 	&	$\surd$	&	 	\\
\citet{Pham2017}	&	646$^\dagger$		&	 	& $\surd$	 	&	$\surd$	&	 	&	 	&	 	\\
\citet{PicazoTadeo2011}	&	129	&	 	&	 	&	$\surd$	&	 	&	 	&	 	\\
\citet{Pinger2017}	&	310	&	$\surd$	&	$\surd$	&	 	&	 	&	 	&	$\surd$	\\
\citet{Pinjisakikool2017}	&	4026$^\dagger$		&	$\surd$	&	 	&	$\surd$	&	 	&	 	&	 	\\
\citet{Qiu2014}	&	432	&	 	&	 	&	$\surd$	&	 	&	 	&	 	\\
\citet{Rieger2015}	&	5758 \& 5573	&	 	&		&	$\surd$	&	 	&	 	&	 	\\
\citet{Rosch2017}	&	25493	&	 	&	 	&	$\surd$	&	 	&	 	&	 	\\
\citet{Rutledge2016}	&	25189$^\dagger$		&	 	&	 	&	$\surd$	&	 	&	 	&	 	\\
\citet{Schneider2020}	&	658	&	 	&	$\surd$ 	&	$\surd$	&	 	&	 	&	 	\\
\citet{Schroeder2020}	&	99$^\dagger$		&	 	&	$\surd$	&	$\surd$	&	 	&	 	&	 	\\
\citet{Schurer2015}	&	135807	&	$\surd$	&	 	&	$\surd$	&	 	&	 	&	 	\\
\citet{Sepahvand2017}	&	31677$^\dagger$		&	$\surd$	&	 	&	$\surd$	&	 	&	 	&	 	\\
\citet{Stanek2018}	&	149	&	 	&	$\surd$	&	 	&	 	&	$\surd$	&	 	\\
\citet{Stanton2011}	&	298	&	 	&	$\surd$	&	$\surd$	&	$\surd$	&	 	&	 	\\
\citet{Stephens2012}	&	732	&	 	&	$\surd$	&	 	&	$\surd$	&	 	&	 	\\
\citet{Sutter2013}	&	639	&	 	&	$\surd$	&	$\surd$	&	 	&	$\surd$	&	 	\\
\citet{Tanaka2010}	&	181	&	 	&	$\surd$	&	$\surd$	&	$\surd$	&	$\surd$	&	$\surd$	\\
\citet{VanderHeijden2011}	&	1102	&	$\surd$	&	$\surd$	&	 	&	 	&	$\surd$	&	 	\\
\citet{Vieider2016}	&	200$^\dagger$		&	 	&	$\surd$	&	$\surd$	&	$\surd$	&	 	&	 	\\
\citet{Vieider2018}	&	493	&	 	&	$\surd$	&	$\surd$	&	 	&	 	&	 	\\
\citet{Vieider2019}	&	197$^\dagger$		&	 	&	$\surd$	&	$\surd$	&	$\surd$	&	 	&	 	\\
\citet{VonGaudecker2011}	&	1422	&	$\surd$	&	$\surd$	&	$\surd$	&	$\surd$	&	 	&	 	\\
\citet{Wahl2020}	&	707$^\dagger$		&	 	&	 	&	$\surd$	&	 	&	 	&	 	\\
\citet{Wang2016}	&	5833	&	 	&	 	&	 	&	 	&	$\surd$	&	$\surd$	\\
\citet{Warneryd1996}	&	1087	&	 	&	 	&	$\surd$	&	 	&	 	&	 	\\
\citet{Yin2020}	&	4754	&	 	&	 	&	$\surd$	&	 	&	 	&	 	\\
\citet{Yoong2019}	&	2379$^\dagger$		&	 	&	 	&	$\surd$	&	 	&	 	&	 	\\ \hline
This study & 12612 & $\surd$	& $\surd$	& $\surd$	& $\surd$	& $\surd$	& $\surd$

    \end{longtable}
\end{scriptsize}
\end{center}

\footnotetext{Papers that reported results for different elicitation tasks and different sub-samples were treated as separate studies. In such cases, we indicate multiple sample sizes for the same paper. Papers that reported results for different elicitation task using the same sample were treated as separate studies and were noted with $^\dagger$.	For \citet{bouchouicha2019gender}, we retained the results from  the most sophisticated specification  (``CPT"). }

\begin{table}
\centering
\scriptsize
\begin{threeparttable}[h]
     \caption{Correlations between individual characteristics and preferences reported in the literature: number of observations}
  \begin{tabular*}{\textwidth}{c @{\extracolsep{\fill}} rccccccccc}
        \toprule
Preferences	& Relationship &	Age&	Male &	Income &	Education&	Children &	Couple &	Cognitive  \\
	&  &	&	 &	&	&	 &	 &	ability \\
\midrule
Risk	&	Positive	&	673209	*	&	8779		&	20537		&	77274		&	411694		&	237801		&	0		\\
aversion	&	n.s.	&	208156		&	63534		&	286372		&	104028 *	&	91068	*	&	236240	*	&	18573		\\
	&	Negative	&	63071		&	657783	*	&	240142	*	&	551373		&	6324		&	127674		&	27261	*	\\ \\
Loss	&	Positive	&	4039		&	3740		&	4066		&	8150		&	0		&	3085		&	3740 *		\\
aversion	&	n.s.	&	9933		&	5737		&	2262		&	8317	*	&	4783		&	2897	*	&	1698		\\
	&	Negative	&	11241	*	&	14106	*	&	7183	*	&	2699		&	4016	*	&	4016		&	3671		\\ \\
Time	&	Positive	&	5171		&	3042	*	&	0		&	0		&	3324	*	&	3740		&	0		\\
discounting	&	n.s.	&	11556		&	27162		&	10621	*	&	6863		&	6548		&	7254	*	&	1118		\\
	&	Negative	&	91825	*	&	95166		&	13273		&	16731	*	&	149		&	3321		&	90268	*	\\ \\
Present	&	Positive	&	1684		&	0	*	&	0		&	947		&	0		&	958		&	0		\\
bias	&	n.s.	&	2509		&	10967		&	2262	*	&	3454	*	&	0	*	&	3740	*	&	3273	*	\\
	&	Negative	&	7929	*	&	0		&	2822		&	0		&	176		&	176		&	1112		\\

 \bottomrule
    \end{tabular*}
\begin{tablenotes}
\item The numbers indicate the aggregated number of participants in all studies finding a specific relationship (at the 10\% level for positively or negatively significant relationships). 
\item  Results from the present study are not included in the numbers reported but are indicated by *.
\item These results should be interpreted with caution, as a few studies using very large datasets can inflate the numbers (e.g. the world values survey database \cite{Mata2016}), and observations are not necessarily independent:  some studies make use of the same or overlapping datasets (e.g. the German Socio-Economic Panel, SOEP), and some use longitudinal datasets, where the same respondents answer multiple times \citep[e.g.][]{Schurer2015}. 
\end{tablenotes}
    \label{tab:litteraturesample}
    \end{threeparttable}
    \end{table}

\newpage

\section{Summary Statistics}
\setcounter{table}{0}
\begin{table}
\footnotesize
\caption{Summary statistics of individual characteristics and controls\label{tab:summarystats}}
\begin{tabular*}{\hsize}{@{\hskip\tabcolsep\extracolsep\fill}l*{9}{c}}
\toprule
            &          FR&          DE&          IT&          PL&          RO&          ES&          SE&          UK&       Total\\
\midrule
Age         &       42.23&       42.76&       43.01&       38.65&       36.50&       41.84&       42.57&       41.44&       41.21\\
            &     (13.49)&     (13.10)&     (12.55)&     (11.73)&     (10.16)&     (12.27)&     (13.48)&     (13.14)&     (12.74)\\
Gender      &       0.506&       0.520&       0.502&       0.526&       0.515&       0.512&       0.506&       0.506&       0.512\\
            &     (0.500)&     (0.500)&     (0.500)&     (0.499)&     (0.500)&     (0.500)&     (0.500)&     (0.500)&     (0.500)\\
Education   &       0.571&       0.271&       0.363&       0.525&       0.659&       0.602&       0.411&       0.606&       0.501\\
            &     (0.495)&     (0.444)&     (0.481)&     (0.499)&     (0.474)&     (0.489)&     (0.492)&     (0.489)&     (0.500)\\
Income      &       29.71&       36.53&       29.04&       14.45&       10.38&       27.47&       42.33&       47.97&       30.11\\
            &     (19.74)&     (21.30)&     (17.49)&     (10.15)&     (10.41)&     (16.82)&     (25.48)&     (28.77)&     (23.14)\\
Children    &       0.596&       0.521&       0.589&       0.628&       0.554&       0.582&       0.521&       0.477&       0.558\\
            &     (0.491)&     (0.500)&     (0.492)&     (0.483)&     (0.497)&     (0.493)&     (0.500)&     (0.499)&     (0.497)\\
Couple      &       0.645&       0.588&       0.690&       0.686&       0.658&       0.705&       0.503&       0.587&       0.634\\
            &     (0.479)&     (0.492)&     (0.463)&     (0.464)&     (0.474)&     (0.456)&     (0.500)&     (0.492)&     (0.482)\\
Urban       &       0.481&       0.504&       0.645&       0.612&       0.671&       0.630&       0.571&       0.653&       0.595\\
            &     (0.500)&     (0.500)&     (0.478)&     (0.487)&     (0.470)&     (0.483)&     (0.495)&     (0.476)&     (0.491)\\
CRT         &       0.881&       1.068&       0.608&       0.785&       0.668&       0.779&       0.992&       0.949&       0.844\\
            &     (1.013)&     (1.055)&     (0.776)&     (1.024)&     (0.934)&     (0.994)&     (1.060)&     (1.053)&     (1.004)\\
Achievement &       0.230&       0.425&       0.509&       0.456&       0.712&       0.462&       0.290&       0.362&       0.427\\
            &     (0.421)&     (0.494)&     (0.500)&     (0.498)&     (0.453)&     (0.499)&     (0.454)&     (0.481)&     (0.495)\\
Benevolence &       0.516&       0.882&       0.859&       0.608&       0.820&       0.771&       0.650&       0.654&       0.719\\
            &     (0.500)&     (0.323)&     (0.348)&     (0.488)&     (0.384)&     (0.420)&     (0.477)&     (0.476)&     (0.449)\\
Conformity  &       0.707&       0.621&       0.744&       0.578&       0.757&       0.642&       0.524&       0.640&       0.653\\
            &     (0.455)&     (0.485)&     (0.436)&     (0.494)&     (0.429)&     (0.479)&     (0.499)&     (0.480)&     (0.476)\\
Hedonism    &       0.512&       0.814&       0.725&       0.295&       0.613&       0.626&       0.528&       0.420&       0.567\\
            &     (0.500)&     (0.389)&     (0.447)&     (0.456)&     (0.487)&     (0.484)&     (0.499)&     (0.494)&     (0.496)\\
Power       &       0.156&       0.322&       0.406&       0.327&       0.324&       0.287&       0.240&       0.262&       0.290\\
            &     (0.362)&     (0.467)&     (0.491)&     (0.469)&     (0.468)&     (0.452)&     (0.427)&     (0.440)&     (0.454)\\
Security    &       0.507&       0.766&       0.830&       0.750&       0.806&       0.745&       0.698&       0.736&       0.728\\
            &     (0.500)&     (0.423)&     (0.376)&     (0.433)&     (0.395)&     (0.436)&     (0.459)&     (0.441)&     (0.445)\\
Self direction&       0.509&       0.696&       0.805&       0.691&       0.800&       0.688&       0.576&       0.608&       0.670\\
            &     (0.500)&     (0.460)&     (0.396)&     (0.462)&     (0.400)&     (0.463)&     (0.494)&     (0.488)&     (0.470)\\
Stimulation &       0.245&       0.382&       0.449&       0.417&       0.462&       0.359&       0.345&       0.366&       0.376\\
            &     (0.430)&     (0.486)&     (0.497)&     (0.493)&     (0.499)&     (0.480)&     (0.475)&     (0.482)&     (0.484)\\
Tradition   &       0.472&       0.569&       0.709&       0.594&       0.656&       0.510&       0.455&       0.518&       0.560\\
            &     (0.499)&     (0.495)&     (0.454)&     (0.491)&     (0.475)&     (0.500)&     (0.498)&     (0.500)&     (0.496)\\
Universalism&       0.648&       0.720&       0.826&       0.598&       0.759&       0.715&       0.584&       0.591&       0.680\\
            &     (0.478)&     (0.449)&     (0.379)&     (0.490)&     (0.428)&     (0.452)&     (0.493)&     (0.492)&     (0.466)\\
ABreversed  &       0.498&       0.496&       0.502&       0.496&       0.508&       0.493&       0.500&       0.510&       0.500\\
            &     (0.500)&     (0.500)&     (0.500)&     (0.500)&     (0.500)&     (0.500)&     (0.500)&     (0.500)&     (0.500)\\
Incentivized&       0.607&       0.446&       0.456&       0.610&       0.720&       0.596&       0.601&       0.448&       0.555\\
            &     (0.488)&     (0.497)&     (0.498)&     (0.488)&     (0.449)&     (0.491)&     (0.490)&     (0.497)&     (0.497)\\
LowStakes   &      0.0988&      0.0499&      0.0537&       0.102&      0.0663&       0.102&      0.0674&      0.0484&      0.0734\\
            &     (0.298)&     (0.218)&     (0.225)&     (0.302)&     (0.249)&     (0.303)&     (0.251)&     (0.215)&     (0.261)\\
HighStakes  &       0.124&      0.0926&      0.0889&       0.122&      0.0852&       0.132&       0.101&      0.0917&       0.105\\
            &     (0.329)&     (0.290)&     (0.285)&     (0.327)&     (0.279)&     (0.339)&     (0.301)&     (0.289)&     (0.306)\\
\bottomrule
\multicolumn{10}{l}{\footnotesize Standard deviation in parentheses}\\
\end{tabular*}

\end{table}
\newpage

\section{Country-level estimates}
\setcounter{table}{0}

\begin{table}[htbp]
\renewcommand{\footnotesize}{\scriptsize}
\scriptsize
\caption{All countries \label{tab:l1t_reduced}}
{
\def\sym#1{\ifmmode^{#1}\else\(^{#1}\)\fi}
\begin{tabular*}{\hsize}{@{\hskip\tabcolsep\extracolsep\fill}l*{6}{c}}
\toprule
                &\makecell{Risk aversion \\ $\alpha$}         &\makecell{Loss aversion \\ $\lambda$}         &\makecell{Discounting \\ $\delta$}         &\makecell{Present bias \\ $\gamma$}         &\makecell{Tremble error \\ $\kappa$}         &\makecell{Fechner error \\ $\mu$}         \\
\midrule
Age             &    0.002\sym{**} &   -0.002\sym{**} &   -0.003\sym{***}&   -0.000         &    0.001         &   -0.007\sym{***}\\
                &  (0.001)         &  (0.001)         &  (0.000)         &  (0.000)         &  (0.000)         &  (0.001)         \\
Male          &   -0.027         &   -0.356\sym{***}&    0.026\sym{**} &    0.004\sym{*}  &   -0.005         &    0.102\sym{**} \\
                &  (0.014)         &  (0.016)         &  (0.009)         &  (0.002)         &  (0.007)         &  (0.032)         \\
Education       &    0.005         &   -0.007         &   -0.022\sym{*}  &   -0.001         &   -0.024\sym{***}&    0.006         \\
                &  (0.015)         &  (0.013)         &  (0.009)         &  (0.002)         &  (0.007)         &  (0.032)         \\
Income          &   -0.003\sym{***}&    0.000         &   -0.001\sym{***}&   -0.000         &   -0.000\sym{**} &    0.002\sym{***}\\
                &  (0.000)         &  (0.000)         &  (0.000)         &  (0.000)         &  (0.000)         &  (0.001)         \\
Children        &   -0.007         &   -0.030\sym{*}  &    0.042\sym{***}&   -0.003         &    0.020\sym{*}  &    0.053         \\
                &  (0.017)         &  (0.014)         &  (0.010)         &  (0.002)         &  (0.008)         &  (0.034)         \\
Couple          &   -0.002         &   -0.006         &    0.023\sym{*}  &    0.000         &    0.002         &    0.043         \\
                &  (0.017)         &  (0.015)         &  (0.010)         &  (0.002)         &  (0.008)         &  (0.034)         \\
Urban           &   -0.017         &   -0.012         &    0.021\sym{*}  &    0.004\sym{*}  &    0.023\sym{***}&    0.037         \\
                &  (0.014)         &  (0.012)         &  (0.009)         &  (0.002)         &  (0.007)         &  (0.029)         \\
CRT             &   -0.020\sym{**} &    0.006         &   -0.033\sym{***}&   -0.001         &   -0.067\sym{***}&   -0.002         \\
                &  (0.007)         &  (0.006)         &  (0.004)         &  (0.001)         &  (0.003)         &  (0.013)         \\
Constant        &    0.526\sym{***}&    2.067\sym{***}&    0.398\sym{***}&    0.012\sym{***}&    0.470\sym{***}&    0.777\sym{***}\\
                &  (0.030)         &  (0.033)         &  (0.019)         &  (0.004)         &  (0.014)         &  (0.063)         \\
\midrule
N               &   443415         &                  &                  &                  &                  &                  \\
Log. Likelihood &  -256038         &                  &                  &                  &                  &                  \\
BIC             &   512777         &                  &                  &                  &                  &                  \\
\bottomrule
\multicolumn{7}{l}{\footnotesize Standard errors (clustered at the subject level) in parentheses}\\
\multicolumn{7}{l}{\footnotesize \sym{*} \(p<0.05\), \sym{**} \(p<0.01\), \sym{***} \(p<0.001\)}\\
\end{tabular*}
}

\renewcommand{\footnotesize}{\footnotesize}
\end{table}

\begin{table}[htbp]
\renewcommand{\footnotesize}{\scriptsize}
\scriptsize
\caption{France \label{tab:l1t_france}}
{
\def\sym#1{\ifmmode^{#1}\else\(^{#1}\)\fi}
\begin{tabular*}{\hsize}{@{\hskip\tabcolsep\extracolsep\fill}l*{6}{c}}
\toprule
                &\makecell{Risk aversion \\ $\alpha$}         &\makecell{Loss aversion \\ $\lambda$}         &\makecell{Discounting \\ $\delta$}         &\makecell{Present bias \\ $\gamma$}         &\makecell{Tremble error \\ $\kappa$}         &\makecell{Fechner error \\ $\mu$}         \\
\midrule
Age             &    0.004\sym{**} &    0.000         &    0.000         &   -0.000\sym{*}  &    0.001         &   -0.007\sym{*}  \\
                &  (0.001)         &  (0.001)         &  (0.001)         &  (0.000)         &  (0.001)         &  (0.003)         \\
Male          &   -0.041         &   -0.051\sym{*}  &    0.041\sym{*}  &    0.000         &   -0.027         &    0.154         \\
                &  (0.034)         &  (0.022)         &  (0.018)         &  (0.003)         &  (0.015)         &  (0.082)         \\
Education       &    0.036         &    0.009         &   -0.045\sym{*}  &    0.000         &   -0.039\sym{*}  &   -0.085         \\
                &  (0.039)         &  (0.022)         &  (0.020)         &  (0.004)         &  (0.016)         &  (0.091)         \\
Income          &   -0.001         &   -0.001         &    0.000         &   -0.000         &   -0.001         &    0.002         \\
                &  (0.001)         &  (0.001)         &  (0.000)         &  (0.000)         &  (0.000)         &  (0.002)         \\
Children        &   -0.028         &   -0.037         &    0.021         &   -0.002         &   -0.013         &   -0.013         \\
                &  (0.037)         &  (0.027)         &  (0.021)         &  (0.004)         &  (0.018)         &  (0.075)         \\
Couple          &   -0.034         &    0.011         &   -0.017         &   -0.001         &   -0.001         &    0.074         \\
                &  (0.037)         &  (0.024)         &  (0.020)         &  (0.004)         &  (0.018)         &  (0.084)         \\
Urban           &   -0.015         &    0.014         &   -0.007         &    0.002         &    0.017         &   -0.016         \\
                &  (0.034)         &  (0.021)         &  (0.018)         &  (0.003)         &  (0.015)         &  (0.077)         \\
CRT             &   -0.011         &    0.009         &   -0.020\sym{*}  &   -0.000         &   -0.042\sym{***}&    0.008         \\
                &  (0.015)         &  (0.010)         &  (0.008)         &  (0.002)         &  (0.007)         &  (0.037)         \\
Constant        &    0.337\sym{***}&    2.007\sym{***}&    0.241\sym{***}&    0.025\sym{***}&    0.439\sym{***}&    0.798\sym{***}\\
                &  (0.073)         &  (0.042)         &  (0.040)         &  (0.007)         &  (0.030)         &  (0.175)         \\
\midrule
N               &    59325         &                  &                  &                  &                  &                  \\
Log. Likelihood &   -32677         &                  &                  &                  &                  &                  \\
BIC             &    65948         &                  &                  &                  &                  &                  \\
\bottomrule
\multicolumn{7}{l}{\footnotesize Standard errors (clustered at the subject level) in parentheses}\\
\multicolumn{7}{l}{\footnotesize \sym{*} \(p<0.05\), \sym{**} \(p<0.01\), \sym{***} \(p<0.001\)}\\
\end{tabular*}
}

\renewcommand{\footnotesize}{\footnotesize}
\end{table}

\begin{table}[htbp]
\centering
\scriptsize
\def\sym#1{\ifmmode^{#1}\else\(^{#1}\)\fi}
\caption{Germany}
\begin{tabular*}{\hsize}{@{\hskip\tabcolsep\extracolsep\fill}l*{6}{c}}
\toprule
                &\makecell{Risk aversion \\ $\alpha$}         &\makecell{Loss aversion \\ $\lambda$}         &\makecell{Discounting \\ $\delta$}         &\makecell{Present bias \\ $\gamma$}         &\makecell{Tremble error \\ $\kappa$}         &\makecell{Fechner error \\ $\mu$}         \\
\midrule
Age             &    0.002         &   -0.002\sym{*}  &   -0.003\sym{***}&   -0.000         &    0.000         &   -0.007\sym{**} \\
                &  (0.002)         &  (0.001)         &  (0.001)         &  (0.000)         &  (0.001)         &  (0.002)         \\
Male          &   -0.043         &   -0.020         &    0.027         &    0.006         &   -0.009         &    0.061         \\
                &  (0.039)         &  (0.023)         &  (0.019)         &  (0.004)         &  (0.018)         &  (0.063)         \\
Education       &   -0.017         &    0.002         &   -0.044\sym{*}  &    0.004         &   -0.039\sym{*}  &    0.062         \\
                &  (0.040)         &  (0.024)         &  (0.018)         &  (0.004)         &  (0.019)         &  (0.085)         \\
Income          &   -0.002\sym{*}  &   -0.000         &   -0.000         &   -0.000         &   -0.000         &    0.002         \\
                &  (0.001)         &  (0.001)         &  (0.001)         &  (0.000)         &  (0.000)         &  (0.002)         \\
Children        &   -0.070         &    0.003         &    0.040\sym{*}  &   -0.005         &    0.052\sym{**} &    0.139\sym{*}  \\
                &  (0.040)         &  (0.024)         &  (0.019)         &  (0.004)         &  (0.019)         &  (0.071)         \\
Couple          &   -0.067         &   -0.387\sym{***}&    0.040         &    0.003         &    0.006         &    0.124         \\
                &  (0.045)         &  (0.031)         &  (0.022)         &  (0.005)         &  (0.020)         &  (0.078)         \\
Urban           &    0.025         &   -0.001         &   -0.015         &    0.006         &    0.007         &   -0.019         \\
                &  (0.034)         &  (0.021)         &  (0.018)         &  (0.004)         &  (0.017)         &  (0.062)         \\
CRT             &   -0.032         &    0.007         &   -0.035\sym{***}&   -0.003         &   -0.077\sym{***}&    0.020         \\
                &  (0.020)         &  (0.010)         &  (0.008)         &  (0.002)         &  (0.008)         &  (0.030)         \\
Constant        &    0.581\sym{***}&    2.056\sym{***}&    0.356\sym{***}&    0.007         &    0.496\sym{***}&    0.630\sym{***}\\
                &  (0.082)         &  (0.057)         &  (0.045)         &  (0.009)         &  (0.040)         &  (0.136)         \\
\midrule
N               &    59115         &                  &                  &                  &                  &                  \\
Log. Likelihood &   -34058         &                  &                  &                  &                  &                  \\
BIC             &    68708         &                  &                  &                  &                  &                  \\
\bottomrule
\multicolumn{7}{l}{\scriptsize Standard errors (clustered at the subject level) in parentheses}\\
\multicolumn{7}{l}{\scriptsize \sym{*} \(p<0.05\), \sym{**} \(p<0.01\), \sym{***} \(p<0.001\)}\\
\end{tabular*}
\end{table}

\begin{table}[htbp]
\centering
\scriptsize
\def\sym#1{\ifmmode^{#1}\else\(^{#1}\)\fi}
\caption{Italy}
\begin{tabular*}{\hsize}{@{\hskip\tabcolsep\extracolsep\fill}l*{6}{c}}
\toprule
                &\makecell{Risk aversion \\ $\alpha$}         &\makecell{Loss aversion \\ $\lambda$}         &\makecell{Discounting \\ $\delta$}         &\makecell{Present bias \\ $\gamma$}         &\makecell{Tremble error \\ $\kappa$}         &\makecell{Fechner error \\ $\mu$}         \\
\midrule
Age             &    0.002         &   -0.005         &   -0.002         &   -0.000         &   -0.001         &   -0.009         \\
                &  (0.002)         &  (0.004)         &  (0.001)         &  (0.000)         &  (0.001)         &  (0.005)         \\
Male          &    0.029         &   -0.041         &   -0.049         &    0.011\sym{*}  &    0.024         &   -0.128         \\
                &  (0.039)         &  (0.071)         &  (0.027)         &  (0.006)         &  (0.018)         &  (0.123)         \\
Education       &   -0.017         &    0.037         &   -0.062\sym{*}  &    0.004         &   -0.008         &   -0.035         \\
                &  (0.039)         &  (0.064)         &  (0.027)         &  (0.006)         &  (0.019)         &  (0.134)         \\
Income          &   -0.002         &   -0.006\sym{***}&   -0.001         &   -0.000         &   -0.001         &    0.003         \\
                &  (0.001)         &  (0.001)         &  (0.001)         &  (0.000)         &  (0.001)         &  (0.004)         \\
Children        &    0.033         &   -0.030         &   -0.004         &   -0.011         &    0.011         &   -0.064         \\
                &  (0.045)         &  (0.079)         &  (0.033)         &  (0.008)         &  (0.024)         &  (0.141)         \\
Couple          &   -0.041         &    0.016         &    0.039         &    0.003         &    0.008         &    0.085         \\
                &  (0.049)         &  (0.079)         &  (0.036)         &  (0.008)         &  (0.024)         &  (0.149)         \\
Urban           &   -0.042         &   -0.027         &    0.011         &    0.009         &    0.024         &    0.075         \\
                &  (0.041)         &  (0.075)         &  (0.028)         &  (0.006)         &  (0.018)         &  (0.130)         \\
CRT             &    0.005         &    0.027         &   -0.031\sym{*}  &   -0.004         &   -0.074\sym{***}&   -0.042         \\
                &  (0.019)         &  (0.034)         &  (0.013)         &  (0.003)         &  (0.011)         &  (0.052)         \\
Constant        &    0.372\sym{***}&    2.301\sym{***}&    0.492\sym{***}&    0.042\sym{**} &    0.474\sym{***}&    1.254\sym{***}\\
                &  (0.090)         &  (0.153)         &  (0.064)         &  (0.014)         &  (0.039)         &  (0.283)         \\
\midrule
N               &    56735         &                  &                  &                  &                  &                  \\
Log. Likelihood &   -32434         &                  &                  &                  &                  &                  \\
BIC             &    65460         &                  &                  &                  &                  &                  \\
\bottomrule
\multicolumn{7}{l}{\scriptsize Standard errors (clustered at the subject level) in parentheses}\\
\multicolumn{7}{l}{\scriptsize \sym{*} \(p<0.05\), \sym{**} \(p<0.01\), \sym{***} \(p<0.001\)}\\
\end{tabular*}
\end{table}

\begin{table}[htbp]
\centering
\scriptsize
\def\sym#1{\ifmmode^{#1}\else\(^{#1}\)\fi}
\caption{Poland}
\begin{tabular*}{\hsize}{@{\hskip\tabcolsep\extracolsep\fill}l*{6}{c}}
\toprule
                &\makecell{Risk aversion \\ $\alpha$}         &\makecell{Loss aversion \\ $\lambda$}         &\makecell{Discounting \\ $\delta$}         &\makecell{Present bias \\ $\gamma$}         &\makecell{Tremble error \\ $\kappa$}         &\makecell{Fechner error \\ $\mu$}         \\
\midrule
Age             &   -0.000         &   -0.015\sym{**} &    0.001         &    0.000         &    0.002         &   -0.000         \\
                &  (0.002)         &  (0.005)         &  (0.002)         &  (0.000)         &  (0.001)         &  (0.004)         \\
Male          &   -0.054         &   -0.218         &    0.038         &   -0.008         &    0.043         &    0.068         \\
                &  (0.038)         &  (0.126)         &  (0.044)         &  (0.006)         &  (0.024)         &  (0.081)         \\
Education       &   -0.041         &   -0.156         &   -0.067         &    0.007         &   -0.006         &   -0.041         \\
                &  (0.051)         &  (0.110)         &  (0.054)         &  (0.006)         &  (0.026)         &  (0.116)         \\
Income          &    0.001         &   -0.005         &   -0.001         &    0.000         &   -0.001         &    0.001         \\
                &  (0.003)         &  (0.006)         &  (0.003)         &  (0.000)         &  (0.003)         &  (0.002)         \\
Children        &    0.074         &    0.104         &   -0.051         &    0.003         &   -0.016         &   -0.069         \\
                &  (0.044)         &  (0.117)         &  (0.045)         &  (0.008)         &  (0.031)         &  (0.086)         \\
Couple          &   -0.017         &    0.068         &    0.011         &   -0.008         &   -0.035         &    0.040         \\
                &  (0.044)         &  (0.120)         &  (0.044)         &  (0.007)         &  (0.028)         &  (0.097)         \\
Urban           &   -0.009         &    0.048         &    0.028         &    0.001         &    0.046         &   -0.043         \\
                &  (0.041)         &  (0.092)         &  (0.039)         &  (0.005)         &  (0.024)         &  (0.088)         \\
CRT             &   -0.047\sym{*}  &    0.025         &   -0.042\sym{*}  &    0.002         &   -0.050\sym{***}&   -0.019         \\
                &  (0.020)         &  (0.046)         &  (0.018)         &  (0.002)         &  (0.011)         &  (0.036)         \\
Constant        &    0.733\sym{***}&    2.485\sym{***}&    0.340\sym{***}&    0.008         &    0.306\sym{***}&    0.625\sym{***}\\
                &  (0.077)         &  (0.171)         &  (0.102)         &  (0.013)         &  (0.050)         &  (0.184)         \\
\midrule
N               &    55195         &                  &                  &                  &                  &                  \\
Log. Likelihood &   -31290         &                  &                  &                  &                  &                  \\
BIC             &    63170         &                  &                  &                  &                  &                  \\
\bottomrule
\multicolumn{7}{l}{\scriptsize Standard errors (clustered at the subject level) in parentheses}\\
\multicolumn{7}{l}{\scriptsize \sym{*} \(p<0.05\), \sym{**} \(p<0.01\), \sym{***} \(p<0.001\)}\\
\end{tabular*}
\end{table}

\begin{table}[htbp]
\centering
\scriptsize
\def\sym#1{\ifmmode^{#1}\else\(^{#1}\)\fi}
\caption{Romania}
\begin{tabular*}{\hsize}{@{\hskip\tabcolsep\extracolsep\fill}l*{6}{c}}
\toprule
                &\makecell{Risk aversion \\ $\alpha$}         &\makecell{Loss aversion \\ $\lambda$}         &\makecell{Discounting \\ $\delta$}         &\makecell{Present bias \\ $\gamma$}         &\makecell{Tremble error \\ $\kappa$}         &\makecell{Fechner error \\ $\mu$}         \\
\midrule
Age             &    0.004         &   -0.001         &    0.001         &   -0.001         &    0.002         &    0.000         \\
                &  (0.004)         &  (0.006)         &  (0.007)         &  (0.001)         &  (0.002)         &  (0.011)         \\
Male          &   -0.021         &   -0.123         &    0.059         &    0.009         &    0.002         &    0.084         \\
                &  (0.112)         &  (0.103)         &  (0.186)         &  (0.015)         &  (0.037)         &  (0.329)         \\
Education       &   -0.089         &   -0.011         &   -0.006         &   -0.004         &    0.032         &    0.066         \\
                &  (0.113)         &  (0.116)         &  (0.138)         &  (0.012)         &  (0.041)         &  (0.275)         \\
Income          &    0.001         &    0.000         &    0.000         &   -0.000         &   -0.004\sym{*}  &    0.003         \\
                &  (0.006)         &  (0.005)         &  (0.004)         &  (0.000)         &  (0.002)         &  (0.015)         \\
Children        &   -0.109         &   -0.097         &    0.151         &   -0.019         &    0.012         &    0.275         \\
                &  (0.114)         &  (0.146)         &  (0.166)         &  (0.017)         &  (0.047)         &  (0.335)         \\
Couple          &   -0.033         &   -0.067         &    0.039         &   -0.001         &    0.009         &    0.084         \\
                &  (0.102)         &  (0.170)         &  (0.092)         &  (0.013)         &  (0.044)         &  (0.209)         \\
Urban           &   -0.000         &   -0.092         &   -0.016         &   -0.018         &    0.038         &   -0.112         \\
                &  (0.084)         &  (0.117)         &  (0.112)         &  (0.012)         &  (0.042)         &  (0.274)         \\
CRT             &    0.004         &    0.081         &   -0.086\sym{*}  &    0.009\sym{*}  &   -0.080\sym{***}&   -0.111         \\
                &  (0.035)         &  (0.048)         &  (0.043)         &  (0.004)         &  (0.016)         &  (0.079)         \\
Constant        &    0.528\sym{***}&    1.833\sym{***}&    0.432\sym{*}  &    0.029         &    0.392\sym{***}&    0.752\sym{*}  \\
                &  (0.153)         &  (0.238)         &  (0.187)         &  (0.023)         &  (0.080)         &  (0.346)         \\
\midrule
N               &    48300         &                  &                  &                  &                  &                  \\
Log. Likelihood &   -29033         &                  &                  &                  &                  &                  \\
BIC             &    58649         &                  &                  &                  &                  &                  \\
\bottomrule
\multicolumn{7}{l}{\scriptsize Standard errors (clustered at the subject level) in parentheses}\\
\multicolumn{7}{l}{\scriptsize \sym{*} \(p<0.05\), \sym{**} \(p<0.01\), \sym{***} \(p<0.001\)}\\
\end{tabular*}
\end{table}

\begin{table}[htbp]
\centering
\scriptsize
\def\sym#1{\ifmmode^{#1}\else\(^{#1}\)\fi}
\caption{Spain}
\begin{tabular*}{\hsize}{@{\hskip\tabcolsep\extracolsep\fill}l*{6}{c}}
\toprule
                &\makecell{Risk aversion \\ $\alpha$}         &\makecell{Loss aversion \\ $\lambda$}         &\makecell{Discounting \\ $\delta$}         &\makecell{Present bias \\ $\gamma$}         &\makecell{Tremble error \\ $\kappa$}         &\makecell{Fechner error \\ $\mu$}         \\
\midrule
Age             &    0.002         &   -0.007         &   -0.001         &   -0.000         &    0.000         &   -0.007         \\
                &  (0.002)         &  (0.007)         &  (0.001)         &  (0.000)         &  (0.001)         &  (0.005)         \\
Male          &   -0.053         &   -0.060         &    0.051         &    0.009         &   -0.001         &    0.263\sym{*}  \\
                &  (0.043)         &  (0.089)         &  (0.029)         &  (0.006)         &  (0.019)         &  (0.129)         \\
Education       &   -0.032         &    0.049         &    0.006         &   -0.003         &   -0.054\sym{*}  &    0.205         \\
                &  (0.044)         &  (0.173)         &  (0.033)         &  (0.006)         &  (0.025)         &  (0.154)         \\
Income          &   -0.001         &   -0.004         &   -0.001         &    0.000         &   -0.001         &   -0.002         \\
                &  (0.001)         &  (0.002)         &  (0.001)         &  (0.000)         &  (0.001)         &  (0.004)         \\
Children        &   -0.025         &   -0.061         &    0.036         &    0.008         &    0.040         &    0.121         \\
                &  (0.045)         &  (0.186)         &  (0.038)         &  (0.007)         &  (0.025)         &  (0.128)         \\
Couple          &    0.040         &   -0.003         &    0.004         &    0.003         &    0.004         &   -0.022         \\
                &  (0.048)         &  (0.232)         &  (0.033)         &  (0.006)         &  (0.023)         &  (0.129)         \\
Urban           &   -0.073         &    0.079         &    0.059         &    0.005         &    0.026         &    0.182         \\
                &  (0.041)         &  (0.178)         &  (0.031)         &  (0.006)         &  (0.020)         &  (0.113)         \\
CRT             &   -0.009         &   -0.011         &   -0.034\sym{**} &   -0.002         &   -0.095\sym{***}&    0.006         \\
                &  (0.018)         &  (0.045)         &  (0.011)         &  (0.002)         &  (0.009)         &  (0.041)         \\
Constant        &    0.441\sym{***}&    2.069\sym{***}&    0.323\sym{***}&    0.006         &    0.517\sym{***}&    0.778\sym{*}  \\
                &  (0.099)         &  (0.481)         &  (0.062)         &  (0.012)         &  (0.048)         &  (0.305)         \\
\midrule
N               &    56385         &                  &                  &                  &                  &                  \\
Log. Likelihood &   -32948         &                  &                  &                  &                  &                  \\
BIC             &    66487         &                  &                  &                  &                  &                  \\
\bottomrule
\multicolumn{7}{l}{\scriptsize Standard errors (clustered at the subject level) in parentheses}\\
\multicolumn{7}{l}{\scriptsize \sym{*} \(p<0.05\), \sym{**} \(p<0.01\), \sym{***} \(p<0.001\)}\\
\end{tabular*}
\end{table}

\begin{table}[htbp]
\centering
\scriptsize
\def\sym#1{\ifmmode^{#1}\else\(^{#1}\)\fi}
\caption{Sweden}
\begin{tabular*}{\hsize}{@{\hskip\tabcolsep\extracolsep\fill}l*{6}{c}}
\toprule
                &\makecell{Risk aversion \\ $\alpha$}         &\makecell{Loss aversion \\ $\lambda$}         &\makecell{Discounting \\ $\delta$}         &\makecell{Present bias \\ $\gamma$}         &\makecell{Tremble error \\ $\kappa$}         &\makecell{Fechner error \\ $\mu$}         \\
\midrule
Age             &    0.001         &   -0.001         &   -0.001         &    0.000         &   -0.001         &   -0.007         \\
                &  (0.002)         &  (0.001)         &  (0.001)         &  (0.000)         &  (0.001)         &  (0.005)         \\
Male          &   -0.052         &   -0.037         &    0.051\sym{*}  &   -0.005         &   -0.003         &    0.170         \\
                &  (0.037)         &  (0.031)         &  (0.021)         &  (0.004)         &  (0.018)         &  (0.132)         \\
Education       &    0.070         &   -0.015         &   -0.062\sym{**} &   -0.006         &   -0.006         &   -0.285\sym{*}  \\
                &  (0.038)         &  (0.028)         &  (0.022)         &  (0.004)         &  (0.018)         &  (0.132)         \\
Income          &   -0.000         &   -0.001         &   -0.000         &    0.000         &   -0.001\sym{**} &    0.001         \\
                &  (0.001)         &  (0.001)         &  (0.001)         &  (0.000)         &  (0.000)         &  (0.003)         \\
Children        &   -0.020         &   -0.038         &    0.041         &   -0.012\sym{*}  &    0.046\sym{*}  &    0.046         \\
                &  (0.053)         &  (0.032)         &  (0.027)         &  (0.005)         &  (0.022)         &  (0.148)         \\
Couple          &   -0.012         &    0.036         &    0.009         &    0.000         &    0.026         &    0.067         \\
                &  (0.047)         &  (0.036)         &  (0.026)         &  (0.005)         &  (0.021)         &  (0.162)         \\
Urban           &   -0.005         &   -0.023         &    0.020         &   -0.001         &    0.052\sym{**} &    0.006         \\
                &  (0.038)         &  (0.028)         &  (0.022)         &  (0.004)         &  (0.018)         &  (0.125)         \\
CRT             &   -0.030         &    0.009         &   -0.025\sym{**} &    0.002         &   -0.068\sym{***}&    0.065         \\
                &  (0.019)         &  (0.013)         &  (0.009)         &  (0.002)         &  (0.009)         &  (0.063)         \\
Constant        &    0.354\sym{***}&    2.108\sym{***}&    0.313\sym{***}&    0.005         &    0.505\sym{***}&    1.060\sym{***}\\
                &  (0.079)         &  (0.079)         &  (0.044)         &  (0.009)         &  (0.039)         &  (0.262)         \\
\midrule
N               &    46025         &                  &                  &                  &                  &                  \\
Log. Likelihood &   -25898         &                  &                  &                  &                  &                  \\
BIC             &    52377         &                  &                  &                  &                  &                  \\
\bottomrule
\multicolumn{7}{l}{\scriptsize Standard errors (clustered at the subject level) in parentheses}\\
\multicolumn{7}{l}{\scriptsize \sym{*} \(p<0.05\), \sym{**} \(p<0.01\), \sym{***} \(p<0.001\)}\\
\end{tabular*}
\end{table}

\begin{table}
\caption{United Kingdom \label{tab:l1t_UK}}
\renewcommand{\footnotesize}{\scriptsize}
\scriptsize

{
\def\sym#1{\ifmmode^{#1}\else\(^{#1}\)\fi}
\begin{tabular*}{\hsize}{@{\hskip\tabcolsep\extracolsep\fill}l*{6}{c}}
\toprule
                &\makecell{Risk aversion \\ $\alpha$}         &\makecell{Loss aversion \\ $\lambda$}         &\makecell{Discounting \\ $\delta$}         &\makecell{Present bias \\ $\gamma$}         &\makecell{Tremble error \\ $\kappa$}         &\makecell{Fechner error \\ $\mu$}         \\
\midrule
Age             &    0.005\sym{***}&   -0.002         &   -0.005\sym{***}&   -0.000         &    0.000         &   -0.009\sym{***}\\
                &  (0.001)         &  (0.002)         &  (0.001)         &  (0.000)         &  (0.001)         &  (0.002)         \\
Male          &   -0.034         &   -0.095\sym{*}  &   -0.009         &    0.006         &   -0.012         &    0.030         \\
                &  (0.034)         &  (0.045)         &  (0.019)         &  (0.003)         &  (0.016)         &  (0.053)         \\
Education       &   -0.031         &    0.040         &   -0.004         &    0.006         &    0.007         &    0.047         \\
                &  (0.039)         &  (0.037)         &  (0.021)         &  (0.004)         &  (0.017)         &  (0.066)         \\
Income          &   -0.001\sym{*}  &   -0.001         &   -0.001         &   -0.000\sym{*}  &   -0.000         &    0.001         \\
                &  (0.001)         &  (0.001)         &  (0.000)         &  (0.000)         &  (0.000)         &  (0.001)         \\
Children        &   -0.039         &   -0.017         &    0.056\sym{**} &    0.004         &    0.047\sym{*}  &    0.019         \\
                &  (0.043)         &  (0.035)         &  (0.020)         &  (0.004)         &  (0.020)         &  (0.063)         \\
Couple          &   -0.018         &    0.019         &    0.008         &    0.002         &    0.001         &   -0.001         \\
                &  (0.045)         &  (0.037)         &  (0.023)         &  (0.004)         &  (0.020)         &  (0.067)         \\
Urban           &   -0.058         &   -0.066         &    0.020         &    0.004         &    0.027         &    0.088         \\
                &  (0.036)         &  (0.041)         &  (0.020)         &  (0.004)         &  (0.016)         &  (0.057)         \\
CRT             &   -0.023         &    0.015         &   -0.014         &   -0.001         &   -0.059\sym{***}&    0.027         \\
                &  (0.016)         &  (0.016)         &  (0.009)         &  (0.002)         &  (0.007)         &  (0.026)         \\
Constant        &    0.497\sym{***}&    2.136\sym{***}&    0.428\sym{***}&    0.016         &    0.422\sym{***}&    0.753\sym{***}\\
                &  (0.074)         &  (0.098)         &  (0.052)         &  (0.009)         &  (0.034)         &  (0.151)         \\
\midrule
N               &    62335         &                  &                  &                  &                  &                  \\
Log. Likelihood &   -34942         &                  &                  &                  &                  &                  \\
BIC             &    70480         &                  &                  &                  &                  &                  \\
\bottomrule
\multicolumn{7}{l}{\footnotesize Standard errors (clustered at the subject level) in parentheses}\\
\multicolumn{7}{l}{\footnotesize \sym{*} \(p<0.05\), \sym{**} \(p<0.01\), \sym{***} \(p<0.001\)}\\
\end{tabular*}
}

\renewcommand{\footnotesize}{\footnotesize}
\end{table}

\newpage
\section{Robustness checks}
\label{sec:robustness}
\setcounter{table}{0}

\subsection{Excluding multiple switchers}

In our main specification, employing maximum likelihood methods to estimate the preference parameters allows us to retain multiple switchers. However, multiple switch points may also signal participant confusion with the experiment or lack of focus. Excluding multiple switchers leads to a reduction of the sample by 2.079 respondents. Table~\ref{tab:l1tnmsp} contains estimates of the main specification for this reduced sample. Qualitatively, the results for the relation between individual characteristics and preference parameters are similar to those reported in Table~\ref{tab:main_full}. For the smaller sample, however, \emph{Power} and \emph{Tradition} are now positively correlated with risk aversion; \emph{FR} is now correlated negatively with risk aversion; \emph{RO} is correlated negatively with loss aversion; \emph{ES} is now correlated positively with present bias. \emph{LowStakes} is no longer significantly related to risk aversion; loss aversion is no longer significantly related to \emph{CRT}; \emph{Stimulation} is not longer related to time discounting; and \emph{Male}, \emph{Urban}, \emph{Conformity} and \emph{RO} are no longer related to present bias. In sum, for the statistically significant coefficients in our main specification, we find no statistically significant coefficient with the opposite sign in Table~\ref{tab:l1tnmsp}, and vice versa. Therefore, our main findings appear robust to excluding multiple switchers.

\begin{table}[htbp]
\centering
\scriptsize
\def\sym#1{\ifmmode^{#1}\else\(^{#1}\)\fi}
\caption{Robustness check: no multiple switchers \label{tab:l1tnmsp}}
\begin{tabular*}{\hsize}{@{\hskip\tabcolsep\extracolsep\fill}l*{6}{c}}
\toprule
                &\makecell{Risk aversion \\ $\alpha$}         &\makecell{Loss aversion \\ $\lambda$}         &\makecell{Discounting \\ $\delta$}         &\makecell{Present bias \\ $\gamma$}         &\makecell{Tremble error \\ $\kappa$}         &\makecell{Fechner error \\ $\mu$}         \\
\midrule
Age             &    0.002\sym{***}&   -0.005\sym{***}&   -0.001\sym{**} &   -0.000\sym{*}  &    0.001\sym{***}&   -0.004\sym{***}\\
                &  (0.001)         &  (0.001)         &  (0.000)         &  (0.000)         &  (0.000)         &  (0.001)         \\
Male          &   -0.039\sym{**} &   -0.132\sym{***}&    0.021\sym{**} &    0.003         &   -0.014\sym{*}  &    0.071\sym{**} \\
                &  (0.013)         &  (0.031)         &  (0.008)         &  (0.002)         &  (0.006)         &  (0.028)         \\
Education       &   -0.023         &    0.007         &   -0.026\sym{**} &    0.000         &   -0.026\sym{***}&    0.029         \\
                &  (0.014)         &  (0.032)         &  (0.009)         &  (0.002)         &  (0.007)         &  (0.030)         \\
Income          &   -0.001\sym{**} &   -0.002\sym{**} &   -0.000         &   -0.000         &   -0.000         &    0.001         \\
                &  (0.000)         &  (0.001)         &  (0.000)         &  (0.000)         &  (0.000)         &  (0.001)         \\
Children        &   -0.009         &   -0.065         &    0.028\sym{**} &   -0.001         &   -0.002         &    0.042         \\
                &  (0.015)         &  (0.037)         &  (0.009)         &  (0.002)         &  (0.008)         &  (0.031)         \\
Couple          &   -0.012         &    0.024         &   -0.001         &   -0.000         &   -0.010         &    0.021         \\
                &  (0.015)         &  (0.036)         &  (0.009)         &  (0.002)         &  (0.008)         &  (0.031)         \\
Urban           &   -0.012         &    0.004         &    0.004         &    0.003         &    0.005         &    0.033         \\
                &  (0.012)         &  (0.030)         &  (0.008)         &  (0.002)         &  (0.006)         &  (0.026)         \\
CRT             &   -0.027\sym{***}&    0.013         &   -0.023\sym{***}&   -0.001         &   -0.040\sym{***}&    0.010         \\
                &  (0.006)         &  (0.014)         &  (0.004)         &  (0.001)         &  (0.003)         &  (0.013)         \\
Achievement     &   -0.012         &    0.023         &    0.010         &   -0.002         &    0.001         &    0.023         \\
                &  (0.016)         &  (0.034)         &  (0.010)         &  (0.002)         &  (0.008)         &  (0.037)         \\
Benevolence     &   -0.006         &    0.032         &    0.002         &    0.001         &    0.004         &   -0.010         \\
                &  (0.015)         &  (0.040)         &  (0.010)         &  (0.002)         &  (0.008)         &  (0.032)         \\
Conformity      &    0.003         &    0.021         &   -0.005         &   -0.003         &    0.010         &   -0.002         \\
                &  (0.014)         &  (0.033)         &  (0.009)         &  (0.002)         &  (0.007)         &  (0.031)         \\
Hedonism        &   -0.026         &   -0.029         &    0.032\sym{***}&    0.002         &   -0.015\sym{*}  &    0.088\sym{*}  \\
                &  (0.016)         &  (0.034)         &  (0.009)         &  (0.002)         &  (0.007)         &  (0.035)         \\
Power           &    0.035\sym{*}  &    0.034         &   -0.012         &    0.000         &    0.019\sym{*}  &   -0.071         \\
                &  (0.017)         &  (0.039)         &  (0.011)         &  (0.002)         &  (0.008)         &  (0.039)         \\
Security        &    0.002         &    0.024         &   -0.006         &   -0.000         &    0.003         &   -0.010         \\
                &  (0.015)         &  (0.035)         &  (0.010)         &  (0.002)         &  (0.007)         &  (0.036)         \\
Self direction  &    0.006         &   -0.019         &    0.003         &   -0.002         &   -0.022\sym{**} &   -0.015         \\
                &  (0.015)         &  (0.036)         &  (0.009)         &  (0.002)         &  (0.007)         &  (0.032)         \\
Stimulation     &   -0.045\sym{**} &   -0.113\sym{***}&    0.019         &   -0.001         &    0.022\sym{**} &    0.113\sym{**} \\
                &  (0.016)         &  (0.033)         &  (0.011)         &  (0.002)         &  (0.007)         &  (0.040)         \\
Tradition       &    0.031\sym{*}  &   -0.035         &   -0.001         &    0.001         &    0.006         &   -0.062\sym{*}  \\
                &  (0.013)         &  (0.031)         &  (0.009)         &  (0.002)         &  (0.007)         &  (0.031)         \\
Universalism    &   -0.012         &   -0.053         &   -0.013         &   -0.003         &   -0.013         &   -0.004         \\
                &  (0.015)         &  (0.037)         &  (0.010)         &  (0.002)         &  (0.008)         &  (0.033)         \\
FR              &   -0.056\sym{*}  &    0.233\sym{***}&    0.017         &    0.006\sym{*}  &   -0.038\sym{**} &    0.067         \\
                &  (0.028)         &  (0.068)         &  (0.015)         &  (0.003)         &  (0.013)         &  (0.055)         \\
IT              &   -0.093\sym{**} &    0.162\sym{**} &    0.083\sym{***}&    0.013\sym{***}&   -0.070\sym{***}&    0.207\sym{**} \\
                &  (0.029)         &  (0.061)         &  (0.016)         &  (0.003)         &  (0.013)         &  (0.064)         \\
PL              &    0.173\sym{***}&    0.052         &    0.067\sym{**} &    0.004         &   -0.117\sym{***}&   -0.061         \\
                &  (0.031)         &  (0.073)         &  (0.022)         &  (0.004)         &  (0.016)         &  (0.051)         \\
RO              &    0.096\sym{*}  &   -0.143\sym{*}  &    0.228\sym{***}&   -0.008         &   -0.025         &    0.090         \\
                &  (0.044)         &  (0.070)         &  (0.044)         &  (0.005)         &  (0.019)         &  (0.094)         \\
ES              &   -0.040         &   -0.042         &    0.071\sym{***}&    0.008\sym{*}  &   -0.021         &    0.164\sym{*}  \\
                &  (0.029)         &  (0.055)         &  (0.017)         &  (0.003)         &  (0.013)         &  (0.065)         \\
SE              &   -0.120\sym{***}&    0.247\sym{***}&    0.019         &    0.003         &   -0.036\sym{**} &    0.226\sym{**} \\
                &  (0.028)         &  (0.057)         &  (0.014)         &  (0.003)         &  (0.013)         &  (0.071)         \\
UK              &    0.030         &    0.245\sym{***}&    0.030\sym{*}  &    0.003         &   -0.037\sym{**} &   -0.042         \\
                &  (0.028)         &  (0.067)         &  (0.015)         &  (0.003)         &  (0.012)         &  (0.051)         \\
ABreversed      &    0.034\sym{**} &    0.021         &   -0.020\sym{*}  &   -0.005\sym{**} &    0.010         &   -0.075\sym{**} \\
                &  (0.012)         &  (0.030)         &  (0.008)         &  (0.002)         &  (0.006)         &  (0.027)         \\
Incentivized    &   -0.049\sym{**} &   -0.176\sym{***}&    0.061\sym{***}&   -0.019\sym{***}&   -0.010         &    0.076\sym{*}  \\
                &  (0.015)         &  (0.038)         &  (0.009)         &  (0.002)         &  (0.007)         &  (0.034)         \\
LowStakes       &   -0.025         &   -0.124\sym{*}  &   -0.011         &    0.026\sym{***}&    0.006         &    0.054         \\
                &  (0.024)         &  (0.050)         &  (0.017)         &  (0.004)         &  (0.012)         &  (0.063)         \\
HighStakes      &    0.025         &    0.218\sym{*}  &   -0.032\sym{*}  &   -0.009\sym{**} &    0.028\sym{*}  &   -0.018         \\
                &  (0.025)         &  (0.085)         &  (0.013)         &  (0.003)         &  (0.011)         &  (0.047)         \\
Constant        &    0.541\sym{***}&    2.211\sym{***}&    0.248\sym{***}&    0.028\sym{***}&    0.392\sym{***}&    0.613\sym{***}\\
                &  (0.043)         &  (0.106)         &  (0.023)         &  (0.005)         &  (0.021)         &  (0.079)         \\
\midrule
N               &   369985         &                  &                  &                  &                  &                  \\
Log. Likelihood &  -200483         &                  &                  &                  &                  &                  \\
BIC             &   403274         &                  &                  &                  &                  &                  \\
\bottomrule
\multicolumn{7}{l}{\scriptsize Standard errors (clustered at the subject level) in parentheses}\\
\multicolumn{7}{l}{\scriptsize \sym{*} \(p<0.05\), \sym{**} \(p<0.01\), \sym{***} \(p<0.001\)}\\
\end{tabular*}
\end{table}

\subsection{Alternative utility specifications}
\label{sec:robustness_utility}

The specification of utility can have a strong impact on the estimated preference parameters and on their correlations with individual characteristics. In this section, we therefore vary several assumptions underlying our modelling of preferences. In particular, there is a vivid discussion in the literature on how to correctly account for structural connections of different preference domains. For instance, the standard time-separable discounted expected utility model assumes that utility over risk and time are connected because risk aversion and the elasticity of intertemporal substitution are governed by the same parameter. In this context, \cite{Andersen2008} argue that a failure to account for utility curvature when eliciting discount rates may bias the resulting estimates. Several studies have subsequently suggested that the curvature of utility over risk is different from the curvature of utility over time \citep{Andreoni2012, Abdellaoui2013}.\footnote{But, see also \cite{meissner2022measuring}, who find similar estimates for curvature of utility over risk and time.} In these cases, risk and time preferences should better be estimated separably. The theoretical framework underlying the main specification of our paper is built on a discounted expected utility model that assumes utility over risk and time to be connected. 

Further, our findings may change if our underlying model allows for probability weighting. If individuals weigh probabilities non linearly, failure to account for probability weighting can bias estimates of risk aversion and other preference parameters \citep{Bouchouicha2019, Abdellaoui2019}. Our setup does not allow to identify probability weighting, because we do not vary the probabilities in our MPLs. For practical reasons, all lotteries in our experiments were 50/50 gambles, conveyed in everyday language as coin flips. As stressed by \citet{Andersen2018}, assessing probability weighting significantly increases task complexity in MPLs. Naturally, our demographically representative sample included many participants with lower levels of education. Therefore, we decided to reduce task complexity and to abstract from probability weighting. 

Below, we first test robustness of our findings with respect to the specified utility function $u(x)$. We then present specifications that estimate time preference separately from risk preferences, and a specification that assumes no present bias.

\paragraph{Different utility curvature in gains and losses:} First, our main model assumes the same curvature of utility in gains and losses (around the reference point of zero). As an alternative, we allow for different curvature in gains and losses, at the cost of restricting the parameter for loss aversion ($\lambda=1$). We therefore assume the following utility function: 

\begin{equation}
u(x)=
\begin{cases}
\frac{x^{1-\alpha^+}}{1-\alpha^+} & \mbox{ if } x\geq 0\\
\frac{-(-x)^{1-\alpha^-}}{1-\alpha^-} & \mbox{ if } x < 0,\\
\end{cases}
\end{equation}

where $\alpha^+$ and $\alpha^-$ denote curvature of utility in gains and losses, respectively. We present the estimation results of this model in Table~\ref{tab:l1t2a}. All correlations of individual characteristics with preference parameters other than the new parameter $\alpha^{-}$ remain almost identical. Many of the individual characteristics that are related to curvature in gains are related in a similar way to curvature in losses. For instance, \emph{Age}, \emph{PL}, \emph{RO}, and \emph{ABreversed} are positively correlated with curvature in both gains and losses. \emph{CRT}, \emph{Stimulation}, \emph{IT}, and \emph{SE} are negatively correlated with curvature in both gains and losses. \emph{Male}, \emph{Income}, \emph{Incentivization}, and \emph{LowStakes} are negatively related to curvature in gains, but not related to curvature in losses. In comparison, \emph{Tradition} (\emph{FR}) is positively (negatively) related to curvature in losses, but is not related to curvature in gains.

\begin{table}[htbp]
\centering
\scriptsize
\def\sym#1{\ifmmode^{#1}\else\(^{#1}\)\fi}
\caption{Robustness check: different utility curvature in gains and losses \label{tab:l1t2a}}
\begin{tabular*}{\hsize}{@{\hskip\tabcolsep\extracolsep\fill}l*{6}{c}}
\toprule
                &\makecell{Gains \\ $\alpha+$}         &\makecell{Losses \\ $\alpha-$}         &\makecell{Discounting \\ $\delta$}         &\makecell{Present bias \\ $\gamma$}         &\makecell{Tremble error \\ $\kappa$}         &\makecell{Fechner error \\ $\mu$}         \\
\midrule
Age             &    0.002\sym{***}&    0.003\sym{***}&   -0.001\sym{**} &   -0.000\sym{*}  &    0.001\sym{*}  &   -0.004\sym{***}\\
                &  (0.001)         &  (0.001)         &  (0.000)         &  (0.000)         &  (0.000)         &  (0.001)         \\
Male          &   -0.037\sym{**} &   -0.013         &    0.019\sym{*}  &    0.003\sym{*}  &   -0.003         &    0.076\sym{**} \\
                &  (0.012)         &  (0.012)         &  (0.008)         &  (0.002)         &  (0.007)         &  (0.028)         \\
Education       &   -0.018         &   -0.020         &   -0.032\sym{***}&    0.001         &   -0.021\sym{**} &    0.019         \\
                &  (0.014)         &  (0.013)         &  (0.009)         &  (0.002)         &  (0.007)         &  (0.032)         \\
Income          &   -0.001\sym{**} &   -0.001         &   -0.000         &   -0.000         &   -0.001\sym{***}&    0.002\sym{*}  \\
                &  (0.000)         &  (0.000)         &  (0.000)         &  (0.000)         &  (0.000)         &  (0.001)         \\
Children        &   -0.014         &   -0.004         &    0.031\sym{***}&   -0.002         &    0.025\sym{**} &    0.030         \\
                &  (0.014)         &  (0.014)         &  (0.009)         &  (0.002)         &  (0.008)         &  (0.031)         \\
Couple          &   -0.013         &   -0.015         &   -0.000         &    0.001         &    0.002         &    0.020         \\
                &  (0.015)         &  (0.014)         &  (0.009)         &  (0.002)         &  (0.008)         &  (0.031)         \\
Urban           &   -0.021         &   -0.020         &    0.008         &    0.003\sym{*}  &    0.020\sym{**} &    0.036         \\
                &  (0.012)         &  (0.012)         &  (0.008)         &  (0.002)         &  (0.007)         &  (0.028)         \\
CRT             &   -0.021\sym{***}&   -0.025\sym{***}&   -0.027\sym{***}&   -0.001         &   -0.064\sym{***}&    0.005         \\
                &  (0.006)         &  (0.006)         &  (0.004)         &  (0.001)         &  (0.003)         &  (0.013)         \\
Achievement     &   -0.012         &   -0.013         &    0.008         &   -0.000         &    0.010         &    0.010         \\
                &  (0.016)         &  (0.016)         &  (0.010)         &  (0.002)         &  (0.008)         &  (0.041)         \\
Benevolence     &    0.001         &   -0.005         &   -0.001         &    0.002         &   -0.006         &   -0.014         \\
                &  (0.015)         &  (0.014)         &  (0.010)         &  (0.002)         &  (0.008)         &  (0.034)         \\
Conformity      &    0.002         &   -0.002         &   -0.004         &   -0.004\sym{*}  &    0.012         &   -0.008         \\
                &  (0.014)         &  (0.013)         &  (0.009)         &  (0.002)         &  (0.007)         &  (0.032)         \\
Hedonism        &   -0.023         &   -0.016         &    0.032\sym{***}&    0.002         &   -0.021\sym{**} &    0.103\sym{**} \\
                &  (0.015)         &  (0.015)         &  (0.009)         &  (0.002)         &  (0.007)         &  (0.036)         \\
Power           &    0.031         &    0.027         &   -0.005         &    0.000         &    0.054\sym{***}&   -0.075         \\
                &  (0.017)         &  (0.017)         &  (0.011)         &  (0.002)         &  (0.009)         &  (0.042)         \\
Security        &   -0.002         &   -0.006         &   -0.005         &   -0.000         &   -0.015         &   -0.008         \\
                &  (0.015)         &  (0.015)         &  (0.010)         &  (0.002)         &  (0.008)         &  (0.036)         \\
Self direction  &   -0.001         &    0.005         &    0.005         &   -0.002         &   -0.031\sym{***}&   -0.012         \\
                &  (0.014)         &  (0.014)         &  (0.009)         &  (0.002)         &  (0.008)         &  (0.033)         \\
Stimulation     &   -0.056\sym{***}&   -0.034\sym{*}  &    0.023\sym{*}  &   -0.001         &    0.041\sym{***}&    0.122\sym{**} \\
                &  (0.016)         &  (0.016)         &  (0.011)         &  (0.002)         &  (0.008)         &  (0.044)         \\
Tradition       &    0.021         &    0.026\sym{*}  &    0.002         &    0.000         &    0.017\sym{*}  &   -0.046         \\
                &  (0.013)         &  (0.013)         &  (0.009)         &  (0.002)         &  (0.007)         &  (0.032)         \\
Universalism    &   -0.013         &   -0.005         &   -0.011         &   -0.004         &   -0.012         &    0.007         \\
                &  (0.014)         &  (0.014)         &  (0.010)         &  (0.002)         &  (0.008)         &  (0.034)         \\
FR              &   -0.045         &   -0.086\sym{**} &    0.017         &    0.006\sym{*}  &   -0.076\sym{***}&    0.102         \\
                &  (0.027)         &  (0.027)         &  (0.014)         &  (0.003)         &  (0.013)         &  (0.056)         \\
IT              &   -0.075\sym{**} &   -0.104\sym{***}&    0.083\sym{***}&    0.016\sym{***}&   -0.066\sym{***}&    0.231\sym{***}\\
                &  (0.028)         &  (0.028)         &  (0.016)         &  (0.003)         &  (0.013)         &  (0.067)         \\
PL              &    0.200\sym{***}&    0.191\sym{***}&    0.067\sym{**} &    0.004         &   -0.139\sym{***}&   -0.033         \\
                &  (0.029)         &  (0.029)         &  (0.022)         &  (0.004)         &  (0.017)         &  (0.052)         \\
RO              &    0.103\sym{*}  &    0.117\sym{**} &    0.246\sym{***}&   -0.018\sym{***}&   -0.035         &    0.195         \\
                &  (0.042)         &  (0.041)         &  (0.043)         &  (0.005)         &  (0.019)         &  (0.105)         \\
ES              &   -0.033         &   -0.032         &    0.075\sym{***}&    0.007         &   -0.019         &    0.206\sym{**} \\
                &  (0.028)         &  (0.029)         &  (0.018)         &  (0.004)         &  (0.013)         &  (0.069)         \\
SE              &   -0.109\sym{***}&   -0.149\sym{***}&    0.019         &    0.003         &   -0.037\sym{**} &    0.255\sym{***}\\
                &  (0.027)         &  (0.027)         &  (0.014)         &  (0.003)         &  (0.013)         &  (0.073)         \\
UK              &    0.045         &    0.003         &    0.031\sym{*}  &    0.003         &   -0.052\sym{***}&   -0.029         \\
                &  (0.027)         &  (0.027)         &  (0.015)         &  (0.003)         &  (0.013)         &  (0.051)         \\
ABreversed      &    0.039\sym{**} &    0.033\sym{**} &   -0.019\sym{*}  &   -0.007\sym{***}&    0.022\sym{***}&   -0.099\sym{***}\\
                &  (0.012)         &  (0.012)         &  (0.008)         &  (0.002)         &  (0.006)         &  (0.028)         \\
Incentivized    &   -0.045\sym{**} &   -0.014         &    0.064\sym{***}&   -0.022\sym{***}&   -0.011         &    0.081\sym{*}  \\
                &  (0.015)         &  (0.015)         &  (0.009)         &  (0.002)         &  (0.007)         &  (0.035)         \\
LowStakes       &   -0.048\sym{*}  &   -0.021         &   -0.011         &    0.028\sym{***}&    0.017         &    0.095         \\
                &  (0.023)         &  (0.024)         &  (0.017)         &  (0.004)         &  (0.013)         &  (0.068)         \\
HighStakes      &    0.025         &    0.005         &   -0.036\sym{**} &   -0.007\sym{**} &    0.030\sym{*}  &   -0.011         \\
                &  (0.025)         &  (0.023)         &  (0.013)         &  (0.003)         &  (0.012)         &  (0.048)         \\
Constant        &    0.528\sym{***}&    0.305\sym{***}&    0.254\sym{***}&    0.030\sym{***}&    0.484\sym{***}&    0.615\sym{***}\\
                &  (0.042)         &  (0.041)         &  (0.023)         &  (0.005)         &  (0.022)         &  (0.082)         \\
\midrule
N               &   443415         &                  &                  &                  &                  &                  \\
Log. Likelihood &  -252970         &                  &                  &                  &                  &                  \\
BIC             &   508280         &                  &                  &                  &                  &                  \\
\bottomrule
\multicolumn{7}{l}{\scriptsize Standard errors (clustered at the subject level) in parentheses}\\
\multicolumn{7}{l}{\scriptsize \sym{*} \(p<0.05\), \sym{**} \(p<0.01\), \sym{***} \(p<0.001\)}\\
\end{tabular*}
\end{table}

\paragraph{CARA utility:} Our main specification assumes that preferences are characterized by constant relative risk aversion (CRRA). To examine whether our results are robust to that assumption, we estimate an alternative specification, which assumes constant absolute risk aversion (CARA) in Table~\ref{tab:l1tCARA}:

\begin{equation}
u(x)=
\begin{cases}
\frac{1-\exp(-\varphi x)}{\varphi} & \mbox{ if } x\geq 0\\
\frac{-\lambda(1-\exp(\varphi x))}{\varphi} & \mbox{ if } x < 0\\
\end{cases}
\end{equation}

Here, $\varphi$ is the parameter of absolute risk aversion. The results are largely robust, with a few exceptions. \emph{Age} and \emph{Male} are no longer correlated with present bias. Similarly, the correlation between \emph{Male}, \emph{Stimulation} and \emph{ABreversed} with discounting are no longer statistically significant. On the other hand, a few relations are statistically significant under this specification but not for our main specification. Risk aversion is now negatively related to living as a couple, and positively related to \emph{Achievement} and \emph{Tradition}. Finally, discounting is now negatively associated with \emph{Income}.

\begin{table}[htbp]
\centering
\scriptsize
\def\sym#1{\ifmmode^{#1}\else\(^{#1}\)\fi}
\caption{Robustness check: CARA utility \label{tab:l1tCARA}}
\begin{tabular*}{\hsize}{@{\hskip\tabcolsep\extracolsep\fill}l*{6}{c}}
\toprule
                &\makecell{Risk aversion \\ $\varphi$}         &\makecell{Loss aversion \\ $\lambda$}         &\makecell{Discounting \\ $\delta$}         &\makecell{Present bias \\ $\gamma$}         &\makecell{Tremble error \\ $\kappa$}         &\makecell{Fechner error \\ $\mu$}         \\
\midrule
Age             &   0.0001\sym{***}&  -0.0056\sym{***}&  -0.0011\sym{**} &  -0.0002         &   0.0006         &  -0.0058\sym{**} \\
                & (0.0000)         & (0.0009)         & (0.0004)         & (0.0001)         & (0.0003)         & (0.0022)         \\
Male          &  -0.0010\sym{**} &  -0.0902\sym{***}&   0.0154         &   0.0034         &  -0.0011         &   0.0379         \\
                & (0.0003)         & (0.0208)         & (0.0083)         & (0.0019)         & (0.0070)         & (0.0506)         \\
Education       &  -0.0005         &   0.0060         &  -0.0406\sym{***}&   0.0004         &  -0.0212\sym{**} &   0.0049         \\
                & (0.0003)         & (0.0220)         & (0.0090)         & (0.0020)         & (0.0072)         & (0.0548)         \\
Income          &  -0.0000\sym{*}  &  -0.0011\sym{*}  &  -0.0004\sym{*}  &  -0.0000         &  -0.0007\sym{***}&   0.0002         \\
                & (0.0000)         & (0.0005)         & (0.0002)         & (0.0000)         & (0.0002)         & (0.0012)         \\
Children        &  -0.0005         &  -0.0423         &   0.0358\sym{***}&  -0.0032         &   0.0294\sym{***}&  -0.0060         \\
                & (0.0004)         & (0.0257)         & (0.0099)         & (0.0022)         & (0.0083)         & (0.0606)         \\
Couple          &  -0.0008\sym{*}  &   0.0245         &  -0.0007         &   0.0016         &   0.0050         &   0.0368         \\
                & (0.0004)         & (0.0255)         & (0.0099)         & (0.0022)         & (0.0081)         & (0.0590)         \\
Urban           &  -0.0006         &   0.0027         &   0.0047         &   0.0039\sym{*}  &   0.0229\sym{***}&  -0.0003         \\
                & (0.0003)         & (0.0205)         & (0.0085)         & (0.0019)         & (0.0069)         & (0.0496)         \\
CRT             &  -0.0004\sym{**} &   0.0297\sym{**} &  -0.0358\sym{***}&  -0.0009         &  -0.0668\sym{***}&  -0.0433         \\
                & (0.0002)         & (0.0098)         & (0.0037)         & (0.0009)         & (0.0033)         & (0.0244)         \\
Achievement     &  -0.0008\sym{*}  &   0.0185         &   0.0091         &  -0.0001         &   0.0134         &  -0.0012         \\
                & (0.0004)         & (0.0236)         & (0.0103)         & (0.0023)         & (0.0080)         & (0.0615)         \\
Benevolence     &  -0.0000         &   0.0253         &  -0.0013         &   0.0026         &  -0.0063         &  -0.0329         \\
                & (0.0004)         & (0.0275)         & (0.0101)         & (0.0023)         & (0.0085)         & (0.0615)         \\
Conformity      &   0.0002         &   0.0106         &  -0.0041         &  -0.0044\sym{*}  &   0.0095         &   0.0209         \\
                & (0.0003)         & (0.0228)         & (0.0091)         & (0.0021)         & (0.0076)         & (0.0547)         \\
Hedonism        &  -0.0003         &  -0.0233         &   0.0294\sym{**} &   0.0016         &  -0.0225\sym{**} &   0.1939\sym{***}\\
                & (0.0004)         & (0.0233)         & (0.0092)         & (0.0021)         & (0.0078)         & (0.0561)         \\
Power           &   0.0004         &   0.0213         &   0.0064         &   0.0007         &   0.0541\sym{***}&  -0.0245         \\
                & (0.0004)         & (0.0261)         & (0.0115)         & (0.0026)         & (0.0089)         & (0.0696)         \\
Security        &  -0.0001         &   0.0203         &  -0.0044         &  -0.0002         &  -0.0175\sym{*}  &   0.0127         \\
                & (0.0004)         & (0.0247)         & (0.0097)         & (0.0022)         & (0.0082)         & (0.0587)         \\
Self direction  &   0.0003         &  -0.0244         &   0.0045         &  -0.0031         &  -0.0344\sym{***}&  -0.0384         \\
                & (0.0004)         & (0.0241)         & (0.0093)         & (0.0021)         & (0.0079)         & (0.0583)         \\
Stimulation     &  -0.0016\sym{***}&  -0.0665\sym{**} &   0.0149         &  -0.0012         &   0.0464\sym{***}&   0.0664         \\
                & (0.0004)         & (0.0224)         & (0.0106)         & (0.0023)         & (0.0079)         & (0.0622)         \\
Tradition       &   0.0008\sym{*}  &  -0.0392         &   0.0063         &  -0.0002         &   0.0103         &  -0.0024         \\
                & (0.0003)         & (0.0216)         & (0.0088)         & (0.0019)         & (0.0073)         & (0.0534)         \\
Universalism    &  -0.0005         &  -0.0203         &  -0.0145         &  -0.0041         &  -0.0063         &  -0.0952         \\
                & (0.0004)         & (0.0258)         & (0.0098)         & (0.0023)         & (0.0082)         & (0.0600)         \\
FR              &  -0.0011         &   0.2004\sym{***}&   0.0124         &   0.0070\sym{*}  &  -0.0743\sym{***}&   0.1347         \\
                & (0.0007)         & (0.0430)         & (0.0148)         & (0.0033)         & (0.0138)         & (0.0929)         \\
IT              &  -0.0020\sym{**} &   0.1698\sym{***}&   0.0838\sym{***}&   0.0167\sym{***}&  -0.0627\sym{***}&   0.3486\sym{***}\\
                & (0.0007)         & (0.0383)         & (0.0166)         & (0.0038)         & (0.0137)         & (0.0967)         \\
PL              &   0.0032\sym{***}&  -0.0038         &   0.2352\sym{***}&   0.0066         &  -0.1310\sym{***}&   0.9789\sym{***}\\
                & (0.0007)         & (0.0433)         & (0.0282)         & (0.0053)         & (0.0172)         & (0.1322)         \\
RO              &   0.0017\sym{*}  &  -0.0760         &   0.4099\sym{***}&  -0.0280\sym{***}&  -0.0365         &   1.3753\sym{***}\\
                & (0.0008)         & (0.0459)         & (0.0426)         & (0.0066)         & (0.0192)         & (0.1913)         \\
ES              &  -0.0008         &   0.0088         &   0.0799\sym{***}&   0.0075         &  -0.0142         &   0.3967\sym{***}\\
                & (0.0007)         & (0.0368)         & (0.0181)         & (0.0039)         & (0.0144)         & (0.1018)         \\
SE              &  -0.0030\sym{***}&   0.2377\sym{***}&   0.0116         &   0.0033         &  -0.0271\sym{*}  &   0.1711         \\
                & (0.0007)         & (0.0425)         & (0.0146)         & (0.0033)         & (0.0137)         & (0.1045)         \\
UK              &   0.0013         &   0.1556\sym{***}&   0.0441\sym{**} &   0.0034         &  -0.0582\sym{***}&   0.1468         \\
                & (0.0007)         & (0.0421)         & (0.0146)         & (0.0033)         & (0.0139)         & (0.0888)         \\
ABreversed      &   0.0009\sym{**} &   0.0037         &  -0.0142         &  -0.0083\sym{***}&   0.0249\sym{***}&  -0.1758\sym{***}\\
                & (0.0003)         & (0.0199)         & (0.0084)         & (0.0019)         & (0.0068)         & (0.0493)         \\
Incentivized    &  -0.0007\sym{*}  &  -0.1264\sym{***}&   0.0641\sym{***}&  -0.0268\sym{***}&  -0.0143         &   0.0156         \\
                & (0.0004)         & (0.0234)         & (0.0094)         & (0.0022)         & (0.0076)         & (0.0556)         \\
LowStakes       &  -0.0016\sym{**} &  -0.0880\sym{**} &  -0.0162         &   0.0315\sym{***}&   0.0171         &   0.1053         \\
                & (0.0006)         & (0.0341)         & (0.0180)         & (0.0047)         & (0.0130)         & (0.1078)         \\
HighStakes      &   0.0006         &   0.1009\sym{*}  &  -0.0421\sym{**} &  -0.0088\sym{**} &   0.0403\sym{**} &  -0.0669         \\
                & (0.0006)         & (0.0497)         & (0.0129)         & (0.0032)         & (0.0129)         & (0.0923)         \\
Constant        &   0.0127\sym{***}&   1.9238\sym{***}&   0.2986\sym{***}&   0.0349\sym{***}&   0.4708\sym{***}&   2.3085\sym{***}\\
                & (0.0011)         & (0.0652)         & (0.0241)         & (0.0057)         & (0.0226)         & (0.1541)         \\
\midrule
N               &   443415         &                  &                  &                  &                  &                  \\
Log. Likelihood &  -252643         &                  &                  &                  &                  &                  \\
BIC             &   507627         &                  &                  &                  &                  &                  \\
\bottomrule
\multicolumn{7}{l}{\scriptsize Standard errors (clustered at the subject level) in parentheses}\\
\multicolumn{7}{l}{\scriptsize \sym{*} \(p<0.05\), \sym{**} \(p<0.01\), \sym{***} \(p<0.001\)}\\
\end{tabular*}
\end{table}

\paragraph{$\varepsilon$-normalization:} While in our main specification the estimate of relative risk aversion is below unity, our multiple price lists allow levels of relative risk aversion above unity (i.e. $\alpha>1$). For such high levels of relative risk aversion, it may be difficult to estimate loss aversion, because the derivatives of the utility function diverge around the reference point (see \citet{Wakker2008} for an illustration). To explore the sensitivity of our findings regarding this issue, we use the following transformation to normalize the utility function:

\begin{equation}
u(x)=
\begin{cases}
\frac{(x+\varepsilon)^{1-\alpha}-\varepsilon^{1-\alpha}}{1-\alpha} & \mbox{ if } x\geq 0\\
\frac{-\lambda((-x+\varepsilon)^{1-\alpha}-\varepsilon^{1-\alpha})}{1-\alpha} & \mbox{ if } x < 0\\
\end{cases}
\end{equation}

This transformation ensures that utility is well behaved for gains and losses for values of $\alpha>1$, while closely approximating CRRA utility for small values of $\varepsilon$ (we set $\varepsilon=0.001$). In particular, this transformation ensures that $u(0)=0$ for all values of $\alpha$, and thus that the derivatives of the utility function around the reference point do not diverge (see also \citet{Vendrik2007} or \citet{Schleich2019} for similar applications). Results for this model appear in Table~\ref{tab:l1te} in the Appendix. They are virtually identical to those obtained from our main specification. 

\begin{table}[htbp]
\caption{Robustness check: $\varepsilon$-normalization \label{tab:l1te}}
\scriptsize
\renewcommand{\footnotesize}{\scriptsize}
{
\def\sym#1{\ifmmode^{#1}\else\(^{#1}\)\fi}
\begin{tabular*}{\hsize}{@{\hskip\tabcolsep\extracolsep\fill}l*{6}{c}}
\toprule
                &\makecell{Risk aversion \\ $\alpha$}         &\makecell{Loss aversion \\ $\lambda$}         &\makecell{Discounting \\ $\delta$}         &\makecell{Present bias \\ $\gamma$}         &\makecell{Tremble error \\ $\kappa$}         &\makecell{Fechner error \\ $\mu$}         \\
\midrule
Age             &    0.002\sym{***}&   -0.006\sym{***}&   -0.001\sym{***}&   -0.000\sym{*}  &    0.001\sym{*}  &   -0.004\sym{***}\\
                &  (0.001)         &  (0.001)         &  (0.000)         &  (0.000)         &  (0.000)         &  (0.001)         \\
Male          &   -0.035\sym{**} &   -0.143\sym{***}&    0.019\sym{*}  &    0.003         &   -0.006         &    0.077\sym{**} \\
                &  (0.012)         &  (0.030)         &  (0.008)         &  (0.002)         &  (0.007)         &  (0.028)         \\
Education       &   -0.018         &    0.001         &   -0.032\sym{***}&    0.001         &   -0.020\sym{**} &    0.017         \\
                &  (0.013)         &  (0.031)         &  (0.009)         &  (0.002)         &  (0.007)         &  (0.031)         \\
Income          &   -0.001\sym{**} &   -0.002\sym{**} &   -0.000         &   -0.000         &   -0.001\sym{***}&    0.002\sym{*}  \\
                &  (0.000)         &  (0.001)         &  (0.000)         &  (0.000)         &  (0.000)         &  (0.001)         \\
Children        &   -0.013         &   -0.066         &    0.031\sym{**} &   -0.002         &    0.024\sym{**} &    0.028         \\
                &  (0.014)         &  (0.035)         &  (0.009)         &  (0.002)         &  (0.008)         &  (0.031)         \\
Couple          &   -0.012         &    0.016         &   -0.001         &    0.001         &    0.001         &    0.021         \\
                &  (0.014)         &  (0.036)         &  (0.009)         &  (0.002)         &  (0.008)         &  (0.031)         \\
Urban           &   -0.020         &   -0.009         &    0.008         &    0.003\sym{*}  &    0.020\sym{**} &    0.033         \\
                &  (0.012)         &  (0.029)         &  (0.008)         &  (0.002)         &  (0.007)         &  (0.028)         \\
CRT             &   -0.022\sym{***}&    0.028\sym{*}  &   -0.026\sym{***}&   -0.001         &   -0.063\sym{***}&    0.004         \\
                &  (0.006)         &  (0.014)         &  (0.004)         &  (0.001)         &  (0.003)         &  (0.013)         \\
Achievement     &   -0.011         &    0.005         &    0.008         &   -0.000         &    0.009         &    0.008         \\
                &  (0.016)         &  (0.034)         &  (0.011)         &  (0.002)         &  (0.008)         &  (0.040)         \\
Benevolence     &    0.001         &    0.035         &   -0.001         &    0.002         &   -0.006         &   -0.015         \\
                &  (0.014)         &  (0.039)         &  (0.010)         &  (0.002)         &  (0.008)         &  (0.033)         \\
Conformity      &    0.002         &    0.019         &   -0.004         &   -0.004\sym{*}  &    0.013         &   -0.010         \\
                &  (0.013)         &  (0.032)         &  (0.009)         &  (0.002)         &  (0.007)         &  (0.032)         \\
Hedonism        &   -0.022         &   -0.033         &    0.032\sym{***}&    0.002         &   -0.021\sym{**} &    0.102\sym{**} \\
                &  (0.015)         &  (0.032)         &  (0.009)         &  (0.002)         &  (0.007)         &  (0.036)         \\
Power           &    0.030         &    0.042         &   -0.005         &    0.000         &    0.053\sym{***}&   -0.071         \\
                &  (0.017)         &  (0.037)         &  (0.011)         &  (0.002)         &  (0.009)         &  (0.041)         \\
Security        &   -0.003         &    0.022         &   -0.004         &   -0.000         &   -0.014         &   -0.006         \\
                &  (0.015)         &  (0.035)         &  (0.010)         &  (0.002)         &  (0.008)         &  (0.036)         \\
Self direction  &   -0.001         &   -0.032         &    0.006         &   -0.002         &   -0.031\sym{***}&   -0.011         \\
                &  (0.014)         &  (0.034)         &  (0.009)         &  (0.002)         &  (0.008)         &  (0.033)         \\
Stimulation     &   -0.054\sym{***}&   -0.126\sym{***}&    0.023\sym{*}  &   -0.001         &    0.039\sym{***}&    0.123\sym{**} \\
                &  (0.016)         &  (0.030)         &  (0.011)         &  (0.002)         &  (0.008)         &  (0.044)         \\
Tradition       &    0.022         &   -0.035         &    0.002         &    0.000         &    0.017\sym{*}  &   -0.046         \\
                &  (0.013)         &  (0.031)         &  (0.009)         &  (0.002)         &  (0.007)         &  (0.032)         \\
Universalism    &   -0.012         &   -0.039         &   -0.011         &   -0.004         &   -0.013         &    0.006         \\
                &  (0.014)         &  (0.038)         &  (0.010)         &  (0.002)         &  (0.008)         &  (0.033)         \\
FR              &   -0.047         &    0.251\sym{***}&    0.018         &    0.006\sym{*}  &   -0.072\sym{***}&    0.094         \\
                &  (0.027)         &  (0.061)         &  (0.014)         &  (0.003)         &  (0.013)         &  (0.056)         \\
IT              &   -0.077\sym{**} &    0.194\sym{***}&    0.083\sym{***}&    0.016\sym{***}&   -0.063\sym{***}&    0.225\sym{***}\\
                &  (0.028)         &  (0.054)         &  (0.016)         &  (0.003)         &  (0.013)         &  (0.067)         \\
PL              &    0.196\sym{***}&    0.044         &    0.064\sym{**} &    0.007         &   -0.134\sym{***}&   -0.035         \\
                &  (0.029)         &  (0.067)         &  (0.023)         &  (0.004)         &  (0.016)         &  (0.052)         \\
RO              &    0.103\sym{*}  &   -0.093         &    0.242\sym{***}&   -0.017\sym{***}&   -0.033         &    0.186         \\
                &  (0.042)         &  (0.065)         &  (0.044)         &  (0.005)         &  (0.018)         &  (0.105)         \\
ES              &   -0.034         &   -0.001         &    0.075\sym{***}&    0.007         &   -0.018         &    0.204\sym{**} \\
                &  (0.028)         &  (0.053)         &  (0.018)         &  (0.004)         &  (0.013)         &  (0.069)         \\
SE              &   -0.110\sym{***}&    0.262\sym{***}&    0.019         &    0.003         &   -0.034\sym{**} &    0.244\sym{***}\\
                &  (0.027)         &  (0.055)         &  (0.015)         &  (0.003)         &  (0.013)         &  (0.074)         \\
UK              &    0.043         &    0.253\sym{***}&    0.031\sym{*}  &    0.003         &   -0.047\sym{***}&   -0.035         \\
                &  (0.027)         &  (0.062)         &  (0.015)         &  (0.003)         &  (0.013)         &  (0.051)         \\
ABreversed      &    0.038\sym{**} &    0.027         &   -0.019\sym{*}  &   -0.007\sym{***}&    0.023\sym{***}&   -0.100\sym{***}\\
                &  (0.012)         &  (0.028)         &  (0.008)         &  (0.002)         &  (0.006)         &  (0.028)         \\
Incentivized    &   -0.044\sym{**} &   -0.189\sym{***}&    0.064\sym{***}&   -0.023\sym{***}&   -0.014         &    0.083\sym{*}  \\
                &  (0.015)         &  (0.036)         &  (0.010)         &  (0.002)         &  (0.007)         &  (0.035)         \\
LowStakes       &   -0.048\sym{*}  &   -0.146\sym{**} &   -0.010         &    0.027\sym{***}&    0.014         &    0.102         \\
                &  (0.023)         &  (0.049)         &  (0.017)         &  (0.004)         &  (0.013)         &  (0.069)         \\
HighStakes      &    0.022         &    0.157\sym{*}  &   -0.035\sym{**} &   -0.007\sym{**} &    0.033\sym{**} &   -0.010         \\
                &  (0.024)         &  (0.073)         &  (0.013)         &  (0.003)         &  (0.012)         &  (0.048)         \\
Constant        &    0.524\sym{***}&    2.270\sym{***}&    0.255\sym{***}&    0.031\sym{***}&    0.490\sym{***}&    0.619\sym{***}\\
                &  (0.041)         &  (0.098)         &  (0.023)         &  (0.005)         &  (0.021)         &  (0.082)         \\
\midrule
N               &   443415         &                  &                  &                  &                  &                  \\
Log. Likelihood &  -253087         &                  &                  &                  &                  &                  \\
BIC             &   508515         &                  &                  &                  &                  &                  \\
\bottomrule
\multicolumn{7}{l}{\footnotesize Standard errors (clustered at the subject level) in parentheses}\\
\multicolumn{7}{l}{\footnotesize \sym{*} \(p<0.05\), \sym{**} \(p<0.01\), \sym{***} \(p<0.001\)}\\
\end{tabular*}
}

\renewcommand{\footnotesize}{\footnotesize}
\end{table}

\paragraph{No present bias:} Further, we estimate a model that abstracts from present bias, and instead includes a single constant discount rate (i.e., assuming exponential discounting). Thus, the intertemporal utility specification corresponds to Equation~\ref{eq:timepreferences}, with the restriction $\gamma=0$. We present the findings of this model in Table~\ref{tab:l1t1d}. The results for all individual characteristics are virtually identical to those found for our main specification. As discount rates may differ across our two time horizons, we also estimate a model that allows the discount rates to vary across the time horizons (Table~\ref{tab:l1t2d}). Because the dates are equally far apart, the two discount rates should be equivalent under exponential discounting. Our data, however, does not support this hypothesis. Instead, we find that discount rates are significantly larger between now and six months than between six and 12 months in the future (Wald test, $p<0.001$). This is, however, merely a restatement of the finding that the present bias parameter is larger than zero in our main specification. Further, we find similar correlations of individual characteristics with both discount rates. Note that the coefficients of the individual characteristics on $\gamma$ in our main specification can be interpreted as the differences in the effects on both discount rates. We will therefore not discuss these differences further here.

\begin{table}
\caption{Robustness check: no present bias \label{tab:l1t1d}}
\renewcommand{\footnotesize}{\scriptsize}
\scriptsize
{
\def\sym#1{\ifmmode^{#1}\else\(^{#1}\)\fi}
\begin{tabular*}{\hsize}{@{\hskip\tabcolsep\extracolsep\fill}l*{5}{c}}
\toprule
                &\makecell{Risk aversion \\ $\alpha$}         &\makecell{Loss aversion \\ $\lambda$}         &\makecell{Discounting \\ $\delta$}         &\makecell{Tremble error \\ $\kappa$}         &\makecell{Fechner error \\ $\mu$}         \\
\midrule
Age             &    0.002\sym{***}&   -0.006\sym{***}&   -0.001\sym{***}&    0.001\sym{*}  &   -0.004\sym{***}\\
                &  (0.001)         &  (0.001)         &  (0.000)         &  (0.000)         &  (0.001)         \\
Male          &   -0.035\sym{**} &   -0.143\sym{***}&    0.022\sym{**} &   -0.006         &    0.077\sym{**} \\
                &  (0.012)         &  (0.030)         &  (0.008)         &  (0.007)         &  (0.028)         \\
Education       &   -0.019         &    0.000         &   -0.031\sym{***}&   -0.019\sym{**} &    0.018         \\
                &  (0.014)         &  (0.031)         &  (0.009)         &  (0.007)         &  (0.032)         \\
Income          &   -0.001\sym{**} &   -0.002\sym{**} &   -0.000         &   -0.001\sym{***}&    0.002\sym{*}  \\
                &  (0.000)         &  (0.001)         &  (0.000)         &  (0.000)         &  (0.001)         \\
Children        &   -0.013         &   -0.065         &    0.028\sym{**} &    0.024\sym{**} &    0.028         \\
                &  (0.014)         &  (0.035)         &  (0.009)         &  (0.008)         &  (0.031)         \\
Couple          &   -0.014         &    0.016         &    0.001         &    0.001         &    0.023         \\
                &  (0.015)         &  (0.035)         &  (0.009)         &  (0.008)         &  (0.032)         \\
Urban           &   -0.021         &   -0.009         &    0.012         &    0.020\sym{**} &    0.035         \\
                &  (0.012)         &  (0.029)         &  (0.008)         &  (0.007)         &  (0.028)         \\
CRT             &   -0.022\sym{***}&    0.028\sym{*}  &   -0.028\sym{***}&   -0.063\sym{***}&    0.003         \\
                &  (0.006)         &  (0.014)         &  (0.004)         &  (0.003)         &  (0.013)         \\
Achievement     &   -0.012         &    0.005         &    0.008         &    0.009         &    0.010         \\
                &  (0.016)         &  (0.034)         &  (0.010)         &  (0.008)         &  (0.041)         \\
Benevolence     &    0.000         &    0.034         &    0.002         &   -0.005         &   -0.014         \\
                &  (0.015)         &  (0.039)         &  (0.010)         &  (0.008)         &  (0.034)         \\
Conformity      &    0.002         &    0.020         &   -0.008         &    0.013         &   -0.008         \\
                &  (0.014)         &  (0.032)         &  (0.009)         &  (0.007)         &  (0.032)         \\
Hedonism        &   -0.022         &   -0.033         &    0.034\sym{***}&   -0.021\sym{**} &    0.104\sym{**} \\
                &  (0.015)         &  (0.032)         &  (0.009)         &  (0.007)         &  (0.036)         \\
Power           &    0.031         &    0.041         &   -0.005         &    0.054\sym{***}&   -0.074         \\
                &  (0.017)         &  (0.037)         &  (0.012)         &  (0.009)         &  (0.042)         \\
Security        &   -0.002         &    0.022         &   -0.005         &   -0.014         &   -0.010         \\
                &  (0.015)         &  (0.035)         &  (0.010)         &  (0.008)         &  (0.036)         \\
Self direction  &   -0.001         &   -0.032         &    0.002         &   -0.031\sym{***}&   -0.013         \\
                &  (0.014)         &  (0.034)         &  (0.009)         &  (0.008)         &  (0.033)         \\
Stimulation     &   -0.056\sym{***}&   -0.125\sym{***}&    0.023\sym{*}  &    0.039\sym{***}&    0.127\sym{**} \\
                &  (0.016)         &  (0.030)         &  (0.011)         &  (0.008)         &  (0.044)         \\
Tradition       &    0.022         &   -0.035         &    0.002         &    0.017\sym{*}  &   -0.046         \\
                &  (0.013)         &  (0.031)         &  (0.009)         &  (0.007)         &  (0.032)         \\
Universalism    &   -0.012         &   -0.038         &   -0.016         &   -0.013         &    0.006         \\
                &  (0.014)         &  (0.038)         &  (0.010)         &  (0.008)         &  (0.033)         \\
FR              &   -0.047         &    0.251\sym{***}&    0.025         &   -0.071\sym{***}&    0.092         \\
                &  (0.027)         &  (0.061)         &  (0.014)         &  (0.013)         &  (0.056)         \\
IT              &   -0.076\sym{**} &    0.194\sym{***}&    0.103\sym{***}&   -0.063\sym{***}&    0.229\sym{***}\\
                &  (0.028)         &  (0.054)         &  (0.017)         &  (0.013)         &  (0.067)         \\
PL              &    0.201\sym{***}&    0.047         &    0.073\sym{**} &   -0.138\sym{***}&   -0.034         \\
                &  (0.029)         &  (0.067)         &  (0.023)         &  (0.016)         &  (0.053)         \\
RO              &    0.105\sym{*}  &   -0.094         &    0.215\sym{***}&   -0.031         &    0.170         \\
                &  (0.042)         &  (0.065)         &  (0.040)         &  (0.018)         &  (0.103)         \\
ES              &   -0.034         &   -0.001         &    0.084\sym{***}&   -0.019         &    0.207\sym{**} \\
                &  (0.028)         &  (0.053)         &  (0.018)         &  (0.013)         &  (0.070)         \\
SE              &   -0.111\sym{***}&    0.262\sym{***}&    0.024         &   -0.033\sym{*}  &    0.247\sym{***}\\
                &  (0.027)         &  (0.055)         &  (0.014)         &  (0.013)         &  (0.074)         \\
UK              &    0.042         &    0.253\sym{***}&    0.035\sym{*}  &   -0.047\sym{***}&   -0.034         \\
                &  (0.027)         &  (0.062)         &  (0.014)         &  (0.013)         &  (0.051)         \\
ABreversed      &    0.036\sym{**} &    0.027         &   -0.026\sym{**} &    0.024\sym{***}&   -0.092\sym{**} \\
                &  (0.012)         &  (0.028)         &  (0.008)         &  (0.006)         &  (0.028)         \\
Incentivized    &   -0.046\sym{**} &   -0.189\sym{***}&    0.037\sym{***}&   -0.015\sym{*}  &    0.084\sym{*}  \\
                &  (0.015)         &  (0.036)         &  (0.009)         &  (0.007)         &  (0.035)         \\
LowStakes       &   -0.046\sym{*}  &   -0.147\sym{**} &    0.024         &    0.015         &    0.100         \\
                &  (0.023)         &  (0.049)         &  (0.018)         &  (0.013)         &  (0.069)         \\
HighStakes      &    0.023         &    0.157\sym{*}  &   -0.044\sym{***}&    0.032\sym{**} &   -0.013         \\
                &  (0.025)         &  (0.073)         &  (0.013)         &  (0.012)         &  (0.049)         \\
Constant        &    0.525\sym{***}&    2.269\sym{***}&    0.293\sym{***}&    0.491\sym{***}&    0.619\sym{***}\\
                &  (0.041)         &  (0.098)         &  (0.023)         &  (0.021)         &  (0.083)         \\
\midrule
N               &   443415         &                  &                  &                  &                  \\
Log. Likelihood &  -253239         &                  &                  &                  &                  \\
BIC             &   508428         &                  &                  &                  &                  \\
\bottomrule
\multicolumn{6}{l}{\footnotesize Standard errors (clustered at the subject level) in parentheses}\\
\multicolumn{6}{l}{\footnotesize \sym{*} \(p<0.05\), \sym{**} \(p<0.01\), \sym{***} \(p<0.001\)}\\
\end{tabular*}
}

\renewcommand{\footnotesize}{\footnotesize}
\end{table}

\paragraph{Separate estimation of time preferences:} Our main specification assumes a discounted utility model, in which the elasticity of intertemporal substitution is determined by the parameter of relative risk aversion. To examine the sensitivity of our findings to this assumption, we estimate time preferences separately. To do so, we impose $\alpha=0$ in Equation~\ref{eq:utility} and only use choices from MPL1.1 and MPL1.2 to estimate $\delta$ and $\gamma$. We report the findings of this model in Table~\ref{tab:l1tbetadelta}. Under the assumption that $\alpha=0$, some correlations are no longer significant. Compared to the main model, discounting and present bias are no longer related to \emph{Age}, and \emph{Male}. Likewise, \emph{Stimulation}, \emph{ABreversed}, and \emph{LowStakes} are no longer related to discounting, and \emph{Urban} is no longer significantly related to present bias. On the other hand, some correlations turn out to be statistically significant, which were not statistically significant in the main specification. \emph{Income} and \emph{Universalism} are now negatively related to discounting, and \emph{PL} is negatively related to present bias. Note however, that while several studies suggest that utility for risk and time preferences is not the same, the assumption that utility over time has no curvature ($\alpha=0$) and is unrelated to utility over risk, appears strong.\footnote{Our setup does not allow to identify curvature of utility over time separately from time discounting. Therefore we cannot directly estimate this curvature.} Indeed, most studies find that utility over time is also concave, but exhibits less curvature than utility over risk. Also, recent findings suggest that utility over risk and time are different but correlated \citep{meissner2022measuring}. Therefore, the results of the robustness check presented above should be interpreted with caution.  

\begin{table}
\caption{Robustness check: two discount rates \label{tab:l1t2d}}
\renewcommand{\footnotesize}{\scriptsize}
\scriptsize
{
\def\sym#1{\ifmmode^{#1}\else\(^{#1}\)\fi}
\begin{tabular*}{\hsize}{@{\hskip\tabcolsep\extracolsep\fill}l*{6}{c}}
\toprule
                &\makecell{Risk aversion \\ $\alpha$}         &\makecell{Loss aversion \\ $\lambda$}         &\makecell{Discounting \\ $\delta_1$}         &\makecell{Discounting \\ $\delta_2$}         &\makecell{Tremble error \\ $\kappa$}         &\makecell{Fechner error \\ $\mu$}         \\
\midrule
Age             &    0.002\sym{***}&   -0.006\sym{***}&   -0.002\sym{***}&   -0.001\sym{**} &    0.001\sym{*}  &   -0.004\sym{***}\\
                &  (0.001)         &  (0.001)         &  (0.000)         &  (0.000)         &  (0.000)         &  (0.001)         \\
Male          &   -0.035\sym{**} &   -0.143\sym{***}&    0.026\sym{**} &    0.019\sym{*}  &   -0.006         &    0.077\sym{**} \\
                &  (0.012)         &  (0.030)         &  (0.008)         &  (0.008)         &  (0.007)         &  (0.028)         \\
Education       &   -0.018         &    0.001         &   -0.031\sym{***}&   -0.033\sym{***}&   -0.019\sym{**} &    0.017         \\
                &  (0.013)         &  (0.031)         &  (0.009)         &  (0.009)         &  (0.007)         &  (0.031)         \\
Income          &   -0.001\sym{**} &   -0.002\sym{**} &   -0.000         &   -0.000         &   -0.001\sym{***}&    0.002\sym{*}  \\
                &  (0.000)         &  (0.001)         &  (0.000)         &  (0.000)         &  (0.000)         &  (0.001)         \\
Children        &   -0.013         &   -0.066         &    0.025\sym{*}  &    0.030\sym{**} &    0.024\sym{**} &    0.027         \\
                &  (0.014)         &  (0.035)         &  (0.010)         &  (0.009)         &  (0.008)         &  (0.031)         \\
Couple          &   -0.014         &    0.016         &    0.002         &    0.000         &    0.001         &    0.023         \\
                &  (0.014)         &  (0.035)         &  (0.010)         &  (0.009)         &  (0.008)         &  (0.031)         \\
Urban           &   -0.021         &   -0.009         &    0.016         &    0.008         &    0.020\sym{**} &    0.034         \\
                &  (0.012)         &  (0.029)         &  (0.009)         &  (0.008)         &  (0.007)         &  (0.028)         \\
CRT             &   -0.022\sym{***}&    0.028\sym{*}  &   -0.028\sym{***}&   -0.027\sym{***}&   -0.063\sym{***}&    0.005         \\
                &  (0.006)         &  (0.014)         &  (0.004)         &  (0.004)         &  (0.003)         &  (0.013)         \\
Achievement     &   -0.013         &    0.005         &    0.007         &    0.009         &    0.010         &    0.012         \\
                &  (0.016)         &  (0.034)         &  (0.011)         &  (0.010)         &  (0.008)         &  (0.040)         \\
Benevolence     &    0.001         &    0.035         &    0.005         &   -0.002         &   -0.006         &   -0.016         \\
                &  (0.015)         &  (0.039)         &  (0.010)         &  (0.010)         &  (0.008)         &  (0.033)         \\
Conformity      &    0.002         &    0.019         &   -0.013         &   -0.004         &    0.013         &   -0.009         \\
                &  (0.014)         &  (0.032)         &  (0.009)         &  (0.009)         &  (0.007)         &  (0.032)         \\
Hedonism        &   -0.022         &   -0.034         &    0.037\sym{***}&    0.031\sym{***}&   -0.021\sym{**} &    0.102\sym{**} \\
                &  (0.015)         &  (0.032)         &  (0.010)         &  (0.009)         &  (0.007)         &  (0.036)         \\
Power           &    0.031         &    0.042         &   -0.005         &   -0.005         &    0.053\sym{***}&   -0.075         \\
                &  (0.017)         &  (0.037)         &  (0.012)         &  (0.011)         &  (0.009)         &  (0.041)         \\
Security        &   -0.003         &    0.022         &   -0.005         &   -0.005         &   -0.014         &   -0.007         \\
                &  (0.015)         &  (0.035)         &  (0.010)         &  (0.010)         &  (0.008)         &  (0.036)         \\
Self direction  &   -0.001         &   -0.032         &   -0.000         &    0.005         &   -0.031\sym{***}&   -0.011         \\
                &  (0.014)         &  (0.034)         &  (0.010)         &  (0.009)         &  (0.008)         &  (0.033)         \\
Stimulation     &   -0.055\sym{***}&   -0.126\sym{***}&    0.020         &    0.024\sym{*}  &    0.039\sym{***}&    0.126\sym{**} \\
                &  (0.016)         &  (0.030)         &  (0.011)         &  (0.011)         &  (0.008)         &  (0.043)         \\
Tradition       &    0.022         &   -0.035         &    0.003         &    0.001         &    0.017\sym{*}  &   -0.047         \\
                &  (0.013)         &  (0.031)         &  (0.009)         &  (0.008)         &  (0.007)         &  (0.032)         \\
Universalism    &   -0.013         &   -0.039         &   -0.021\sym{*}  &   -0.011         &   -0.013         &    0.008         \\
                &  (0.014)         &  (0.038)         &  (0.010)         &  (0.009)         &  (0.008)         &  (0.033)         \\
FR              &   -0.047         &    0.251\sym{***}&    0.034\sym{*}  &    0.016         &   -0.071\sym{***}&    0.092         \\
                &  (0.027)         &  (0.061)         &  (0.015)         &  (0.014)         &  (0.013)         &  (0.056)         \\
IT              &   -0.076\sym{**} &    0.194\sym{***}&    0.126\sym{***}&    0.081\sym{***}&   -0.063\sym{***}&    0.222\sym{***}\\
                &  (0.028)         &  (0.054)         &  (0.019)         &  (0.016)         &  (0.013)         &  (0.067)         \\
PL              &    0.199\sym{***}&    0.045         &    0.081\sym{***}&    0.067\sym{**} &   -0.135\sym{***}&   -0.039         \\
                &  (0.029)         &  (0.067)         &  (0.024)         &  (0.022)         &  (0.016)         &  (0.052)         \\
RO              &    0.105\sym{*}  &   -0.093         &    0.192\sym{***}&    0.241\sym{***}&   -0.033         &    0.180         \\
                &  (0.042)         &  (0.065)         &  (0.038)         &  (0.043)         &  (0.018)         &  (0.104)         \\
ES              &   -0.034         &   -0.001         &    0.093\sym{***}&    0.075\sym{***}&   -0.018         &    0.203\sym{**} \\
                &  (0.028)         &  (0.053)         &  (0.019)         &  (0.018)         &  (0.013)         &  (0.069)         \\
SE              &   -0.110\sym{***}&    0.262\sym{***}&    0.027         &    0.020         &   -0.033\sym{**} &    0.244\sym{***}\\
                &  (0.027)         &  (0.055)         &  (0.015)         &  (0.014)         &  (0.013)         &  (0.073)         \\
UK              &    0.042         &    0.252\sym{***}&    0.039\sym{*}  &    0.031\sym{*}  &   -0.047\sym{***}&   -0.035         \\
                &  (0.027)         &  (0.062)         &  (0.015)         &  (0.014)         &  (0.013)         &  (0.051)         \\
ABreversed      &    0.038\sym{**} &    0.027         &   -0.037\sym{***}&   -0.020\sym{*}  &    0.023\sym{***}&   -0.098\sym{***}\\
                &  (0.012)         &  (0.028)         &  (0.008)         &  (0.008)         &  (0.006)         &  (0.028)         \\
Incentivized    &   -0.046\sym{**} &   -0.189\sym{***}&    0.011         &    0.064\sym{***}&   -0.014         &    0.087\sym{*}  \\
                &  (0.015)         &  (0.036)         &  (0.010)         &  (0.009)         &  (0.007)         &  (0.035)         \\
LowStakes       &   -0.047\sym{*}  &   -0.147\sym{**} &    0.059\sym{**} &   -0.010         &    0.014         &    0.099         \\
                &  (0.023)         &  (0.050)         &  (0.019)         &  (0.017)         &  (0.013)         &  (0.068)         \\
HighStakes      &    0.023         &    0.158\sym{*}  &   -0.053\sym{***}&   -0.036\sym{**} &    0.032\sym{**} &   -0.010         \\
                &  (0.025)         &  (0.073)         &  (0.014)         &  (0.013)         &  (0.012)         &  (0.048)         \\
Constant        &    0.527\sym{***}&    2.270\sym{***}&    0.329\sym{***}&    0.255\sym{***}&    0.489\sym{***}&    0.615\sym{***}\\
                &  (0.041)         &  (0.098)         &  (0.025)         &  (0.023)         &  (0.021)         &  (0.082)         \\
\midrule
N               &   443415         &                  &                  &                  &                  &                  \\
Log. Likelihood &  -253092         &                  &                  &                  &                  &                  \\
BIC             &   508525         &                  &                  &                  &                  &                  \\
\bottomrule
\multicolumn{7}{l}{\footnotesize Standard errors (clustered at the subject level) in parentheses}\\
\multicolumn{7}{l}{\footnotesize \sym{*} \(p<0.05\), \sym{**} \(p<0.01\), \sym{***} \(p<0.001\)}\\
\end{tabular*}
}

\renewcommand{\footnotesize}{\footnotesize}
\end{table}

\begin{table}[htbp]
\centering
\scriptsize
\def\sym#1{\ifmmode^{#1}\else\(^{#1}\)\fi}
\caption{Robustness check: seperate estimation of time preferences \label{tab:l1tbetadelta}}
\begin{tabular*}{\hsize}{@{\hskip\tabcolsep\extracolsep\fill}l*{5}{c}}
\toprule
                &\makecell{Discounting \\ $\delta$}         &\makecell{Present bias \\ $\gamma$}         &\makecell{Tremble error \\ $\kappa$}         &\makecell{Fechner error \\ $\mu$}         \\
\midrule
Age             &   -0.001         &   -0.000         &    0.000         &   -0.005         \\
                &  (0.001)         &  (0.000)         &  (0.001)         &  (0.012)         \\
Male          &    0.014         &    0.002         &   -0.003         &    0.050         \\
                &  (0.016)         &  (0.003)         &  (0.014)         &  (0.278)         \\
Education       &   -0.077\sym{***}&   -0.001         &   -0.016         &    0.021         \\
                &  (0.018)         &  (0.003)         &  (0.015)         &  (0.335)         \\
Income          &   -0.001\sym{***}&   -0.000         &   -0.001\sym{**} &   -0.004         \\
                &  (0.000)         &  (0.000)         &  (0.000)         &  (0.006)         \\
Children        &    0.055\sym{**} &   -0.007         &    0.055\sym{**} &   -0.605         \\
                &  (0.019)         &  (0.004)         &  (0.017)         &  (0.346)         \\
Couple          &   -0.022         &    0.004         &   -0.026         &    0.300         \\
                &  (0.019)         &  (0.004)         &  (0.017)         &  (0.345)         \\
Urban           &    0.004         &    0.005         &    0.003         &    0.299         \\
                &  (0.016)         &  (0.003)         &  (0.014)         &  (0.275)         \\
CRT             &   -0.071\sym{***}&    0.000         &   -0.052\sym{***}&   -0.400\sym{**} \\
                &  (0.007)         &  (0.001)         &  (0.007)         &  (0.143)         \\
Achievement     &   -0.013         &    0.002         &    0.015         &   -0.434         \\
                &  (0.019)         &  (0.004)         &  (0.017)         &  (0.333)         \\
Benevolence     &   -0.007         &    0.005         &   -0.014         &   -0.111         \\
                &  (0.018)         &  (0.004)         &  (0.017)         &  (0.319)         \\
Conformity      &    0.001         &   -0.007\sym{*}  &   -0.009         &    0.480         \\
                &  (0.017)         &  (0.004)         &  (0.015)         &  (0.298)         \\
Hedonism        &    0.050\sym{**} &   -0.002         &   -0.011         &    0.243         \\
                &  (0.017)         &  (0.003)         &  (0.016)         &  (0.305)         \\
Power           &    0.030         &    0.001         &    0.026         &    0.945\sym{*}  \\
                &  (0.023)         &  (0.004)         &  (0.021)         &  (0.447)         \\
Security        &   -0.011         &   -0.001         &    0.006         &   -0.104         \\
                &  (0.017)         &  (0.004)         &  (0.016)         &  (0.298)         \\
Self direction  &    0.023         &   -0.003         &   -0.048\sym{**} &    0.342         \\
                &  (0.017)         &  (0.004)         &  (0.015)         &  (0.288)         \\
Stimulation     &   -0.016         &   -0.002         &    0.058\sym{***}&   -0.546         \\
                &  (0.019)         &  (0.004)         &  (0.016)         &  (0.307)         \\
Tradition       &    0.032         &   -0.000         &    0.025         &   -0.046         \\
                &  (0.017)         &  (0.003)         &  (0.014)         &  (0.281)         \\
Universalism    &   -0.043\sym{*}  &   -0.007         &    0.004         &   -0.755\sym{*}  \\
                &  (0.018)         &  (0.004)         &  (0.016)         &  (0.314)         \\
FR              &    0.005         &    0.011\sym{*}  &   -0.196\sym{***}&    1.043\sym{*}  \\
                &  (0.028)         &  (0.005)         &  (0.022)         &  (0.448)         \\
IT              &    0.158\sym{**} &    0.021\sym{***}&   -0.295\sym{**} &    3.456         \\
                &  (0.055)         &  (0.006)         &  (0.100)         &  (1.982)         \\
PL              &    0.941\sym{***}&   -0.063\sym{***}&   -1.013\sym{**} &   21.751\sym{**} \\
                &  (0.110)         &  (0.015)         &  (0.348)         &  (6.921)         \\
RO              &    1.075\sym{***}&   -0.099\sym{***}&   -0.543\sym{***}&   12.902\sym{***}\\
                &  (0.112)         &  (0.014)         &  (0.127)         &  (2.702)         \\
ES              &    0.145\sym{***}&    0.011         &   -0.063\sym{*}  &    1.733\sym{**} \\
                &  (0.038)         &  (0.006)         &  (0.030)         &  (0.661)         \\
SE              &   -0.034         &    0.004         &   -0.132\sym{***}&    0.463         \\
                &  (0.026)         &  (0.005)         &  (0.024)         &  (0.463)         \\
UK              &    0.110\sym{***}&    0.006         &   -0.142\sym{***}&    1.432\sym{**} \\
                &  (0.029)         &  (0.006)         &  (0.025)         &  (0.517)         \\
ABreversed      &    0.002         &   -0.013\sym{***}&    0.021         &    0.160         \\
                &  (0.016)         &  (0.003)         &  (0.014)         &  (0.296)         \\
Incentivized    &    0.119\sym{***}&   -0.044\sym{***}&   -0.031\sym{*}  &    0.433         \\
                &  (0.019)         &  (0.004)         &  (0.015)         &  (0.309)         \\
LowStakes       &   -0.055         &    0.044\sym{***}&    0.008         &    0.327         \\
                &  (0.034)         &  (0.007)         &  (0.029)         &  (0.588)         \\
HighStakes      &   -0.073\sym{***}&   -0.014\sym{*}  &    0.013         &   -0.119         \\
                &  (0.021)         &  (0.005)         &  (0.024)         &  (0.441)         \\
Constant        &    0.567\sym{***}&    0.052\sym{***}&    0.464\sym{***}&    5.620\sym{***}\\
                &  (0.048)         &  (0.009)         &  (0.044)         &  (0.896)         \\
\midrule
N               &   177366         &                  &                  &                  \\
Log. Likelihood &   -99503         &                  &                  &                  \\
BIC             &   200455         &                  &                  &                  \\
\bottomrule
\multicolumn{5}{l}{\scriptsize Standard errors (clustered at the subject level) in parentheses}\\
\multicolumn{5}{l}{\scriptsize \sym{*} \(p<0.05\), \sym{**} \(p<0.01\), \sym{***} \(p<0.001\)}\\
\end{tabular*}
\end{table}

\subsection{Alternative error term specifications}
\label{sec:robustness_error}
As noted by \citet{Drichoutis2014}, the specification of the error process may impact the preference estimates. We therefore change the error process in three ways compared to our main specification which employs a logit-link Fechner error and tremble. First, we use a probit-link Fechner error and tremble. Second, we run a model without a tremble error. Finally, we allow the Fechner error to differ between our MPLs. 

\paragraph{Probit-link Fechner error and tremble:} First, we estimate the model assuming the difference in errors to be normally distributed: $F(\xi)=\Phi(\xi)$, where $\Phi$ represents the standard normal CDF. The results (see Table~\ref{tab:p1t}) are virtually identical to those obtained with our main specification. \emph{Male}, however, is no longer related to present bias.

\begin{table}[htbp]
\centering
\scriptsize
\def\sym#1{\ifmmode^{#1}\else\(^{#1}\)\fi}
\caption{Robustness check: probit link \label{tab:p1t}}
\begin{tabular*}{\hsize}{@{\hskip\tabcolsep\extracolsep\fill}l*{6}{c}}
\toprule
                &\makecell{Risk aversion \\ $\alpha$}         &\makecell{Loss aversion \\ $\lambda$}         &\makecell{Discounting \\ $\delta$}         &\makecell{Present bias \\ $\gamma$}         &\makecell{Tremble error \\ $\kappa$}         &\makecell{Fechner error \\ $\mu$}         \\
\midrule
Age             &    0.002\sym{***}&   -0.006\sym{***}&   -0.001\sym{**} &   -0.000\sym{*}  &    0.001\sym{*}  &   -0.006\sym{***}\\
                &  (0.001)         &  (0.001)         &  (0.000)         &  (0.000)         &  (0.000)         &  (0.002)         \\
Male          &   -0.034\sym{**} &   -0.146\sym{***}&    0.018\sym{*}  &    0.003         &   -0.006         &    0.127\sym{**} \\
                &  (0.012)         &  (0.030)         &  (0.008)         &  (0.002)         &  (0.006)         &  (0.045)         \\
Education       &   -0.017         &   -0.000         &   -0.033\sym{***}&    0.001         &   -0.020\sym{**} &    0.027         \\
                &  (0.013)         &  (0.032)         &  (0.009)         &  (0.002)         &  (0.007)         &  (0.049)         \\
Income          &   -0.001\sym{**} &   -0.002\sym{**} &   -0.000         &   -0.000         &   -0.001\sym{***}&    0.003\sym{*}  \\
                &  (0.000)         &  (0.001)         &  (0.000)         &  (0.000)         &  (0.000)         &  (0.001)         \\
Children        &   -0.013         &   -0.066         &    0.031\sym{***}&   -0.003         &    0.024\sym{**} &    0.050         \\
                &  (0.014)         &  (0.037)         &  (0.009)         &  (0.002)         &  (0.008)         &  (0.049)         \\
Couple          &   -0.013         &    0.014         &   -0.000         &    0.001         &    0.001         &    0.038         \\
                &  (0.014)         &  (0.036)         &  (0.009)         &  (0.002)         &  (0.007)         &  (0.050)         \\
Urban           &   -0.020         &   -0.007         &    0.008         &    0.003\sym{*}  &    0.020\sym{**} &    0.051         \\
                &  (0.012)         &  (0.030)         &  (0.008)         &  (0.002)         &  (0.006)         &  (0.044)         \\
CRT             &   -0.021\sym{***}&    0.028\sym{*}  &   -0.028\sym{***}&   -0.000         &   -0.063\sym{***}&    0.008         \\
                &  (0.006)         &  (0.014)         &  (0.004)         &  (0.001)         &  (0.003)         &  (0.021)         \\
Achievement     &   -0.012         &    0.007         &    0.008         &   -0.000         &    0.009         &    0.017         \\
                &  (0.016)         &  (0.034)         &  (0.010)         &  (0.002)         &  (0.008)         &  (0.063)         \\
Benevolence     &    0.000         &    0.033         &   -0.001         &    0.002         &   -0.006         &   -0.018         \\
                &  (0.014)         &  (0.040)         &  (0.010)         &  (0.002)         &  (0.008)         &  (0.053)         \\
Conformity      &    0.002         &    0.020         &   -0.004         &   -0.004\sym{*}  &    0.013         &   -0.016         \\
                &  (0.013)         &  (0.032)         &  (0.009)         &  (0.002)         &  (0.007)         &  (0.050)         \\
Hedonism        &   -0.022         &   -0.036         &    0.032\sym{***}&    0.001         &   -0.019\sym{**} &    0.161\sym{**} \\
                &  (0.015)         &  (0.033)         &  (0.009)         &  (0.002)         &  (0.007)         &  (0.058)         \\
Power           &    0.029         &    0.039         &   -0.004         &    0.000         &    0.053\sym{***}&   -0.111         \\
                &  (0.017)         &  (0.038)         &  (0.011)         &  (0.002)         &  (0.008)         &  (0.065)         \\
Security        &   -0.002         &    0.023         &   -0.005         &   -0.000         &   -0.015\sym{*}  &   -0.011         \\
                &  (0.014)         &  (0.035)         &  (0.010)         &  (0.002)         &  (0.008)         &  (0.057)         \\
Self direction  &   -0.000         &   -0.033         &    0.005         &   -0.002         &   -0.032\sym{***}&   -0.016         \\
                &  (0.014)         &  (0.034)         &  (0.009)         &  (0.002)         &  (0.007)         &  (0.052)         \\
Stimulation     &   -0.053\sym{***}&   -0.125\sym{***}&    0.023\sym{*}  &   -0.001         &    0.039\sym{***}&    0.194\sym{**} \\
                &  (0.016)         &  (0.031)         &  (0.011)         &  (0.002)         &  (0.007)         &  (0.067)         \\
Tradition       &    0.021         &   -0.034         &    0.002         &    0.000         &    0.016\sym{*}  &   -0.072         \\
                &  (0.013)         &  (0.031)         &  (0.009)         &  (0.002)         &  (0.007)         &  (0.050)         \\
Universalism    &   -0.013         &   -0.035         &   -0.011         &   -0.003         &   -0.013         &    0.015         \\
                &  (0.014)         &  (0.038)         &  (0.009)         &  (0.002)         &  (0.008)         &  (0.052)         \\
FR              &   -0.047         &    0.259\sym{***}&    0.017         &    0.006\sym{*}  &   -0.072\sym{***}&    0.160         \\
                &  (0.026)         &  (0.060)         &  (0.014)         &  (0.003)         &  (0.012)         &  (0.092)         \\
IT              &   -0.073\sym{**} &    0.194\sym{***}&    0.082\sym{***}&    0.014\sym{***}&   -0.062\sym{***}&    0.365\sym{***}\\
                &  (0.027)         &  (0.056)         &  (0.016)         &  (0.003)         &  (0.012)         &  (0.108)         \\
PL              &    0.198\sym{***}&    0.040         &    0.067\sym{**} &    0.004         &   -0.123\sym{***}&   -0.106         \\
                &  (0.029)         &  (0.067)         &  (0.022)         &  (0.004)         &  (0.016)         &  (0.085)         \\
RO              &    0.103\sym{*}  &   -0.092         &    0.239\sym{***}&   -0.018\sym{***}&   -0.022         &    0.242         \\
                &  (0.041)         &  (0.067)         &  (0.042)         &  (0.005)         &  (0.017)         &  (0.163)         \\
ES              &   -0.034         &   -0.002         &    0.075\sym{***}&    0.006         &   -0.016         &    0.317\sym{**} \\
                &  (0.028)         &  (0.053)         &  (0.018)         &  (0.003)         &  (0.013)         &  (0.111)         \\
SE              &   -0.111\sym{***}&    0.272\sym{***}&    0.019         &    0.003         &   -0.033\sym{**} &    0.410\sym{***}\\
                &  (0.026)         &  (0.058)         &  (0.014)         &  (0.003)         &  (0.013)         &  (0.118)         \\
UK              &    0.041         &    0.255\sym{***}&    0.032\sym{*}  &    0.002         &   -0.046\sym{***}&   -0.055         \\
                &  (0.026)         &  (0.064)         &  (0.015)         &  (0.003)         &  (0.012)         &  (0.083)         \\
ABreversed      &    0.037\sym{**} &    0.030         &   -0.018\sym{*}  &   -0.007\sym{***}&    0.023\sym{***}&   -0.160\sym{***}\\
                &  (0.012)         &  (0.029)         &  (0.008)         &  (0.002)         &  (0.006)         &  (0.044)         \\
Incentivized    &   -0.043\sym{**} &   -0.195\sym{***}&    0.063\sym{***}&   -0.022\sym{***}&   -0.014\sym{*}  &    0.135\sym{*}  \\
                &  (0.015)         &  (0.037)         &  (0.009)         &  (0.002)         &  (0.007)         &  (0.055)         \\
LowStakes       &   -0.047\sym{*}  &   -0.147\sym{**} &   -0.011         &    0.027\sym{***}&    0.014         &    0.163         \\
                &  (0.023)         &  (0.049)         &  (0.017)         &  (0.004)         &  (0.013)         &  (0.108)         \\
HighStakes      &    0.021         &    0.163\sym{*}  &   -0.035\sym{**} &   -0.007\sym{*}  &    0.034\sym{**} &   -0.023         \\
                &  (0.024)         &  (0.069)         &  (0.013)         &  (0.003)         &  (0.011)         &  (0.077)         \\
Constant        &    0.524\sym{***}&    2.277\sym{***}&    0.259\sym{***}&    0.029\sym{***}&    0.496\sym{***}&    1.026\sym{***}\\
                &  (0.040)         &  (0.097)         &  (0.023)         &  (0.005)         &  (0.020)         &  (0.133)         \\
\midrule
N               &   443415         &                  &                  &                  &                  &                  \\
Log. Likelihood &  -252908         &                  &                  &                  &                  &                  \\
BIC             &   508157         &                  &                  &                  &                  &                  \\
\bottomrule
\multicolumn{7}{l}{\scriptsize Standard errors (clustered at the subject level) in parentheses}\\
\multicolumn{7}{l}{\scriptsize \sym{*} \(p<0.05\), \sym{**} \(p<0.01\), \sym{***} \(p<0.001\)}\\
\end{tabular*}
\end{table}

\paragraph{No tremble error:} Second, we estimate the model without a tremble error, using only a logit-link Fechner error. Because this specification of the error terms is often applied in the literature, the results from this specification may be more easily compared with previous literature than the results from our main specification. As can be seen in Table~\ref{tab:l1}, most of the individual characteristics that were significantly related with the preference parameters in our main specification are also statistically significant for the specification with the simpler error terms. Yet, \emph{Stimulation}, \emph{RO}, and \emph{LowStakes} are no longer significantly related to risk aversion. \emph{Age}, \emph{Male}, \emph{Urban} and \emph{HighStakes} are no longer significantly related to present bias, and \emph{Male} and \emph{Stimulation} are no longer significantly associated with discounting. Some variables are statistically significant for this specification but not for our main specification. \emph{Hedonism}, \emph{FR} and \emph{ES} are now negatively related to risk aversion. \emph{Power} and \emph{Tradition} are now positively related to risk aversion. \emph{Children} and \emph{RO} are now negatively related to loss aversion.  

\begin{table}[htbp]
\centering
\scriptsize
\def\sym#1{\ifmmode^{#1}\else\(^{#1}\)\fi}
\caption{Robustness check: no tremble error \label{tab:l1}}
\begin{tabular*}{\hsize}{@{\hskip\tabcolsep\extracolsep\fill}l*{5}{c}}
\toprule
                &\makecell{Risk aversion \\ $\alpha$}         &\makecell{Loss aversion \\ $\lambda$}         &\makecell{Discounting \\ $\delta$}         &\makecell{Present bias \\ $\gamma$}         &\makecell{Fechner error \\ $\mu$}         \\
\midrule
Age             &    0.001\sym{**} &   -0.006\sym{***}&   -0.001\sym{*}  &   -0.000         &   -0.003         \\
                &  (0.000)         &  (0.001)         &  (0.001)         &  (0.000)         &  (0.003)         \\
Male          &   -0.036\sym{***}&   -0.134\sym{***}&    0.019         &    0.005         &    0.159\sym{*}  \\
                &  (0.010)         &  (0.024)         &  (0.011)         &  (0.002)         &  (0.063)         \\
Education       &   -0.023         &   -0.007         &   -0.047\sym{***}&    0.003         &    0.016         \\
                &  (0.012)         &  (0.025)         &  (0.012)         &  (0.002)         &  (0.074)         \\
Income          &   -0.001\sym{***}&   -0.002\sym{**} &   -0.000         &   -0.000         &    0.003         \\
                &  (0.000)         &  (0.001)         &  (0.000)         &  (0.000)         &  (0.002)         \\
Children        &    0.001         &   -0.069\sym{*}  &    0.039\sym{**} &   -0.004         &    0.091         \\
                &  (0.011)         &  (0.028)         &  (0.013)         &  (0.003)         &  (0.071)         \\
Couple          &    0.002         &    0.035         &   -0.011         &    0.001         &   -0.031         \\
                &  (0.011)         &  (0.028)         &  (0.013)         &  (0.003)         &  (0.071)         \\
Urban           &   -0.008         &   -0.010         &    0.009         &    0.004         &    0.126         \\
                &  (0.010)         &  (0.023)         &  (0.012)         &  (0.002)         &  (0.066)         \\
CRT             &   -0.033\sym{***}&    0.044\sym{***}&   -0.035\sym{***}&    0.002         &   -0.122\sym{***}\\
                &  (0.005)         &  (0.011)         &  (0.005)         &  (0.001)         &  (0.029)         \\
Achievement     &    0.009         &    0.010         &    0.001         &   -0.000         &   -0.051         \\
                &  (0.016)         &  (0.028)         &  (0.016)         &  (0.003)         &  (0.104)         \\
Benevolence     &   -0.004         &    0.044         &    0.004         &    0.004         &    0.011         \\
                &  (0.012)         &  (0.030)         &  (0.014)         &  (0.003)         &  (0.077)         \\
Conformity      &    0.010         &    0.028         &   -0.011         &   -0.006\sym{*}  &    0.001         \\
                &  (0.013)         &  (0.026)         &  (0.013)         &  (0.003)         &  (0.081)         \\
Hedonism        &   -0.040\sym{**} &   -0.027         &    0.047\sym{***}&   -0.001         &    0.178         \\
                &  (0.014)         &  (0.027)         &  (0.013)         &  (0.003)         &  (0.095)         \\
Power           &    0.044\sym{*}  &   -0.012         &   -0.012         &    0.000         &   -0.025         \\
                &  (0.018)         &  (0.030)         &  (0.017)         &  (0.003)         &  (0.121)         \\
Security        &   -0.006         &    0.037         &   -0.002         &    0.000         &   -0.050         \\
                &  (0.014)         &  (0.028)         &  (0.014)         &  (0.003)         &  (0.088)         \\
Self direction  &   -0.006         &   -0.009         &    0.008         &   -0.003         &   -0.115         \\
                &  (0.013)         &  (0.027)         &  (0.013)         &  (0.003)         &  (0.085)         \\
Stimulation     &   -0.030         &   -0.108\sym{***}&    0.022         &   -0.001         &    0.350\sym{*}  \\
                &  (0.020)         &  (0.027)         &  (0.018)         &  (0.003)         &  (0.144)         \\
Tradition       &    0.026\sym{*}  &   -0.036         &    0.004         &   -0.002         &   -0.083         \\
                &  (0.013)         &  (0.025)         &  (0.013)         &  (0.002)         &  (0.087)         \\
Universalism    &   -0.009         &   -0.041         &   -0.019         &   -0.004         &   -0.027         \\
                &  (0.012)         &  (0.028)         &  (0.014)         &  (0.003)         &  (0.077)         \\
FR              &   -0.078\sym{***}&    0.185\sym{***}&    0.016         &    0.009\sym{*}  &    0.223         \\
                &  (0.023)         &  (0.047)         &  (0.021)         &  (0.004)         &  (0.153)         \\
IT              &   -0.087\sym{**} &    0.095\sym{*}  &    0.109\sym{***}&    0.015\sym{**} &    0.321         \\
                &  (0.029)         &  (0.044)         &  (0.026)         &  (0.005)         &  (0.211)         \\
PL              &    0.065\sym{**} &    0.018         &    0.133\sym{***}&   -0.005         &   -0.423\sym{***}\\
                &  (0.023)         &  (0.052)         &  (0.029)         &  (0.005)         &  (0.128)         \\
RO              &    0.015         &   -0.133\sym{*}  &    0.356\sym{***}&   -0.041\sym{***}&    0.256         \\
                &  (0.039)         &  (0.056)         &  (0.067)         &  (0.008)         &  (0.280)         \\
ES              &   -0.055\sym{*}  &   -0.029         &    0.090\sym{***}&    0.005         &    0.349         \\
                &  (0.027)         &  (0.045)         &  (0.026)         &  (0.005)         &  (0.193)         \\
SE              &   -0.111\sym{***}&    0.179\sym{***}&   -0.001         &    0.003         &    0.646\sym{**} \\
                &  (0.028)         &  (0.047)         &  (0.022)         &  (0.005)         &  (0.228)         \\
UK              &    0.015         &    0.172\sym{***}&    0.045\sym{*}  &   -0.001         &   -0.191         \\
                &  (0.023)         &  (0.048)         &  (0.021)         &  (0.004)         &  (0.138)         \\
ABreversed      &    0.044\sym{***}&   -0.003         &   -0.026\sym{*}  &   -0.013\sym{***}&   -0.193\sym{**} \\
                &  (0.010)         &  (0.023)         &  (0.011)         &  (0.002)         &  (0.067)         \\
Incentivized    &   -0.044\sym{**} &   -0.142\sym{***}&    0.097\sym{***}&   -0.036\sym{***}&    0.153         \\
                &  (0.014)         &  (0.026)         &  (0.013)         &  (0.003)         &  (0.086)         \\
LowStakes       &   -0.041         &   -0.124\sym{**} &   -0.020         &    0.038\sym{***}&    0.304         \\
                &  (0.022)         &  (0.041)         &  (0.025)         &  (0.005)         &  (0.178)         \\
HighStakes      &    0.030         &    0.136\sym{**} &   -0.060\sym{***}&   -0.008         &    0.028         \\
                &  (0.019)         &  (0.048)         &  (0.018)         &  (0.004)         &  (0.111)         \\
Constant        &    0.686\sym{***}&    2.235\sym{***}&    0.249\sym{***}&    0.040\sym{***}&    1.696\sym{***}\\
                &  (0.028)         &  (0.070)         &  (0.031)         &  (0.007)         &  (0.175)         \\
\midrule
N               &   443415         &                  &                  &                  &                  \\
Log. Likelihood &  -261314         &                  &                  &                  &                  \\
BIC             &   524579         &                  &                  &                  &                  \\
\bottomrule
\multicolumn{6}{l}{\scriptsize Standard errors (clustered at the subject level) in parentheses}\\
\multicolumn{6}{l}{\scriptsize \sym{*} \(p<0.05\), \sym{**} \(p<0.001\), \sym{***} \(p<0.001\)}\\
\end{tabular*}
\end{table}

\paragraph{Three Fechner errors:} Finally, different price lists, designed to elicit different domains of preference, may induce individuals to make different kinds of errors.\footnote{For instance, MPLs may be ``coarse'' in the sense that the difference in expected utility between two adjacent rows may be larger or smaller in different MPLs. A larger difference might lead to larger estimates of the noise parameters.} Therefore, we allow the Fechner error to vary between our three price list designs.\footnote{Because the design of the first two price lists (MPL1.1 and MPL1.2) is identical, we restrict the Fechner error to be the same across these lists.} Because price list design should not affect the tremble error, we restrict the tremble error to be constant across all decisions. This leaves us with three Fechner error parameters $(\mu_1,\mu_2,\mu_3)$ for MPL1, MPL2 and MPL3, respectively. We display the findings of this model in Table~\ref{tab:l3t}. Compared to our main specification of the error terms, we observe only few differences.  \emph{IT}, \emph{PL}, and \emph{RO} are no longer related to risk aversion, \emph{CRT} and \emph{IT} are no longer related to loss aversion, and\emph{Education}, \emph{Stimulation} and \emph{UK} are no longer related to discounting. Allowing the Fechner error to vary across MPLs leads to two new statistically significant relations: \emph{Power} is now positively related with risk aversion, \emph{Hedonism} is now negatively related with risk aversion, and \emph{Children} is negatively correlated with loss aversion.

\begin{table}[htbp]
\centering
\tiny
\def\sym#1{\ifmmode^{#1}\else\(^{#1}\)\fi}
\caption{Robustness check: three fechner errors \label{tab:l3t}}
\begin{tabular*}{\hsize}{@{\hskip\tabcolsep\extracolsep\fill}l*{9}{c}}
\toprule
                &\makecell{Risk aversion \\ $\alpha$}         &\makecell{Loss aversion \\ $\lambda$}         &\makecell{Discounting \\ $\delta$}         &\makecell{Present bias \\ $\gamma$}         &\makecell{Tremble error \\ $\kappa$}         &\makecell{F. error 1\\ $\mu_1$}         &\makecell{F. error 2\\ $\mu_2$}         &\makecell{F. error 3\\ $\mu_3$}         \\
\midrule
Age             &    0.002\sym{***}&   -0.007\sym{***}&   -0.001\sym{***}&   -0.000\sym{*}  &   -0.001         &   -0.004\sym{*}  &   -0.002         &   -0.013\sym{*}  \\
                &  (0.001)         &  (0.002)         &  (0.000)         &  (0.000)         &  (0.001)         &  (0.001)         &  (0.001)         &  (0.005)         \\
Male          &   -0.041\sym{**} &   -0.205\sym{***}&    0.019\sym{*}  &    0.004\sym{**} &    0.022         &    0.074\sym{*}  &    0.031         &   -0.017         \\
                &  (0.013)         &  (0.038)         &  (0.008)         &  (0.001)         &  (0.014)         &  (0.035)         &  (0.029)         &  (0.113)         \\
Education       &   -0.029         &   -0.031         &   -0.018         &    0.001         &    0.005         &    0.020         &   -0.027         &   -0.047         \\
                &  (0.016)         &  (0.039)         &  (0.010)         &  (0.001)         &  (0.017)         &  (0.041)         &  (0.035)         &  (0.130)         \\
Income          &   -0.002\sym{***}&   -0.003\sym{**} &    0.000         &   -0.000         &    0.001         &    0.001         &    0.001         &    0.007\sym{*}  \\
                &  (0.000)         &  (0.001)         &  (0.000)         &  (0.000)         &  (0.000)         &  (0.001)         &  (0.001)         &  (0.004)         \\
Children        &   -0.013         &   -0.092\sym{*}  &    0.026\sym{**} &   -0.002         &    0.027         &    0.032         &    0.052         &    0.051         \\
                &  (0.017)         &  (0.044)         &  (0.010)         &  (0.002)         &  (0.020)         &  (0.038)         &  (0.034)         &  (0.127)         \\
Couple          &    0.007         &    0.041         &   -0.010         &   -0.000         &    0.001         &   -0.058         &    0.022         &   -0.104         \\
                &  (0.014)         &  (0.045)         &  (0.010)         &  (0.002)         &  (0.014)         &  (0.037)         &  (0.032)         &  (0.130)         \\
Urban           &   -0.018         &   -0.002         &    0.008         &    0.003\sym{*}  &    0.014         &    0.060         &    0.061\sym{*}  &    0.300\sym{**} \\
                &  (0.013)         &  (0.036)         &  (0.008)         &  (0.001)         &  (0.013)         &  (0.032)         &  (0.030)         &  (0.111)         \\
CRT             &   -0.033\sym{*}  &    0.025         &   -0.014\sym{*}  &    0.000         &   -0.039\sym{**} &   -0.006         &   -0.040\sym{**} &   -0.016         \\
                &  (0.013)         &  (0.018)         &  (0.006)         &  (0.001)         &  (0.013)         &  (0.020)         &  (0.014)         &  (0.066)         \\
Achievement     &    0.009         &    0.021         &   -0.002         &   -0.001         &   -0.011         &   -0.023         &    0.044         &    0.060         \\
                &  (0.018)         &  (0.043)         &  (0.012)         &  (0.002)         &  (0.015)         &  (0.054)         &  (0.045)         &  (0.186)         \\
Benevolence     &   -0.002         &    0.056         &   -0.002         &    0.002         &    0.004         &   -0.044         &   -0.007         &   -0.008         \\
                &  (0.015)         &  (0.048)         &  (0.010)         &  (0.002)         &  (0.017)         &  (0.043)         &  (0.039)         &  (0.160)         \\
Conformity      &   -0.003         &    0.030         &   -0.003         &   -0.003\sym{*}  &    0.024\sym{*}  &   -0.019         &   -0.022         &   -0.005         \\
                &  (0.014)         &  (0.041)         &  (0.009)         &  (0.002)         &  (0.012)         &  (0.041)         &  (0.035)         &  (0.146)         \\
Hedonism        &   -0.036\sym{*}  &   -0.051         &    0.035\sym{***}&    0.001         &   -0.008         &    0.121\sym{**} &    0.083\sym{*}  &    0.149         \\
                &  (0.018)         &  (0.042)         &  (0.010)         &  (0.002)         &  (0.016)         &  (0.045)         &  (0.038)         &  (0.159)         \\
Power           &    0.046\sym{*}  &    0.025         &   -0.014         &   -0.001         &    0.030         &   -0.063         &   -0.043         &   -0.057         \\
                &  (0.020)         &  (0.046)         &  (0.013)         &  (0.002)         &  (0.018)         &  (0.058)         &  (0.046)         &  (0.187)         \\
Security        &   -0.003         &    0.041         &   -0.001         &    0.001         &    0.000         &    0.004         &   -0.058         &   -0.095         \\
                &  (0.015)         &  (0.043)         &  (0.010)         &  (0.002)         &  (0.014)         &  (0.043)         &  (0.039)         &  (0.167)         \\
Self direction  &   -0.011         &   -0.026         &    0.011         &   -0.002         &   -0.018         &    0.013         &   -0.064         &   -0.080         \\
                &  (0.015)         &  (0.043)         &  (0.010)         &  (0.002)         &  (0.014)         &  (0.039)         &  (0.037)         &  (0.139)         \\
Stimulation     &   -0.043\sym{*}  &   -0.145\sym{***}&    0.017         &   -0.001         &    0.037\sym{*}  &    0.108         &    0.194\sym{***}&    0.417\sym{*}  \\
                &  (0.019)         &  (0.042)         &  (0.013)         &  (0.002)         &  (0.016)         &  (0.060)         &  (0.052)         &  (0.204)         \\
Tradition       &    0.023         &   -0.054         &    0.002         &   -0.000         &    0.007         &   -0.021         &   -0.015         &   -0.234         \\
                &  (0.014)         &  (0.039)         &  (0.010)         &  (0.001)         &  (0.013)         &  (0.043)         &  (0.035)         &  (0.159)         \\
Universalism    &   -0.011         &   -0.049         &   -0.011         &   -0.003         &   -0.018         &   -0.002         &    0.013         &    0.134         \\
                &  (0.015)         &  (0.045)         &  (0.010)         &  (0.002)         &  (0.013)         &  (0.042)         &  (0.036)         &  (0.161)         \\
FR              &   -0.078         &    0.284\sym{***}&    0.017         &    0.005\sym{*}  &   -0.048         &    0.007         &    0.087         &    1.103\sym{**} \\
                &  (0.047)         &  (0.078)         &  (0.019)         &  (0.003)         &  (0.047)         &  (0.083)         &  (0.059)         &  (0.337)         \\
IT              &   -0.084         &    0.131         &    0.068\sym{**} &    0.013\sym{***}&   -0.020         &    0.039         &    0.191\sym{**} &    0.615         \\
                &  (0.048)         &  (0.072)         &  (0.022)         &  (0.003)         &  (0.044)         &  (0.090)         &  (0.071)         &  (0.359)         \\
PL              &    0.106         &    0.010         &    0.076\sym{*}  &    0.002         &   -0.177\sym{**} &    0.095         &   -0.011         &   -0.633\sym{*}  \\
                &  (0.060)         &  (0.084)         &  (0.031)         &  (0.003)         &  (0.068)         &  (0.078)         &  (0.056)         &  (0.274)         \\
RO              &    0.081         &   -0.164         &    0.139\sym{***}&   -0.015\sym{***}&   -0.176         &    0.203         &    0.463\sym{***}&    0.152         \\
                &  (0.062)         &  (0.088)         &  (0.040)         &  (0.004)         &  (0.090)         &  (0.116)         &  (0.131)         &  (0.365)         \\
ES              &   -0.069         &   -0.045         &    0.078\sym{**} &    0.006         &    0.001         &    0.283\sym{*}  &    0.202\sym{**} &    0.484         \\
                &  (0.045)         &  (0.070)         &  (0.025)         &  (0.003)         &  (0.035)         &  (0.122)         &  (0.073)         &  (0.354)         \\
SE              &   -0.130\sym{**} &    0.259\sym{***}&    0.014         &    0.001         &    0.001         &    0.076         &    0.267\sym{***}&    1.501\sym{***}\\
                &  (0.045)         &  (0.075)         &  (0.019)         &  (0.003)         &  (0.041)         &  (0.108)         &  (0.079)         &  (0.445)         \\
UK              &    0.058         &    0.294\sym{***}&    0.004         &    0.001         &   -0.085         &   -0.113         &   -0.048         &    0.053         \\
                &  (0.048)         &  (0.082)         &  (0.020)         &  (0.003)         &  (0.046)         &  (0.071)         &  (0.049)         &  (0.285)         \\
ABreversed      &    0.037\sym{**} &   -0.002         &   -0.025\sym{**} &   -0.006\sym{***}&    0.025         &   -0.112\sym{***}&   -0.089\sym{**} &   -0.430\sym{***}\\
                &  (0.013)         &  (0.035)         &  (0.009)         &  (0.002)         &  (0.013)         &  (0.034)         &  (0.029)         &  (0.119)         \\
Incentivized    &   -0.070\sym{***}&   -0.234\sym{***}&    0.072\sym{***}&   -0.019\sym{***}&    0.036\sym{*}  &    0.101\sym{*}  &    0.026         &    0.159         \\
                &  (0.017)         &  (0.043)         &  (0.011)         &  (0.002)         &  (0.016)         &  (0.046)         &  (0.037)         &  (0.146)         \\
LowStakes       &   -0.068\sym{*}  &   -0.157\sym{**} &    0.008         &    0.027\sym{***}&    0.032         &    0.191         &    0.131         &    0.519         \\
                &  (0.028)         &  (0.058)         &  (0.020)         &  (0.004)         &  (0.021)         &  (0.109)         &  (0.078)         &  (0.321)         \\
HighStakes      &    0.023         &    0.221\sym{*}  &   -0.031\sym{**} &   -0.006\sym{**} &   -0.004         &   -0.006         &    0.054         &    0.244         \\
                &  (0.021)         &  (0.086)         &  (0.012)         &  (0.002)         &  (0.024)         &  (0.051)         &  (0.052)         &  (0.198)         \\
Constant        &    0.667\sym{***}&    2.567\sym{***}&    0.202\sym{***}&    0.025\sym{***}&    0.229\sym{*}  &    0.687\sym{***}&    0.762\sym{***}&    3.040\sym{***}\\
                &  (0.075)         &  (0.115)         &  (0.030)         &  (0.004)         &  (0.097)         &  (0.100)         &  (0.088)         &  (0.348)         \\
\midrule
N               &   443415         &                  &                  &                  &                  &                  &                  &                  \\
Log. Likelihood &  -248914         &                  &                  &                  &                  &                  &                  &                  \\
BIC             &   500949         &                  &                  &                  &                  &                  &                  &                  \\
\bottomrule
\multicolumn{9}{l}{\tiny Standard errors (clustered at the subject level) in parentheses}\\
\multicolumn{9}{l}{\tiny \sym{*} \(p<0.05\), \sym{**} \(p<0.01\), \sym{***} \(p<0.001\)}\\
\end{tabular*}
\end{table}

%A measure often used in the elicitation of preferences, particularly in MPL designs, are switch points. Switch points indicate where a respondent switches from one option in a multiple price list to another. For instance, in MPL2, a risk-averse respondent would choose Option A in the upper half of the MPL, and then switch to Option B somewhere in the lower half of the MPL. Switch points are an easy and convenient measure of preferences. They also allow to measure preferences independently from any particular underlying model.\footnote{Though they would allow the calculation of a preference parameter range (under a specific model) that is consistent with the observed switch point.} 

\subsection{Alternative measures of preferences}
\label{sec:nA}

To derive our main results, we use maximum likelihood methods to estimate the preference parameters for a structural model capturing risk and time preferences. An alternative measure of preferences pertains to counting how often a participant makes specific choices in an MPL. For instance, when eliciting risk aversion, several studies use the number of less risky choices to measure risk aversion \citep[see e.g.][and references therein]{Andersson2016}. Likewise, discounting may be measured by the number of times an individual prefers an early payment. In a similar vein, preferences may by measured by switch points, i.e. the point in an MPL where a respondent switches from choosing Option A to choosing Option B. Respondents who switch multiple times, however, must be excluded with this measure. There are a few apparent advantages of using such non-parametric measures. First, counting the number of e.g. less risky choices is simple and quick. Moreover, this method provides preference measures at an individual level, whereas the strategy employed for our main specification estimates preferences at the aggregate level only.\footnote{In principle, structural maximum likelihood estimations can also be employed to estimate preferences at an individual level. However, this typically requires multiple different MPLs for each elicited preference parameter.}

However, these alternative measures of preferences also have some disadvantages. First, they cannot properly take into account structural dependencies between different domains of preference. For instance, the number of times someone chooses the less risky option (Option A) in MPL3 should be determined by \emph{both} loss aversion and risk aversion. Therefore, the number of times Option A is chosen does not properly identify loss aversion. Second, using these alternative measures typically does not allow to sufficiently control for decision noise. This argument is made convincingly in \cite{Andersson2016, Andersson2020}, who show that a failure to control for observed heterogeneity in decision noise leads to spurious correlations between preference measures and individual characteristics that are correlated with decision noise. To test whether using these alternative measures leads to spurious correlations in our setting, we conduct a simulation exercise in Appendix~\ref{sec:simulation}. This exercise confirms that using, for example, the number of risky choices to measure risk aversion leads to spurious correlations with individual characteristics that are correlated with decision noise. 

To facilitate comparability of our findings with those in the literature using these alternative measures of preferences, we have included regressions of these measures on our set of individual characteristics further below.
We use the number of times Option A was chosen in MPL2 to measure risk aversion. Loss aversion is measured by the number of times Option A was chosen in MPL3, with the caveat highlighted above: this measures risk and loss aversion jointly. Time discounting is measured by the amount of impatient choices in MPL1.1 and MPL1.2, i.e. the number of times Option A was chosen. Present bias is measured by how discounting differs in the two time horizons of MPL1.1 and MPL1.2. We therefore measure present bias as the difference of the number of times Option A is chosen in MPL1.2 and MPL1.1. In Table~\ref{tab:nA}, we report regressions of our individual characteristics on these measures. In Table~\ref{tab:nAnmsp}, we report analogous specifications, dropping subjects who switch multiple times between Option A and B. In these regressions, the number of times a respondent chooses Option A can also be interpreted as the switch point.

As expected, results differ considerably between our main specification and the specifications using these alternative measures. These differences pertain, in particular, to covariates that can be expected to correlate with decision noise, such as age, education and CRT. For the reasons stated above and in Section~\ref{sec:simulation}, we will not discuss these differences further. We interpret these differences as indicating decision noise, which must be controlled for, rather than as evidence that our main specification is not robust.

\begin{table}[htbp]
\centering
\scriptsize
\def\sym#1{\ifmmode^{#1}\else\(^{#1}\)\fi}
\caption{Robustness check: number of times Option A is chosen \label{tab:nA}}
\begin{tabular*}{\hsize}{@{\hskip\tabcolsep\extracolsep\fill}l*{5}{c}}
\toprule
          &\multicolumn{1}{c}{\makecell{Risk aversion \\MPL2}}&\multicolumn{1}{c}{\makecell{Loss aversion \\ MPL3}}&\multicolumn{1}{c}{\makecell{Discounting \\ MPL1.1+MPL1.2}}&\multicolumn{1}{c}{\makecell{Present bias \\MPL1.2-MPL1.1}}\\
\midrule
Age       &    0.005         &   -0.010\sym{***}&   -0.005         &   -0.003\sym{*}  \\
          &  (0.003)         &  (0.002)         &  (0.003)         &  (0.001)         \\
Male    &   -0.197\sym{**} &   -0.231\sym{***}&    0.023         &    0.061\sym{*}  \\
          &  (0.072)         &  (0.043)         &  (0.076)         &  (0.026)         \\
Education &    0.023         &   -0.012         &   -0.441\sym{***}&    0.017         \\
          &  (0.076)         &  (0.045)         &  (0.080)         &  (0.028)         \\
Income    &   -0.002         &   -0.003\sym{**} &   -0.007\sym{***}&   -0.001         \\
          &  (0.002)         &  (0.001)         &  (0.002)         &  (0.001)         \\
Children  &   -0.185\sym{*}  &   -0.111\sym{*}  &    0.246\sym{**} &   -0.042         \\
          &  (0.086)         &  (0.052)         &  (0.092)         &  (0.032)         \\
Couple    &   -0.209\sym{*}  &    0.057         &   -0.083         &   -0.004         \\
          &  (0.086)         &  (0.052)         &  (0.091)         &  (0.032)         \\
Urban     &   -0.153\sym{*}  &    0.005         &    0.055         &    0.028         \\
          &  (0.073)         &  (0.044)         &  (0.078)         &  (0.027)         \\
CRT       &    0.223\sym{***}&    0.064\sym{**} &   -0.365\sym{***}&    0.019         \\
          &  (0.034)         &  (0.021)         &  (0.038)         &  (0.012)         \\
Achievement&   -0.205\sym{*}  &    0.040         &    0.066         &   -0.013         \\
          &  (0.086)         &  (0.051)         &  (0.092)         &  (0.032)         \\
Benevolence&    0.033         &    0.061         &    0.022         &    0.064         \\
          &  (0.090)         &  (0.055)         &  (0.098)         &  (0.034)         \\
Conformity&    0.034         &    0.030         &   -0.153         &   -0.079\sym{**} \\
          &  (0.081)         &  (0.049)         &  (0.088)         &  (0.031)         \\
Hedonism  &   -0.039         &   -0.052         &    0.127         &    0.027         \\
          &  (0.083)         &  (0.050)         &  (0.088)         &  (0.031)         \\
Power     &   -0.085         &    0.020         &    0.025         &   -0.004         \\
          &  (0.090)         &  (0.054)         &  (0.096)         &  (0.034)         \\
Security  &    0.040         &    0.034         &   -0.012         &    0.012         \\
          &  (0.086)         &  (0.052)         &  (0.093)         &  (0.034)         \\
Self direction&    0.229\sym{**} &   -0.025         &    0.116         &   -0.018         \\
          &  (0.083)         &  (0.050)         &  (0.090)         &  (0.032)         \\
Stimulation&   -0.510\sym{***}&   -0.169\sym{***}&   -0.105         &   -0.049         \\
          &  (0.084)         &  (0.050)         &  (0.090)         &  (0.031)         \\
Tradition &    0.143         &   -0.053         &    0.147         &   -0.009         \\
          &  (0.078)         &  (0.047)         &  (0.084)         &  (0.029)         \\
Universalism&   -0.065         &   -0.055         &   -0.229\sym{*}  &   -0.058         \\
          &  (0.086)         &  (0.052)         &  (0.093)         &  (0.033)         \\
FR        &    0.115         &    0.358\sym{***}&   -0.214         &    0.162\sym{**} \\
          &  (0.146)         &  (0.090)         &  (0.158)         &  (0.054)         \\
IT        &   -0.292\sym{*}  &    0.168\sym{*}  &    0.669\sym{***}&    0.258\sym{***}\\
          &  (0.141)         &  (0.084)         &  (0.150)         &  (0.054)         \\
PL        &    0.497\sym{**} &   -0.016         &    1.779\sym{***}&    0.071         \\
          &  (0.159)         &  (0.094)         &  (0.175)         &  (0.060)         \\
RO        &   -0.335         &   -0.245\sym{*}  &    1.685\sym{***}&   -0.154\sym{*}  \\
          &  (0.174)         &  (0.100)         &  (0.181)         &  (0.062)         \\
ES        &   -0.242         &   -0.042         &    0.462\sym{**} &    0.101         \\
          &  (0.145)         &  (0.087)         &  (0.161)         &  (0.057)         \\
SE        &   -0.489\sym{***}&    0.340\sym{***}&   -0.579\sym{***}&    0.004         \\
          &  (0.148)         &  (0.089)         &  (0.160)         &  (0.056)         \\
UK        &    0.432\sym{**} &    0.305\sym{***}&    0.414\sym{**} &    0.065         \\
          &  (0.142)         &  (0.087)         &  (0.155)         &  (0.055)         \\
ABreversed&    0.007         &   -0.039         &   -0.155\sym{*}  &   -0.187\sym{***}\\
          &  (0.070)         &  (0.042)         &  (0.075)         &  (0.026)         \\
Incentivized&   -0.105         &   -0.261\sym{***}&    0.153         &   -0.382\sym{***}\\
          &  (0.080)         &  (0.048)         &  (0.085)         &  (0.031)         \\
LowStakes &   -0.445\sym{**} &   -0.205\sym{*}  &    0.049         &    0.388\sym{***}\\
          &  (0.139)         &  (0.082)         &  (0.146)         &  (0.053)         \\
HighStakes&   -0.027         &    0.184\sym{*}  &   -0.568\sym{***}&   -0.136\sym{**} \\
          &  (0.132)         &  (0.077)         &  (0.136)         &  (0.050)         \\
\_cons    &    9.176\sym{***}&    4.401\sym{***}&    8.572\sym{***}&    0.596\sym{***}\\
          &  (0.216)         &  (0.127)         &  (0.230)         &  (0.081)         \\
\midrule
\(N\)     &    12669         &    12669         &    12669         &    12669         \\
adj. R2   &    0.019         &    0.018         &    0.057         &    0.027         \\
BIC       &71080.359         &57903.621         &72572.135         &45885.025         \\
\bottomrule
\multicolumn{5}{l}{\scriptsize Standard errors (clustered at the subject level) in parentheses}\\
\multicolumn{5}{l}{\scriptsize \sym{*} \(p<0.05\), \sym{**} \(p<0.01\), \sym{***} \(p<0.001\)}\\
\end{tabular*}
\end{table}

\begin{table}[htbp]
\centering
\scriptsize
\def\sym#1{\ifmmode^{#1}\else\(^{#1}\)\fi}
\caption{Robustness check: switch points \label{tab:nAnmsp}}
\begin{tabular*}{\hsize}{@{\hskip\tabcolsep\extracolsep\fill}l*{5}{c}}
\toprule
          &\multicolumn{1}{c}{\makecell{Risk aversion \\MPL2}}&\multicolumn{1}{c}{\makecell{Loss aversion \\ MPL3}}&\multicolumn{1}{c}{\makecell{Discounting \\ MPL1.1+MPL1.2}}&\multicolumn{1}{c}{\makecell{Present bias \\MPL1.2-MPL1.1}}\\
\midrule
Age       &    0.003         &   -0.009\sym{***}&   -0.006         &   -0.004\sym{**} \\
          &  (0.004)         &  (0.002)         &  (0.004)         &  (0.001)         \\
Male    &   -0.210\sym{**} &   -0.218\sym{***}&    0.073         &    0.047         \\
          &  (0.081)         &  (0.049)         &  (0.085)         &  (0.027)         \\
Education &   -0.022         &    0.015         &   -0.438\sym{***}&    0.018         \\
          &  (0.086)         &  (0.052)         &  (0.089)         &  (0.029)         \\
Income    &   -0.005\sym{*}  &   -0.003\sym{**} &   -0.009\sym{***}&   -0.001         \\
          &  (0.002)         &  (0.001)         &  (0.002)         &  (0.001)         \\
Children  &   -0.106         &   -0.133\sym{*}  &    0.324\sym{**} &   -0.015         \\
          &  (0.097)         &  (0.059)         &  (0.102)         &  (0.033)         \\
Couple    &   -0.129         &    0.062         &   -0.098         &   -0.012         \\
          &  (0.096)         &  (0.058)         &  (0.101)         &  (0.033)         \\
Urban     &   -0.110         &    0.011         &    0.062         &    0.019         \\
          &  (0.082)         &  (0.050)         &  (0.086)         &  (0.028)         \\
CRT       &    0.098\sym{**} &    0.046\sym{*}  &   -0.418\sym{***}&    0.011         \\
          &  (0.038)         &  (0.023)         &  (0.041)         &  (0.013)         \\
Achievement&   -0.196\sym{*}  &    0.073         &    0.076         &   -0.034         \\
          &  (0.096)         &  (0.058)         &  (0.101)         &  (0.033)         \\
Benevolence&   -0.029         &    0.064         &    0.015         &    0.036         \\
          &  (0.100)         &  (0.061)         &  (0.108)         &  (0.035)         \\
Conformity&    0.013         &    0.015         &   -0.177         &   -0.087\sym{**} \\
          &  (0.089)         &  (0.055)         &  (0.096)         &  (0.031)         \\
Hedonism  &   -0.038         &   -0.053         &    0.192\sym{*}  &    0.051         \\
          &  (0.092)         &  (0.056)         &  (0.098)         &  (0.031)         \\
Power     &    0.122         &    0.052         &    0.035         &    0.003         \\
          &  (0.103)         &  (0.063)         &  (0.108)         &  (0.035)         \\
Security  &   -0.011         &    0.042         &   -0.049         &   -0.003         \\
          &  (0.096)         &  (0.059)         &  (0.102)         &  (0.034)         \\
Self direction&    0.207\sym{*}  &   -0.033         &    0.107         &   -0.017         \\
          &  (0.093)         &  (0.056)         &  (0.098)         &  (0.032)         \\
Stimulation&   -0.485\sym{***}&   -0.190\sym{***}&   -0.081         &   -0.038         \\
          &  (0.094)         &  (0.057)         &  (0.100)         &  (0.032)         \\
Tradition &    0.252\sym{**} &   -0.046         &    0.223\sym{*}  &    0.013         \\
          &  (0.085)         &  (0.052)         &  (0.091)         &  (0.029)         \\
Universalism&   -0.060         &   -0.052         &   -0.250\sym{*}  &   -0.038         \\
          &  (0.095)         &  (0.058)         &  (0.102)         &  (0.033)         \\
FR        &    0.017         &    0.335\sym{***}&   -0.219         &    0.175\sym{**} \\
          &  (0.160)         &  (0.099)         &  (0.173)         &  (0.055)         \\
IT        &   -0.311\sym{*}  &    0.106         &    0.660\sym{***}&    0.247\sym{***}\\
          &  (0.158)         &  (0.095)         &  (0.167)         &  (0.056)         \\
PL        &    0.526\sym{**} &   -0.041         &    1.929\sym{***}&    0.072         \\
          &  (0.180)         &  (0.108)         &  (0.196)         &  (0.061)         \\
RO        &   -0.312         &   -0.392\sym{***}&    2.085\sym{***}&   -0.066         \\
          &  (0.208)         &  (0.118)         &  (0.206)         &  (0.064)         \\
ES        &   -0.195         &   -0.140         &    0.574\sym{**} &    0.143\sym{*}  \\
          &  (0.162)         &  (0.098)         &  (0.180)         &  (0.058)         \\
SE        &   -0.522\sym{**} &    0.346\sym{***}&   -0.565\sym{**} &    0.048         \\
          &  (0.164)         &  (0.100)         &  (0.177)         &  (0.056)         \\
UK        &    0.424\sym{**} &    0.304\sym{**} &    0.434\sym{*}  &    0.096         \\
          &  (0.155)         &  (0.096)         &  (0.170)         &  (0.057)         \\
ABreversed&    0.033         &   -0.041         &   -0.139         &   -0.122\sym{***}\\
          &  (0.079)         &  (0.047)         &  (0.083)         &  (0.027)         \\
Incentivized&   -0.089         &   -0.281\sym{***}&    0.144         &   -0.347\sym{***}\\
          &  (0.089)         &  (0.054)         &  (0.093)         &  (0.032)         \\
LowStakes &   -0.373\sym{*}  &   -0.223\sym{*}  &    0.163         &    0.377\sym{***}\\
          &  (0.155)         &  (0.093)         &  (0.162)         &  (0.053)         \\
HighStakes&    0.060         &    0.249\sym{**} &   -0.531\sym{***}&   -0.170\sym{***}\\
          &  (0.149)         &  (0.088)         &  (0.153)         &  (0.051)         \\
\_cons    &    9.603\sym{***}&    4.440\sym{***}&    8.649\sym{***}&    0.605\sym{***}\\
          &  (0.247)         &  (0.147)         &  (0.259)         &  (0.084)         \\
\midrule
\(N\)     &    10571         &    10571         &    10571         &    10571         \\
adj. R2   &    0.013         &    0.019         &    0.069         &    0.022         \\
BIC       &59782.565         &49068.389         &60810.559         &36902.283         \\
\bottomrule
\multicolumn{5}{l}{\scriptsize Standard errors (clustered at the subject level) in parentheses}\\
\multicolumn{5}{l}{\scriptsize \sym{*} \(p<0.05\), \sym{**} \(p<0.01\), \sym{***} \(p<0.001\)}\\
\end{tabular*}
\end{table}

\subsection{Simulations}

\label{sec:simulation}

The analysis in \cite{Andersson2016} revealed potential spurious correlations of risk preferences and individual characteristics that are correlated with decision noise. Building on these concerns, \cite{Andersson2020} propose elicitation and estimation methods that are robust to these spurious effects. They propose a balanced design of the multiple price lists, as well as structural estimations to model heterogeneity of noise. The design of our MPLs as well as the structural estimations in Section~\ref{sec:MLE} reflect these principles. 

To test whether our MPLs and estimation mitigate the spurious correlation effects, we simulate decision data: For each simulated individual, we randomly draw preference and error parameters, according to the following data generating process:\\

\begin{figure}
\caption{Distributions of simulated preference parameters\label{fig:sim}}
\includegraphics[width=1.0\textwidth]{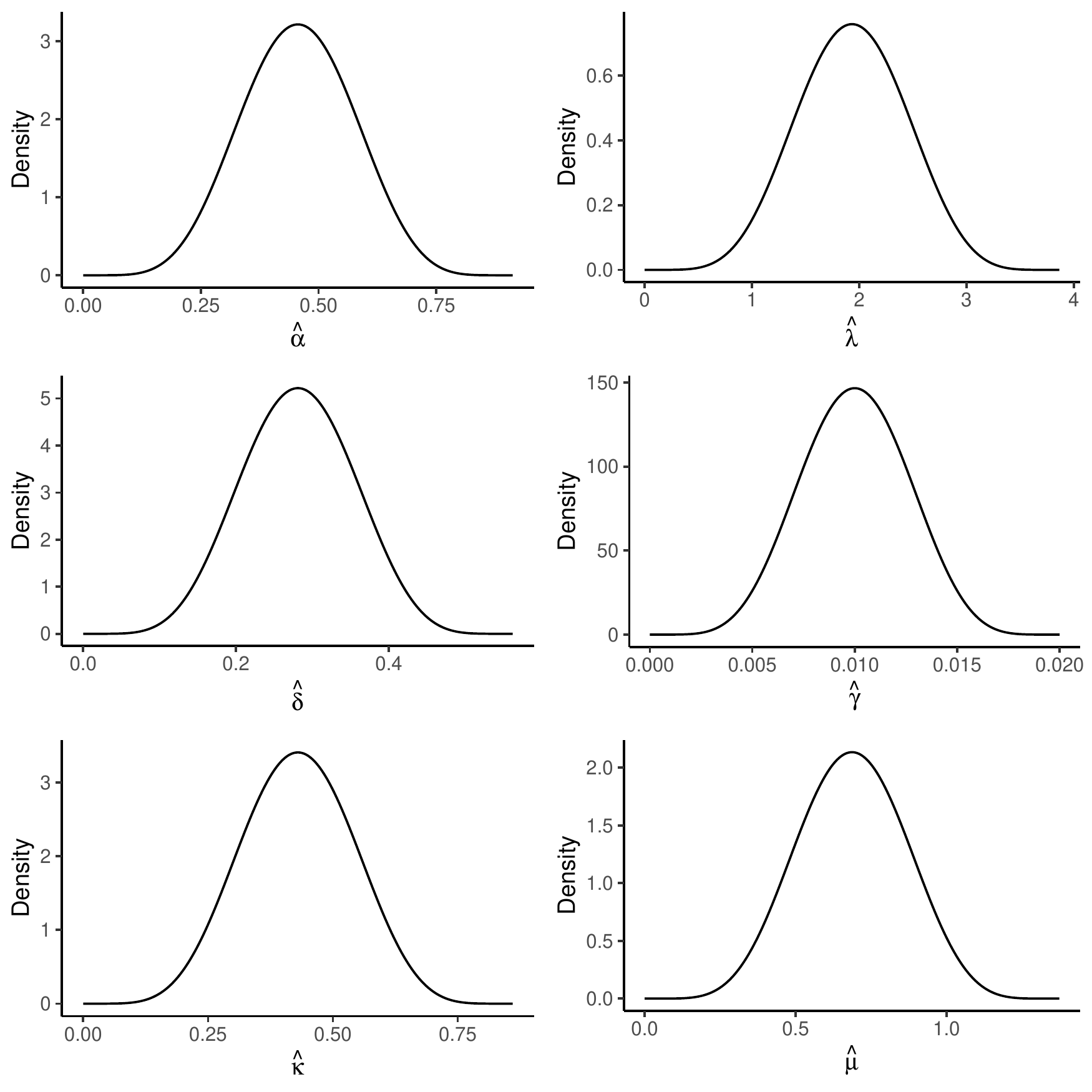}

\end{figure}

\noindent
$\hat{\alpha}\sim B(0,0.912,7,7)$\\
$\hat{\lambda}\sim B(0,3.866,7,7)$\\
$\hat{\delta}\sim B(0,0.020,7,7)$\\
$\hat{\gamma}\sim B(0,0.562,7,7)$\\
$\hat{\kappa}\sim B(0,0.860,7,7)$\\
$\hat{\mu}\sim B(0,1.374,7,7)$,\\

where $B$ is the four parameter beta distribution. The first two arguments indicate the minimum and maximum value, and the last two arguments indicate the $\alpha$ and $\beta$ parameters of the beta distribution. Figure~\ref{fig:sim} illustrates the probability density functions of the simulated preference parameters. We chose beta distributions because some of the parameters require distributions with limited support.\footnote{The tremble error $\kappa$, for instance, is a probability and therefore has to lie on the unit interval.} The distributions were parameterized such that the means matched the point estimates in Table~\ref{tab:main}. Lower and upper limits were set to $-/+100\%$ of the respective mean, and the $\alpha$ and $\beta$ parameters were set to 7, generating symmetric distributions.\footnote{Note that the main result outlined below is not sensitive to the parameterization of the distributions of preference parameters. Spurious correlations are also generated with no variance in the preference parameters (Though variance in the error parameters is necessary of course).}

We then simulate individual choice profiles, based on the drawn preference parameters and the random utility model described in section~\ref{sec:MLE}, using the decision situations from our MPLs. We then employ the maximum likelihood estimation from section~\ref{sec:MLE} to estimate preferences of the simulated subjects. 

To test whether variables correlated with the error term may generate a spurious correlation, we used the randomly drawn realizations of the error parameters ($\hat{\kappa}$ and $\hat{\mu}$ for the tremble and Fechner error, respectively) as covariates. These covariates are, of course, perfectly correlated with the respective error parameters. In contrast, the correlation between other covariates (such as measures of cognitive ability) and the error parameters should be much lower. In that sense, our approach is a rather strict test of potential spurious effects. 

We simulate 20 experiments, with $n=12.000$ subjects each, totaling $N=420.000$ choices per experiment. Table~\ref{tab:sim_l1t} summarizes the estimations using our preferred model (including a tremble error). Table~\ref{tab:sim_l1} displays estimations on the same data using a simplified model without the tremble error. Finally, Table~\ref{tab:sim_reg} contains the results of regressing the error parameters on an alternative measure of preferences: the number of times option A was chosen. Each row in the tables indicates one simulated experiment. 

In the following, we will focus on spurious correlations of noise and risk preferences, though the argument is also relevant for other domains of preference. As is apparent from Tables~\ref{tab:sim_l1t}--\ref{tab:sim_reg}, not (or imperfectly) controlling for heterogeneity of decision noise leads to a serious bias of estimates. Using OLS to estimate the impact of noise on `the number of times Option A was chosen' in MPL2 reveals a consistent and large negative effect of decision noise on risk aversion. This confirms the findings in \cite{Andersson2016} and is consistent with our findings when using the number of times Option A is chosen as a measure of risk aversion in our data. This finding therefore underlines the notion that this measure of risk aversion is not robust to spurious correlation effects. 

Using structural maximum likelihood estimations, but omitting the tremble error also leads to spurious effects. In this case, we find consistent and large positive spurious correlations between risk aversion and the tremble error. This effect goes in the same direction as the effect we find using our survey data and estimations without the tremble error. 

Finally, using our preferred model, i.e. a structural estimation with logit-link Fechner error and tremble error, removes any consistent spurious effects. Note that this finding is not self-evident, because design features of the used MPLs may still lead to spurious effects.

The results of this simulation exercise strengthens our belief that our main specification is best suited to control for decision noise, and that the remaining negative correlation of risk aversion and cognitive ability documented in Section~\ref{sec:MLE} is not spurious. This interpretation, however, depends on the underlying structural model of decision noise being correct. Correctness of the underlying model can be ensured in a simulation exercise, but it is of course possible that our structural model is still misspecified, when applied to our survey data.

\begin{sidewaystable}
\caption{Simulations: main specification \label{tab:sim_l1t}}
\tabcolsep1.2mm
\tiny
\renewcommand{\footnotesize}{\tiny}

{
\def\sym#1{\ifmmode^{#1}\else\(^{#1}\)\fi}
\begin{tabular}{l*{20}{c}}
\toprule
          &\multicolumn{1}{c}{(1)}         &\multicolumn{1}{c}{(2)}         &\multicolumn{1}{c}{(3)}         &\multicolumn{1}{c}{(4)}         &\multicolumn{1}{c}{(5)}         &\multicolumn{1}{c}{(6)}         &\multicolumn{1}{c}{(7)}         &\multicolumn{1}{c}{(8)}         &\multicolumn{1}{c}{(9)}         &\multicolumn{1}{c}{(10)}         &\multicolumn{1}{c}{(11)}         &\multicolumn{1}{c}{(12)}         &\multicolumn{1}{c}{(13)}         &\multicolumn{1}{c}{(14)}         &\multicolumn{1}{c}{(15)}         &\multicolumn{1}{c}{(16)}         &\multicolumn{1}{c}{(17)}         &\multicolumn{1}{c}{(18)}         &\multicolumn{1}{c}{(19)}         &\multicolumn{1}{c}{(20)}         \\
\midrule
Risk aversion: $\alpha$&                  &                  &                  &                  &                  &                  &                  &                  &                  &                  &                  &                  &                  &                  &                  &                  &                  &                  &                  &                  \\
$\hat{\kappa}$&    -0.02         &     0.04         &     0.03         &    -0.02         &    -0.03         &     0.06         &    -0.05         &     0.00         &    -0.04         &    -0.04         &    -0.02         &    -0.02         &    -0.02         &    -0.02         &     0.01         &     0.01         &     0.01         &    -0.00         &    -0.05         &    -0.04         \\
          &   (0.03)         &   (0.03)         &   (0.03)         &   (0.03)         &   (0.03)         &   (0.03)         &   (0.03)         &   (0.03)         &   (0.03)         &   (0.03)         &   (0.03)         &   (0.03)         &   (0.03)         &   (0.03)         &   (0.03)         &   (0.03)         &   (0.03)         &   (0.03)         &   (0.03)         &   (0.03)         \\
$\hat{\mu}$&     0.02         &     0.01         &    -0.00         &    -0.05\sym{**} &    -0.03         &    -0.00         &    -0.03         &    -0.02         &    -0.00         &    -0.02         &    -0.02         &    -0.01         &    -0.01         &    -0.03         &    -0.02         &    -0.03         &    -0.05\sym{**} &    -0.01         &    -0.02         &    -0.01         \\
          &   (0.02)         &   (0.02)         &   (0.02)         &   (0.02)         &   (0.02)         &   (0.02)         &   (0.02)         &   (0.02)         &   (0.02)         &   (0.02)         &   (0.02)         &   (0.02)         &   (0.02)         &   (0.02)         &   (0.02)         &   (0.02)         &   (0.02)         &   (0.02)         &   (0.02)         &   (0.02)         \\
Constant  &     0.44\sym{***}&     0.42\sym{***}&     0.44\sym{***}&     0.49\sym{***}&     0.47\sym{***}&     0.42\sym{***}&     0.49\sym{***}&     0.46\sym{***}&     0.46\sym{***}&     0.48\sym{***}&     0.47\sym{***}&     0.47\sym{***}&     0.46\sym{***}&     0.47\sym{***}&     0.45\sym{***}&     0.46\sym{***}&     0.48\sym{***}&     0.45\sym{***}&     0.48\sym{***}&     0.47\sym{***}\\
          &   (0.02)         &   (0.02)         &   (0.02)         &   (0.02)         &   (0.02)         &   (0.02)         &   (0.02)         &   (0.02)         &   (0.02)         &   (0.02)         &   (0.02)         &   (0.02)         &   (0.02)         &   (0.02)         &   (0.02)         &   (0.02)         &   (0.02)         &   (0.02)         &   (0.02)         &   (0.02)         \\
\midrule
Loss aversion: $\lambda$&                  &                  &                  &                  &                  &                  &                  &                  &                  &                  &                  &                  &                  &                  &                  &                  &                  &                  &                  &                  \\
$\hat{\kappa}$&     0.02         &     0.05         &     0.03         &    -0.06         &     0.05         &    -0.05         &     0.02         &    -0.04         &    -0.09\sym{*}  &     0.03         &     0.05         &     0.12\sym{**} &    -0.01         &    -0.06         &    -0.01         &    -0.08         &     0.06         &    -0.12\sym{**} &     0.02         &     0.06         \\
          &   (0.04)         &   (0.04)         &   (0.04)         &   (0.04)         &   (0.05)         &   (0.04)         &   (0.04)         &   (0.05)         &   (0.04)         &   (0.04)         &   (0.04)         &   (0.05)         &   (0.04)         &   (0.04)         &   (0.04)         &   (0.04)         &   (0.04)         &   (0.04)         &   (0.04)         &   (0.04)         \\
$\hat{\mu}$&    -0.05\sym{*}  &    -0.04         &    -0.03         &    -0.05         &    -0.03         &    -0.03         &    -0.05         &    -0.07\sym{**} &    -0.03         &    -0.03         &    -0.03         &    -0.07\sym{*}  &    -0.08\sym{**} &    -0.05\sym{*}  &    -0.06\sym{*}  &    -0.04         &    -0.04         &    -0.03         &    -0.02         &    -0.07\sym{**} \\
          &   (0.03)         &   (0.03)         &   (0.02)         &   (0.03)         &   (0.03)         &   (0.03)         &   (0.03)         &   (0.03)         &   (0.03)         &   (0.02)         &   (0.03)         &   (0.03)         &   (0.03)         &   (0.03)         &   (0.02)         &   (0.02)         &   (0.03)         &   (0.03)         &   (0.03)         &   (0.03)         \\
Constant  &     1.99\sym{***}&     1.97\sym{***}&     1.97\sym{***}&     2.02\sym{***}&     1.96\sym{***}&     2.01\sym{***}&     1.99\sym{***}&     2.02\sym{***}&     2.02\sym{***}&     1.98\sym{***}&     1.96\sym{***}&     1.96\sym{***}&     2.02\sym{***}&     2.03\sym{***}&     2.00\sym{***}&     2.02\sym{***}&     1.96\sym{***}&     2.03\sym{***}&     1.98\sym{***}&     1.98\sym{***}\\
          &   (0.02)         &   (0.02)         &   (0.02)         &   (0.02)         &   (0.02)         &   (0.02)         &   (0.02)         &   (0.02)         &   (0.02)         &   (0.02)         &   (0.02)         &   (0.02)         &   (0.02)         &   (0.02)         &   (0.02)         &   (0.02)         &   (0.02)         &   (0.02)         &   (0.02)         &   (0.02)         \\
\midrule
Discounting: $\delta$&                  &                  &                  &                  &                  &                  &                  &                  &                  &                  &                  &                  &                  &                  &                  &                  &                  &                  &                  &                  \\
$\hat{\kappa}$&     0.02         &     0.00         &     0.00         &    -0.02         &     0.01         &    -0.03         &     0.02         &    -0.02         &    -0.02         &    -0.01         &     0.02         &     0.03         &     0.00         &     0.01         &     0.03         &    -0.05         &    -0.02         &    -0.04         &     0.08\sym{**} &     0.01         \\
          &   (0.03)         &   (0.03)         &   (0.03)         &   (0.03)         &   (0.03)         &   (0.03)         &   (0.03)         &   (0.03)         &   (0.03)         &   (0.03)         &   (0.03)         &   (0.03)         &   (0.03)         &   (0.03)         &   (0.03)         &   (0.03)         &   (0.03)         &   (0.03)         &   (0.03)         &   (0.03)         \\
$\hat{\mu}$&    -0.01         &     0.00         &     0.02         &     0.02         &    -0.01         &     0.00         &     0.01         &     0.01         &    -0.01         &     0.01         &     0.00         &     0.01         &     0.01         &    -0.01         &     0.00         &     0.02         &     0.03\sym{*}  &    -0.01         &     0.02         &    -0.01         \\
          &   (0.02)         &   (0.02)         &   (0.02)         &   (0.02)         &   (0.02)         &   (0.02)         &   (0.02)         &   (0.02)         &   (0.02)         &   (0.02)         &   (0.02)         &   (0.02)         &   (0.02)         &   (0.02)         &   (0.02)         &   (0.02)         &   (0.01)         &   (0.02)         &   (0.02)         &   (0.02)         \\
Constant  &     0.27\sym{***}&     0.28\sym{***}&     0.27\sym{***}&     0.28\sym{***}&     0.28\sym{***}&     0.29\sym{***}&     0.26\sym{***}&     0.28\sym{***}&     0.30\sym{***}&     0.28\sym{***}&     0.27\sym{***}&     0.26\sym{***}&     0.28\sym{***}&     0.28\sym{***}&     0.27\sym{***}&     0.28\sym{***}&     0.27\sym{***}&     0.30\sym{***}&     0.24\sym{***}&     0.28\sym{***}\\
          &   (0.01)         &   (0.02)         &   (0.01)         &   (0.02)         &   (0.01)         &   (0.01)         &   (0.01)         &   (0.01)         &   (0.02)         &   (0.01)         &   (0.02)         &   (0.02)         &   (0.01)         &   (0.01)         &   (0.01)         &   (0.01)         &   (0.01)         &   (0.01)         &   (0.01)         &   (0.02)         \\
\midrule
Present bias: $\gamma$&                  &                  &                  &                  &                  &                  &                  &                  &                  &                  &                  &                  &                  &                  &                  &                  &                  &                  &                  &                  \\
$\hat{\kappa}$&    -0.02         &    -0.01         &    -0.02         &     0.01         &     0.01         &    -0.01         &     0.00         &     0.02         &     0.02         &    -0.01         &    -0.02         &    -0.00         &    -0.01         &    -0.02         &    -0.01         &     0.00         &    -0.01         &     0.01         &    -0.03\sym{*}  &     0.00         \\
          &   (0.01)         &   (0.01)         &   (0.01)         &   (0.01)         &   (0.01)         &   (0.01)         &   (0.01)         &   (0.01)         &   (0.01)         &   (0.01)         &   (0.01)         &   (0.01)         &   (0.01)         &   (0.01)         &   (0.01)         &   (0.01)         &   (0.01)         &   (0.01)         &   (0.01)         &   (0.01)         \\
$\hat{\mu}$&     0.01         &    -0.01         &     0.01         &     0.00         &     0.01\sym{*}  &     0.00         &    -0.00         &     0.00         &     0.01         &     0.01         &     0.00         &    -0.00         &     0.00         &     0.01         &     0.01         &     0.00         &     0.01         &     0.01         &    -0.00         &     0.00         \\
          &   (0.01)         &   (0.01)         &   (0.01)         &   (0.01)         &   (0.01)         &   (0.01)         &   (0.01)         &   (0.01)         &   (0.01)         &   (0.01)         &   (0.01)         &   (0.01)         &   (0.01)         &   (0.01)         &   (0.01)         &   (0.01)         &   (0.01)         &   (0.01)         &   (0.01)         &   (0.01)         \\
Constant  &     0.01         &     0.02\sym{***}&     0.01         &     0.00         &    -0.00         &     0.01         &     0.01\sym{*}  &     0.00         &    -0.00         &     0.01         &     0.02\sym{*}  &     0.01         &     0.01\sym{*}  &     0.01\sym{*}  &     0.01         &     0.01         &     0.01         &     0.00         &     0.02\sym{***}&     0.01         \\
          &   (0.01)         &   (0.01)         &   (0.01)         &   (0.01)         &   (0.01)         &   (0.01)         &   (0.01)         &   (0.01)         &   (0.01)         &   (0.01)         &   (0.01)         &   (0.01)         &   (0.01)         &   (0.01)         &   (0.01)         &   (0.01)         &   (0.01)         &   (0.01)         &   (0.01)         &   (0.01)         \\
\midrule
Tremble error: $\kappa$&                  &                  &                  &                  &                  &                  &                  &                  &                  &                  &                  &                  &                  &                  &                  &                  &                  &                  &                  &                  \\
$\hat{\kappa}$&     0.98\sym{***}&     0.95\sym{***}&     0.94\sym{***}&     0.95\sym{***}&     0.95\sym{***}&     0.93\sym{***}&     0.94\sym{***}&     0.92\sym{***}&     0.92\sym{***}&     0.98\sym{***}&     0.97\sym{***}&     0.96\sym{***}&     0.93\sym{***}&     0.94\sym{***}&     0.93\sym{***}&     0.94\sym{***}&     0.96\sym{***}&     0.95\sym{***}&     0.97\sym{***}&     0.98\sym{***}\\
          &   (0.02)         &   (0.02)         &   (0.02)         &   (0.02)         &   (0.02)         &   (0.02)         &   (0.02)         &   (0.02)         &   (0.02)         &   (0.02)         &   (0.02)         &   (0.02)         &   (0.02)         &   (0.02)         &   (0.02)         &   (0.02)         &   (0.02)         &   (0.02)         &   (0.02)         &   (0.02)         \\
$\hat{\mu}$&     0.03\sym{*}  &     0.03\sym{*}  &     0.00         &     0.06\sym{***}&     0.02         &     0.02         &     0.05\sym{***}&     0.03\sym{*}  &     0.01         &     0.03\sym{*}  &     0.02         &     0.02         &     0.00         &     0.04\sym{**} &     0.02         &     0.04\sym{**} &     0.03\sym{*}  &     0.03\sym{*}  &     0.02         &     0.04\sym{**} \\
          &   (0.01)         &   (0.01)         &   (0.01)         &   (0.01)         &   (0.01)         &   (0.01)         &   (0.01)         &   (0.01)         &   (0.01)         &   (0.01)         &   (0.01)         &   (0.01)         &   (0.01)         &   (0.01)         &   (0.01)         &   (0.01)         &   (0.01)         &   (0.01)         &   (0.01)         &   (0.01)         \\
Constant  &     0.03\sym{*}  &     0.04\sym{**} &     0.06\sym{***}&     0.02         &     0.04\sym{**} &     0.05\sym{***}&     0.02         &     0.05\sym{***}&     0.06\sym{***}&     0.02\sym{*}  &     0.03\sym{*}  &     0.04\sym{**} &     0.05\sym{***}&     0.04\sym{**} &     0.06\sym{***}&     0.04\sym{**} &     0.03\sym{*}  &     0.04\sym{**} &     0.04\sym{**} &     0.02         \\
          &   (0.01)         &   (0.01)         &   (0.01)         &   (0.01)         &   (0.01)         &   (0.01)         &   (0.01)         &   (0.01)         &   (0.01)         &   (0.01)         &   (0.01)         &   (0.01)         &   (0.01)         &   (0.01)         &   (0.01)         &   (0.01)         &   (0.01)         &   (0.01)         &   (0.01)         &   (0.01)         \\
\midrule
Fechner error: $\mu$&                  &                  &                  &                  &                  &                  &                  &                  &                  &                  &                  &                  &                  &                  &                  &                  &                  &                  &                  &                  \\
$\hat{\kappa}$&    -0.15         &    -0.19         &    -0.01         &    -0.22         &    -0.02         &    -0.06         &     0.02         &     0.15         &     0.09         &    -0.14         &    -0.05         &    -0.04         &     0.19         &    -0.04         &     0.09         &    -0.11         &    -0.20         &    -0.11         &     0.06         &    -0.11         \\
          &   (0.12)         &   (0.11)         &   (0.12)         &   (0.12)         &   (0.12)         &   (0.12)         &   (0.12)         &   (0.12)         &   (0.11)         &   (0.12)         &   (0.11)         &   (0.13)         &   (0.11)         &   (0.12)         &   (0.12)         &   (0.10)         &   (0.12)         &   (0.11)         &   (0.12)         &   (0.12)         \\
$\hat{\mu}$&     0.87\sym{***}&     0.76\sym{***}&     0.98\sym{***}&     0.81\sym{***}&     0.95\sym{***}&     0.92\sym{***}&     0.76\sym{***}&     0.90\sym{***}&     0.90\sym{***}&     0.96\sym{***}&     0.89\sym{***}&     0.91\sym{***}&     0.99\sym{***}&     0.84\sym{***}&     0.96\sym{***}&     0.86\sym{***}&     0.97\sym{***}&     0.92\sym{***}&     0.88\sym{***}&     0.84\sym{***}\\
          &   (0.07)         &   (0.07)         &   (0.08)         &   (0.07)         &   (0.07)         &   (0.07)         &   (0.07)         &   (0.07)         &   (0.07)         &   (0.07)         &   (0.07)         &   (0.07)         &   (0.07)         &   (0.07)         &   (0.07)         &   (0.07)         &   (0.07)         &   (0.06)         &   (0.07)         &   (0.07)         \\
Constant  &     0.22\sym{***}&     0.30\sym{***}&     0.09         &     0.28\sym{***}&     0.15\sym{*}  &     0.17\sym{**} &     0.22\sym{***}&     0.09         &     0.13\sym{*}  &     0.15\sym{*}  &     0.17\sym{**} &     0.17\sym{*}  &     0.04         &     0.20\sym{***}&     0.06         &     0.19\sym{**} &     0.16\sym{*}  &     0.17\sym{**} &     0.13\sym{*}  &     0.23\sym{***}\\
          &   (0.06)         &   (0.06)         &   (0.06)         &   (0.06)         &   (0.06)         &   (0.06)         &   (0.06)         &   (0.06)         &   (0.06)         &   (0.06)         &   (0.06)         &   (0.07)         &   (0.06)         &   (0.06)         &   (0.06)         &   (0.06)         &   (0.06)         &   (0.05)         &   (0.06)         &   (0.06)         \\
\midrule
\(N\)     &   420000         &   420000         &   420000         &   420000         &   420000         &   420000         &   420000         &   420000         &   420000         &   420000         &   420000         &   420000         &   420000         &   420000         &   420000         &   420000         &   420000         &   420000         &   420000         &   420000         \\
\bottomrule
\multicolumn{21}{l}{\footnotesize Standard errors in parentheses}\\
\multicolumn{21}{l}{\footnotesize \sym{*} \(p<0.05\), \sym{**} \(p<0.01\), \sym{***} \(p<0.001\)}\\
\end{tabular}
}

\renewcommand{\footnotesize}{\footnotesize}
\end{sidewaystable}

\begin{sidewaystable}
\tabcolsep1.2mm
\caption{Simulations: no tremble error \label{tab:sim_l1}}
\tiny
\renewcommand{\footnotesize}{\tiny}

{
\def\sym#1{\ifmmode^{#1}\else\(^{#1}\)\fi}
\begin{tabular}{l*{20}{c}}
\toprule
          &\multicolumn{1}{c}{(1)}         &\multicolumn{1}{c}{(2)}         &\multicolumn{1}{c}{(3)}         &\multicolumn{1}{c}{(4)}         &\multicolumn{1}{c}{(5)}         &\multicolumn{1}{c}{(6)}         &\multicolumn{1}{c}{(7)}         &\multicolumn{1}{c}{(8)}         &\multicolumn{1}{c}{(9)}         &\multicolumn{1}{c}{(10)}         &\multicolumn{1}{c}{(11)}         &\multicolumn{1}{c}{(12)}         &\multicolumn{1}{c}{(13)}         &\multicolumn{1}{c}{(14)}         &\multicolumn{1}{c}{(15)}         &\multicolumn{1}{c}{(16)}         &\multicolumn{1}{c}{(17)}         &\multicolumn{1}{c}{(18)}         &\multicolumn{1}{c}{(19)}         &\multicolumn{1}{c}{(20)}         \\
\midrule
Risk aversion: $\alpha$&                  &                  &                  &                  &                  &                  &                  &                  &                  &                  &                  &                  &                  &                  &                  &                  &                  &                  &                  &                  \\
$\hat{\kappa}$&     0.21\sym{***}&     0.25\sym{***}&     0.20\sym{***}&     0.22\sym{***}&     0.21\sym{***}&     0.20\sym{***}&     0.18\sym{***}&     0.18\sym{***}&     0.16\sym{***}&     0.19\sym{***}&     0.21\sym{***}&     0.23\sym{***}&     0.17\sym{***}&     0.21\sym{***}&     0.19\sym{***}&     0.21\sym{***}&     0.23\sym{***}&     0.20\sym{***}&     0.20\sym{***}&     0.18\sym{***}\\
          &   (0.02)         &   (0.02)         &   (0.02)         &   (0.02)         &   (0.02)         &   (0.02)         &   (0.02)         &   (0.02)         &   (0.02)         &   (0.02)         &   (0.02)         &   (0.02)         &   (0.02)         &   (0.02)         &   (0.02)         &   (0.02)         &   (0.02)         &   (0.02)         &   (0.02)         &   (0.02)         \\
$\hat{\mu}$&    -0.01         &    -0.01         &    -0.03\sym{*}  &    -0.01         &    -0.03         &     0.00         &     0.00         &    -0.03\sym{*}  &    -0.00         &    -0.04\sym{*}  &    -0.03         &    -0.02         &    -0.03         &    -0.03\sym{*}  &    -0.02         &    -0.02         &    -0.05\sym{**} &    -0.02         &    -0.05\sym{**} &    -0.02         \\
          &   (0.02)         &   (0.02)         &   (0.02)         &   (0.02)         &   (0.02)         &   (0.02)         &   (0.02)         &   (0.02)         &   (0.02)         &   (0.02)         &   (0.02)         &   (0.02)         &   (0.02)         &   (0.01)         &   (0.02)         &   (0.02)         &   (0.02)         &   (0.02)         &   (0.02)         &   (0.01)         \\
Constant  &     0.43\sym{***}&     0.41\sym{***}&     0.45\sym{***}&     0.42\sym{***}&     0.43\sym{***}&     0.42\sym{***}&     0.43\sym{***}&     0.45\sym{***}&     0.43\sym{***}&     0.45\sym{***}&     0.44\sym{***}&     0.43\sym{***}&     0.45\sym{***}&     0.44\sym{***}&     0.44\sym{***}&     0.43\sym{***}&     0.45\sym{***}&     0.44\sym{***}&     0.45\sym{***}&     0.44\sym{***}\\
          &   (0.01)         &   (0.01)         &   (0.01)         &   (0.01)         &   (0.01)         &   (0.01)         &   (0.01)         &   (0.01)         &   (0.01)         &   (0.01)         &   (0.01)         &   (0.01)         &   (0.01)         &   (0.01)         &   (0.01)         &   (0.01)         &   (0.01)         &   (0.01)         &   (0.01)         &   (0.01)         \\
\midrule
Loss aversion: $\lambda$&                  &                  &                  &                  &                  &                  &                  &                  &                  &                  &                  &                  &                  &                  &                  &                  &                  &                  &                  &                  \\
$\hat{\kappa}$&    -0.26\sym{***}&    -0.23\sym{***}&    -0.25\sym{***}&    -0.33\sym{***}&    -0.17\sym{**} &    -0.32\sym{***}&    -0.27\sym{***}&    -0.30\sym{***}&    -0.40\sym{***}&    -0.29\sym{***}&    -0.24\sym{***}&    -0.14\sym{*}  &    -0.27\sym{***}&    -0.32\sym{***}&    -0.35\sym{***}&    -0.30\sym{***}&    -0.24\sym{***}&    -0.42\sym{***}&    -0.25\sym{***}&    -0.23\sym{***}\\
          &   (0.06)         &   (0.06)         &   (0.06)         &   (0.06)         &   (0.06)         &   (0.06)         &   (0.06)         &   (0.06)         &   (0.06)         &   (0.06)         &   (0.06)         &   (0.06)         &   (0.06)         &   (0.06)         &   (0.06)         &   (0.06)         &   (0.06)         &   (0.06)         &   (0.06)         &   (0.06)         \\
$\hat{\mu}$&     0.03         &     0.04         &    -0.00         &    -0.01         &     0.03         &     0.06         &     0.01         &     0.02         &     0.05         &     0.07         &     0.04         &    -0.02         &    -0.01         &     0.02         &     0.03         &     0.00         &     0.01         &     0.01         &     0.02         &     0.02         \\
          &   (0.04)         &   (0.04)         &   (0.04)         &   (0.04)         &   (0.04)         &   (0.04)         &   (0.04)         &   (0.04)         &   (0.04)         &   (0.04)         &   (0.04)         &   (0.04)         &   (0.04)         &   (0.04)         &   (0.04)         &   (0.04)         &   (0.04)         &   (0.04)         &   (0.04)         &   (0.04)         \\
Constant  &     1.85\sym{***}&     1.84\sym{***}&     1.87\sym{***}&     1.90\sym{***}&     1.82\sym{***}&     1.85\sym{***}&     1.86\sym{***}&     1.86\sym{***}&     1.90\sym{***}&     1.83\sym{***}&     1.83\sym{***}&     1.84\sym{***}&     1.89\sym{***}&     1.88\sym{***}&     1.89\sym{***}&     1.89\sym{***}&     1.86\sym{***}&     1.93\sym{***}&     1.85\sym{***}&     1.85\sym{***}\\
          &   (0.04)         &   (0.04)         &   (0.04)         &   (0.04)         &   (0.04)         &   (0.04)         &   (0.04)         &   (0.04)         &   (0.04)         &   (0.04)         &   (0.04)         &   (0.03)         &   (0.04)         &   (0.03)         &   (0.03)         &   (0.03)         &   (0.04)         &   (0.04)         &   (0.04)         &   (0.04)         \\
\midrule
Discounting: $\delta$&                  &                  &                  &                  &                  &                  &                  &                  &                  &                  &                  &                  &                  &                  &                  &                  &                  &                  &                  &                  \\
$\hat{\kappa}$&    -0.00         &    -0.06         &    -0.02         &    -0.09\sym{*}  &    -0.04         &    -0.04         &    -0.04         &    -0.08\sym{*}  &    -0.09\sym{**} &    -0.02         &    -0.01         &    -0.04         &    -0.04         &    -0.07         &     0.00         &    -0.09\sym{*}  &    -0.07\sym{*}  &    -0.12\sym{***}&     0.04         &    -0.03         \\
          &   (0.03)         &   (0.03)         &   (0.03)         &   (0.04)         &   (0.03)         &   (0.03)         &   (0.03)         &   (0.03)         &   (0.03)         &   (0.03)         &   (0.04)         &   (0.04)         &   (0.03)         &   (0.03)         &   (0.03)         &   (0.04)         &   (0.03)         &   (0.03)         &   (0.04)         &   (0.03)         \\
$\hat{\mu}$&    -0.06\sym{**} &    -0.05\sym{*}  &    -0.03         &    -0.07\sym{**} &    -0.08\sym{***}&    -0.09\sym{***}&    -0.06\sym{**} &    -0.06\sym{**} &    -0.10\sym{***}&    -0.04         &    -0.06\sym{**} &    -0.04         &    -0.07\sym{**} &    -0.08\sym{***}&    -0.07\sym{***}&    -0.05\sym{*}  &    -0.04         &    -0.07\sym{**} &    -0.02         &    -0.08\sym{***}\\
          &   (0.02)         &   (0.02)         &   (0.02)         &   (0.02)         &   (0.02)         &   (0.02)         &   (0.02)         &   (0.02)         &   (0.02)         &   (0.02)         &   (0.02)         &   (0.02)         &   (0.02)         &   (0.02)         &   (0.02)         &   (0.02)         &   (0.02)         &   (0.02)         &   (0.02)         &   (0.02)         \\
Constant  &     0.36\sym{***}&     0.37\sym{***}&     0.34\sym{***}&     0.40\sym{***}&     0.39\sym{***}&     0.40\sym{***}&     0.37\sym{***}&     0.39\sym{***}&     0.43\sym{***}&     0.36\sym{***}&     0.36\sym{***}&     0.36\sym{***}&     0.38\sym{***}&     0.40\sym{***}&     0.37\sym{***}&     0.38\sym{***}&     0.37\sym{***}&     0.42\sym{***}&     0.32\sym{***}&     0.38\sym{***}\\
          &   (0.02)         &   (0.02)         &   (0.02)         &   (0.02)         &   (0.02)         &   (0.02)         &   (0.02)         &   (0.02)         &   (0.02)         &   (0.02)         &   (0.02)         &   (0.02)         &   (0.02)         &   (0.02)         &   (0.02)         &   (0.02)         &   (0.02)         &   (0.02)         &   (0.02)         &   (0.02)         \\
\midrule
Present bias: $\gamma$&                  &                  &                  &                  &                  &                  &                  &                  &                  &                  &                  &                  &                  &                  &                  &                  &                  &                  &                  &                  \\
$\hat{\kappa}$&    -0.03\sym{*}  &    -0.02         &    -0.04\sym{*}  &     0.02         &    -0.00         &    -0.02         &     0.01         &     0.01         &     0.02         &    -0.02         &    -0.04\sym{*}  &    -0.00         &    -0.01         &    -0.01         &    -0.01         &     0.00         &    -0.02         &     0.01         &    -0.04\sym{**} &    -0.01         \\
          &   (0.01)         &   (0.01)         &   (0.01)         &   (0.01)         &   (0.01)         &   (0.01)         &   (0.01)         &   (0.01)         &   (0.01)         &   (0.01)         &   (0.01)         &   (0.01)         &   (0.01)         &   (0.01)         &   (0.01)         &   (0.01)         &   (0.01)         &   (0.01)         &   (0.01)         &   (0.01)         \\
$\hat{\mu}$&     0.01         &    -0.00         &     0.01         &     0.01         &     0.03\sym{**} &     0.02         &     0.01         &     0.01         &     0.03\sym{**} &     0.01         &     0.00         &     0.00         &     0.01         &     0.02\sym{*}  &     0.01         &     0.01         &     0.01         &     0.01         &     0.00         &     0.01         \\
          &   (0.01)         &   (0.01)         &   (0.01)         &   (0.01)         &   (0.01)         &   (0.01)         &   (0.01)         &   (0.01)         &   (0.01)         &   (0.01)         &   (0.01)         &   (0.01)         &   (0.01)         &   (0.01)         &   (0.01)         &   (0.01)         &   (0.01)         &   (0.01)         &   (0.01)         &   (0.01)         \\
Constant  &     0.01         &     0.02         &     0.02         &    -0.01         &    -0.01         &     0.00         &     0.00         &    -0.00         &    -0.02\sym{*}  &     0.01         &     0.02\sym{*}  &     0.01         &    -0.00         &    -0.00         &     0.00         &    -0.00         &     0.00         &    -0.01         &     0.02\sym{**} &     0.00         \\
          &   (0.01)         &   (0.01)         &   (0.01)         &   (0.01)         &   (0.01)         &   (0.01)         &   (0.01)         &   (0.01)         &   (0.01)         &   (0.01)         &   (0.01)         &   (0.01)         &   (0.01)         &   (0.01)         &   (0.01)         &   (0.01)         &   (0.01)         &   (0.01)         &   (0.01)         &   (0.01)         \\
\midrule
Tremble error: $\kappa$&                  &                  &                  &                  &                  &                  &                  &                  &                  &                  &                  &                  &                  &                  &                  &                  &                  &                  &                  &                  \\
$\hat{\kappa}$&     4.58\sym{***}&     4.22\sym{***}&     4.70\sym{***}&     4.38\sym{***}&     4.53\sym{***}&     4.68\sym{***}&     4.72\sym{***}&     4.81\sym{***}&     4.85\sym{***}&     4.83\sym{***}&     4.69\sym{***}&     4.54\sym{***}&     5.06\sym{***}&     4.53\sym{***}&     4.77\sym{***}&     4.53\sym{***}&     4.25\sym{***}&     4.61\sym{***}&     4.87\sym{***}&     4.96\sym{***}\\
          &   (0.22)         &   (0.22)         &   (0.25)         &   (0.24)         &   (0.22)         &   (0.24)         &   (0.22)         &   (0.22)         &   (0.22)         &   (0.22)         &   (0.23)         &   (0.24)         &   (0.22)         &   (0.23)         &   (0.23)         &   (0.23)         &   (0.22)         &   (0.23)         &   (0.22)         &   (0.22)         \\
$\hat{\mu}$&     1.01\sym{***}&     0.93\sym{***}&     1.19\sym{***}&     1.10\sym{***}&     1.16\sym{***}&     0.87\sym{***}&     0.88\sym{***}&     1.20\sym{***}&     0.74\sym{***}&     1.32\sym{***}&     1.14\sym{***}&     1.05\sym{***}&     1.18\sym{***}&     1.21\sym{***}&     1.08\sym{***}&     1.10\sym{***}&     1.32\sym{***}&     1.15\sym{***}&     1.26\sym{***}&     1.14\sym{***}\\
          &   (0.15)         &   (0.16)         &   (0.16)         &   (0.17)         &   (0.16)         &   (0.16)         &   (0.18)         &   (0.16)         &   (0.16)         &   (0.16)         &   (0.17)         &   (0.16)         &   (0.16)         &   (0.15)         &   (0.16)         &   (0.16)         &   (0.17)         &   (0.17)         &   (0.16)         &   (0.15)         \\
Constant  &     0.14         &     0.40\sym{**} &     0.01         &     0.23         &     0.12         &     0.23         &     0.20         &    -0.05         &     0.31\sym{*}  &    -0.12         &     0.03         &     0.16         &    -0.06         &     0.10         &     0.10         &     0.15         &     0.04         &     0.11         &    -0.06         &    -0.01         \\
          &   (0.12)         &   (0.13)         &   (0.13)         &   (0.14)         &   (0.12)         &   (0.13)         &   (0.12)         &   (0.12)         &   (0.12)         &   (0.12)         &   (0.12)         &   (0.13)         &   (0.12)         &   (0.12)         &   (0.12)         &   (0.12)         &   (0.12)         &   (0.13)         &   (0.11)         &   (0.12)         \\
\midrule
\(N\)     &   420000         &   420000         &   420000         &   420000         &   420000         &   420000         &   420000         &   420000         &   420000         &   420000         &   420000         &   420000         &   420000         &   420000         &   420000         &   420000         &   420000         &   420000         &   420000         &   420000         \\
\bottomrule
\multicolumn{21}{l}{\footnotesize Standard errors in parentheses}\\
\multicolumn{21}{l}{\footnotesize \sym{*} \(p<0.05\), \sym{**} \(p<0.01\), \sym{***} \(p<0.001\)}\\
\end{tabular}
}

\renewcommand{\footnotesize}{\footnotesize}
\end{sidewaystable}

\begin{sidewaystable}
\tabcolsep1.2mm
\tiny
\caption{Simulations: number of times option A is chosen \label{tab:sim_reg}}
\renewcommand{\footnotesize}{\tiny}
{
\def\sym#1{\ifmmode^{#1}\else\(^{#1}\)\fi}
\begin{tabular}{l*{20}{c}}
\toprule
          &\multicolumn{1}{c}{(1)}         &\multicolumn{1}{c}{(2)}         &\multicolumn{1}{c}{(3)}         &\multicolumn{1}{c}{(4)}         &\multicolumn{1}{c}{(5)}         &\multicolumn{1}{c}{(6)}         &\multicolumn{1}{c}{(7)}         &\multicolumn{1}{c}{(8)}         &\multicolumn{1}{c}{(9)}         &\multicolumn{1}{c}{(10)}         &\multicolumn{1}{c}{(11)}         &\multicolumn{1}{c}{(12)}         &\multicolumn{1}{c}{(13)}         &\multicolumn{1}{c}{(14)}         &\multicolumn{1}{c}{(15)}         &\multicolumn{1}{c}{(16)}         &\multicolumn{1}{c}{(17)}         &\multicolumn{1}{c}{(18)}         &\multicolumn{1}{c}{(19)}         &\multicolumn{1}{c}{(20)}         \\
\midrule
Risk aversion:\\
MPL2\\
$\hat{\kappa}$&    -2.17\sym{***}&    -2.10\sym{***}&    -2.12\sym{***}&    -1.94\sym{***}&    -2.09\sym{***}&    -2.06\sym{***}&    -2.22\sym{***}&    -2.30\sym{***}&    -2.36\sym{***}&    -2.28\sym{***}&    -2.30\sym{***}&    -2.18\sym{***}&    -2.28\sym{***}&    -2.12\sym{***}&    -2.21\sym{***}&    -2.07\sym{***}&    -2.11\sym{***}&    -2.27\sym{***}&    -2.46\sym{***}&    -2.31\sym{***}\\
          &   (0.13)         &   (0.14)         &   (0.14)         &   (0.14)         &   (0.13)         &   (0.13)         &   (0.14)         &   (0.14)         &   (0.14)         &   (0.14)         &   (0.14)         &   (0.14)         &   (0.14)         &   (0.13)         &   (0.13)         &   (0.13)         &   (0.14)         &   (0.13)         &   (0.13)         &   (0.13)         \\
$\hat{\mu}$&    -0.76\sym{***}&    -0.63\sym{***}&    -0.70\sym{***}&    -0.67\sym{***}&    -0.79\sym{***}&    -0.61\sym{***}&    -0.72\sym{***}&    -0.78\sym{***}&    -0.75\sym{***}&    -0.77\sym{***}&    -0.82\sym{***}&    -0.73\sym{***}&    -0.65\sym{***}&    -0.85\sym{***}&    -0.78\sym{***}&    -0.80\sym{***}&    -0.88\sym{***}&    -0.71\sym{***}&    -0.87\sym{***}&    -0.72\sym{***}\\
          &   (0.08)         &   (0.09)         &   (0.09)         &   (0.08)         &   (0.08)         &   (0.09)         &   (0.09)         &   (0.09)         &   (0.09)         &   (0.08)         &   (0.09)         &   (0.09)         &   (0.09)         &   (0.08)         &   (0.09)         &   (0.09)         &   (0.09)         &   (0.08)         &   (0.08)         &   (0.08)         \\
Constant  &     9.76\sym{***}&     9.61\sym{***}&     9.68\sym{***}&     9.57\sym{***}&     9.71\sym{***}&     9.58\sym{***}&     9.75\sym{***}&     9.81\sym{***}&     9.78\sym{***}&     9.81\sym{***}&     9.83\sym{***}&     9.72\sym{***}&     9.69\sym{***}&     9.78\sym{***}&     9.76\sym{***}&     9.74\sym{***}&     9.82\sym{***}&     9.75\sym{***}&     9.93\sym{***}&     9.76\sym{***}\\
          &   (0.08)         &   (0.08)         &   (0.08)         &   (0.08)         &   (0.08)         &   (0.08)         &   (0.08)         &   (0.08)         &   (0.08)         &   (0.08)         &   (0.08)         &   (0.08)         &   (0.08)         &   (0.08)         &   (0.08)         &   (0.08)         &   (0.08)         &   (0.08)         &   (0.08)         &   (0.08)         \\

\midrule
Loss aversion:\\
MPL3\\

$\hat{\kappa}$&     0.20         &     0.20         &     0.07         &    -0.05         &     0.22\sym{*}  &     0.10         &     0.06         &     0.09         &    -0.05         &     0.10         &     0.16         &     0.22\sym{*}  &     0.12         &     0.05         &    -0.10         &     0.09         &     0.17         &    -0.06         &     0.17         &     0.18         \\
          &   (0.10)         &   (0.10)         &   (0.10)         &   (0.10)         &   (0.10)         &   (0.10)         &   (0.10)         &   (0.10)         &   (0.10)         &   (0.10)         &   (0.10)         &   (0.10)         &   (0.10)         &   (0.10)         &   (0.10)         &   (0.10)         &   (0.10)         &   (0.10)         &   (0.10)         &   (0.10)         \\
$\hat{\mu}$&     0.07         &     0.10         &    -0.03         &    -0.01         &     0.04         &     0.11         &     0.07         &     0.05         &     0.05         &     0.10         &     0.07         &     0.01         &     0.01         &     0.07         &     0.06         &     0.00         &     0.02         &     0.06         &     0.05         &     0.08         \\
          &   (0.06)         &   (0.07)         &   (0.06)         &   (0.06)         &   (0.06)         &   (0.07)         &   (0.07)         &   (0.07)         &   (0.06)         &   (0.06)         &   (0.06)         &   (0.06)         &   (0.06)         &   (0.06)         &   (0.06)         &   (0.06)         &   (0.07)         &   (0.06)         &   (0.07)         &   (0.06)         \\
Constant  &     3.30\sym{***}&     3.30\sym{***}&     3.43\sym{***}&     3.46\sym{***}&     3.33\sym{***}&     3.32\sym{***}&     3.35\sym{***}&     3.35\sym{***}&     3.43\sym{***}&     3.32\sym{***}&     3.32\sym{***}&     3.34\sym{***}&     3.39\sym{***}&     3.37\sym{***}&     3.44\sym{***}&     3.41\sym{***}&     3.35\sym{***}&     3.41\sym{***}&     3.34\sym{***}&     3.32\sym{***}\\
          &   (0.06)         &   (0.06)         &   (0.06)         &   (0.06)         &   (0.06)         &   (0.06)         &   (0.06)         &   (0.06)         &   (0.06)         &   (0.06)         &   (0.06)         &   (0.06)         &   (0.06)         &   (0.06)         &   (0.06)         &   (0.06)         &   (0.06)         &   (0.06)         &   (0.06)         &   (0.06)         \\

\midrule
Discounting:\\
MPL1.1+MPL1.2\\

$\hat{\kappa}$&    -1.23\sym{***}&    -1.34\sym{***}&    -1.42\sym{***}&    -1.26\sym{***}&    -1.27\sym{***}&    -1.44\sym{***}&    -1.25\sym{***}&    -1.48\sym{***}&    -1.62\sym{***}&    -1.37\sym{***}&    -1.45\sym{***}&    -1.20\sym{***}&    -1.43\sym{***}&    -1.49\sym{***}&    -1.18\sym{***}&    -1.47\sym{***}&    -1.47\sym{***}&    -1.68\sym{***}&    -1.23\sym{***}&    -1.33\sym{***}\\
          &   (0.15)         &   (0.15)         &   (0.15)         &   (0.16)         &   (0.15)         &   (0.15)         &   (0.15)         &   (0.15)         &   (0.15)         &   (0.15)         &   (0.15)         &   (0.15)         &   (0.15)         &   (0.15)         &   (0.15)         &   (0.16)         &   (0.15)         &   (0.15)         &   (0.15)         &   (0.15)         \\
$\hat{\mu}$&    -0.66\sym{***}&    -0.71\sym{***}&    -0.52\sym{***}&    -0.69\sym{***}&    -0.69\sym{***}&    -0.73\sym{***}&    -0.64\sym{***}&    -0.74\sym{***}&    -0.71\sym{***}&    -0.62\sym{***}&    -0.82\sym{***}&    -0.61\sym{***}&    -0.76\sym{***}&    -0.79\sym{***}&    -0.75\sym{***}&    -0.60\sym{***}&    -0.64\sym{***}&    -0.79\sym{***}&    -0.61\sym{***}&    -0.81\sym{***}\\
          &   (0.10)         &   (0.09)         &   (0.10)         &   (0.10)         &   (0.09)         &   (0.10)         &   (0.10)         &   (0.10)         &   (0.10)         &   (0.09)         &   (0.09)         &   (0.10)         &   (0.10)         &   (0.09)         &   (0.09)         &   (0.10)         &   (0.10)         &   (0.09)         &   (0.10)         &   (0.10)         \\
Constant  &     8.80\sym{***}&     8.87\sym{***}&     8.79\sym{***}&     8.84\sym{***}&     8.83\sym{***}&     8.94\sym{***}&     8.77\sym{***}&     8.95\sym{***}&     8.98\sym{***}&     8.84\sym{***}&     8.99\sym{***}&     8.73\sym{***}&     8.95\sym{***}&     8.98\sym{***}&     8.85\sym{***}&     8.82\sym{***}&     8.88\sym{***}&     9.06\sym{***}&     8.76\sym{***}&     8.92\sym{***}\\
          &   (0.09)         &   (0.09)         &   (0.09)         &   (0.10)         &   (0.09)         &   (0.10)         &   (0.10)         &   (0.09)         &   (0.09)         &   (0.09)         &   (0.10)         &   (0.09)         &   (0.09)         &   (0.09)         &   (0.09)         &   (0.09)         &   (0.09)         &   (0.09)         &   (0.09)         &   (0.09)         \\

\midrule
Present bias:\\
MPL1.2-MPL1.1\\

$\hat{\kappa}$&    -0.39\sym{**} &    -0.49\sym{***}&    -0.65\sym{***}&    -0.01         &    -0.22         &    -0.45\sym{***}&    -0.18         &    -0.20         &    -0.09         &    -0.35\sym{*}  &    -0.48\sym{***}&    -0.21         &    -0.30\sym{*}  &    -0.35\sym{**} &    -0.31\sym{*}  &    -0.16         &    -0.42\sym{**} &    -0.08         &    -0.60\sym{***}&    -0.27\sym{*}  \\
          &   (0.14)         &   (0.14)         &   (0.14)         &   (0.14)         &   (0.14)         &   (0.14)         &   (0.14)         &   (0.13)         &   (0.13)         &   (0.14)         &   (0.14)         &   (0.14)         &   (0.14)         &   (0.14)         &   (0.14)         &   (0.14)         &   (0.14)         &   (0.14)         &   (0.14)         &   (0.14)         \\
$\hat{\mu}$&    -0.00         &    -0.13         &     0.00         &     0.04         &     0.15         &     0.05         &    -0.02         &    -0.02         &     0.11         &    -0.01         &    -0.04         &    -0.07         &     0.04         &     0.11         &    -0.03         &     0.08         &     0.02         &    -0.03         &    -0.06         &     0.02         \\
          &   (0.09)         &   (0.08)         &   (0.09)         &   (0.09)         &   (0.09)         &   (0.09)         &   (0.09)         &   (0.09)         &   (0.09)         &   (0.08)         &   (0.09)         &   (0.09)         &   (0.09)         &   (0.08)         &   (0.09)         &   (0.09)         &   (0.09)         &   (0.09)         &   (0.09)         &   (0.09)         \\
Constant  &     0.30\sym{***}&     0.44\sym{***}&     0.41\sym{***}&     0.12         &     0.14         &     0.32\sym{***}&     0.28\sym{***}&     0.24\sym{**} &     0.10         &     0.31\sym{***}&     0.37\sym{***}&     0.30\sym{***}&     0.24\sym{**} &     0.22\sym{**} &     0.32\sym{***}&     0.16         &     0.30\sym{***}&     0.18\sym{*}  &     0.45\sym{***}&     0.24\sym{**} \\
          &   (0.08)         &   (0.08)         &   (0.08)         &   (0.08)         &   (0.08)         &   (0.08)         &   (0.08)         &   (0.08)         &   (0.08)         &   (0.08)         &   (0.08)         &   (0.08)         &   (0.08)         &   (0.08)         &   (0.08)         &   (0.08)         &   (0.08)         &   (0.08)         &   (0.08)         &   (0.08)         \\
\midrule
\(N\)     &    12000         &    12000         &    12000         &    12000         &    12000         &    12000         &    12000         &    12000         &    12000         &    12000         &    12000         &    12000         &    12000         &    12000         &    12000         &    12000         &    12000         &    12000         &    12000         &    12000         \\
\bottomrule
\multicolumn{21}{l}{\footnotesize Standard errors in parentheses}\\
\multicolumn{21}{l}{\footnotesize \sym{*} \(p<0.05\), \sym{**} \(p<0.01\), \sym{***} \(p<0.001\)}\\
\end{tabular}
}

\renewcommand{\footnotesize}{\footnotesize}
\end{sidewaystable}

\end{appendices}

\end{document}